\documentclass[journal]{vgtc}                     

\ifpdf
  \pdfoutput=1\relax                   
	\usepackage{dblfloatfix}    
  \usepackage{graphicx}                
  \DeclareGraphicsExtensions{.pdf,.png,.jpg,.jpeg} 
\else
  \usepackage{graphicx}                
  \DeclareGraphicsExtensions{.eps}     
\fi%
\graphicspath{{figures/}{pictures/}{images/}{./images/}} 
\usepackage{microtype}                 
\PassOptionsToPackage{warn}{textcomp}  
\usepackage{textcomp}                  
\usepackage{mathptmx}                  
\usepackage{amsfonts}
\usepackage{times}                     
\usepackage{cite}                      
\usepackage{tabu}                      
\usepackage{booktabs}                  
\usepackage{makecell}
\usepackage{xspace}
\usepackage{enumitem}
\usepackage{amsmath}
\usepackage{subfiles}
\usepackage{ccicons}
\usepackage{textcomp}
\usepackage{multirow}
\usepackage{makecell}

\usepackage{cuted}
\usepackage{overpic}

\usepackage{subfigure}
\usepackage[dvipsnames,svgnames]{xcolor}
\newcommand{\method}{SpatialTouch\xspace}
\newcommand{\etal}{et~al.\xspace}
\newcommand{\eg}{e.\,g.}

\definecolor{colorbelow}{HTML}{926060}
\definecolor{colorabove}{HTML}{6294AD}
\definecolor{coloracross}{HTML}{724C83}
\newcommand{\belowS}{\textbf{\textcolor{colorbelow}{Below}}\xspace}
\newcommand{\aboveS}{\textbf{\textcolor{colorabove}{Above}}\xspace}
\newcommand{\acrossS}{\textbf{\textcolor{coloracross}{Across}}\xspace}

\newcommand{\www}[1]{\textbf{\textcolor{coloracross}{#1}}}
\renewcommand{\www}[1]{#1}

\onlineid{1626}

\vgtccategory{Research}
\vgtcpapertype{Representations/Interaction}

\title{\method: Exploring Spatial Data Visualizations in Cross-reality}

\author{%
   \authororcid{Lixiang Zhao}{0000-0001-6181-1673},
   \authororcid{Tobias Isenberg}{0000-0001-7953-8644},
   \authororcid{Fuqi Xie}{0009-0008-4728-9346},
   \authororcid{Hai-Ning Liang}{0000-0003-3600-8955}, 
   \authororcid{Lingyun Yu}{0000-0002-3152-2587}
}

\authorfooter{
\item  L. Zhao (\raisebox{-.5pt}{\includegraphics[height=6pt]{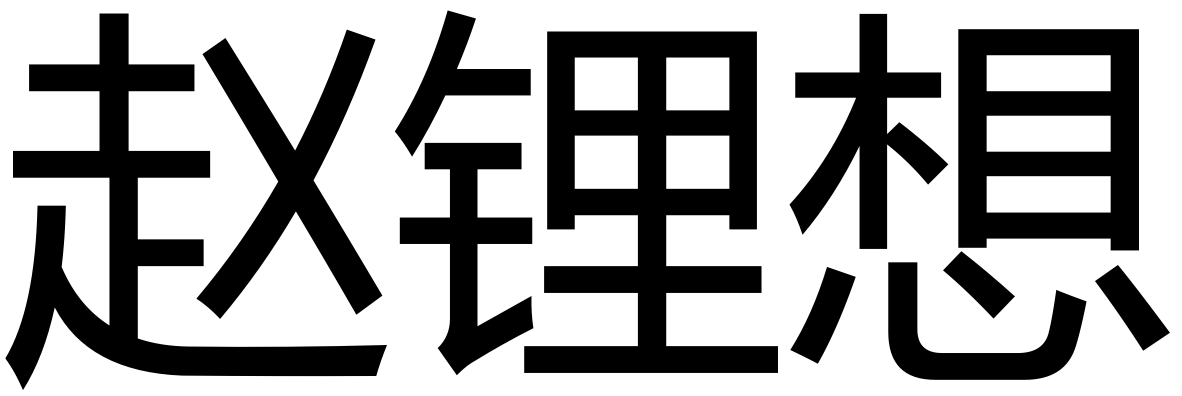}}), F. Xie (\raisebox{-.5pt}{\includegraphics[height=6pt]{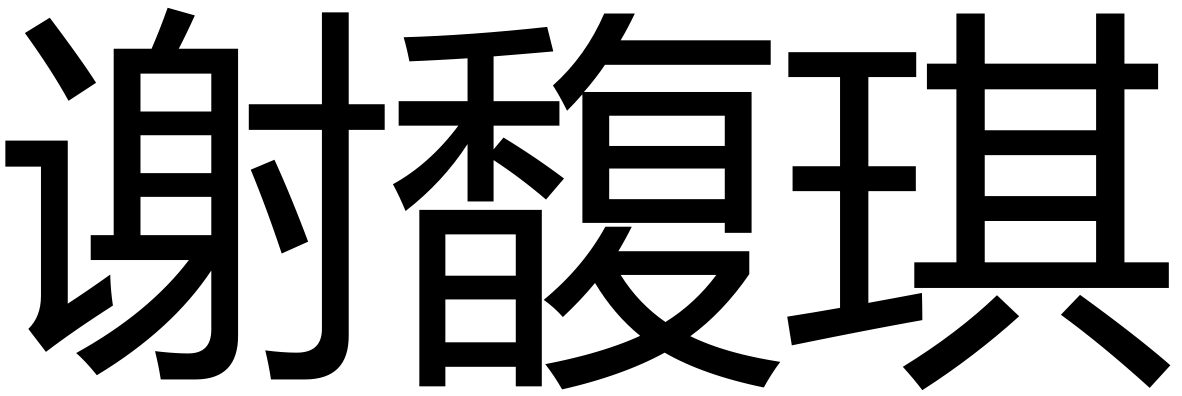}}), and L. Yu (\raisebox{-.5pt}{\includegraphics[height=6pt]{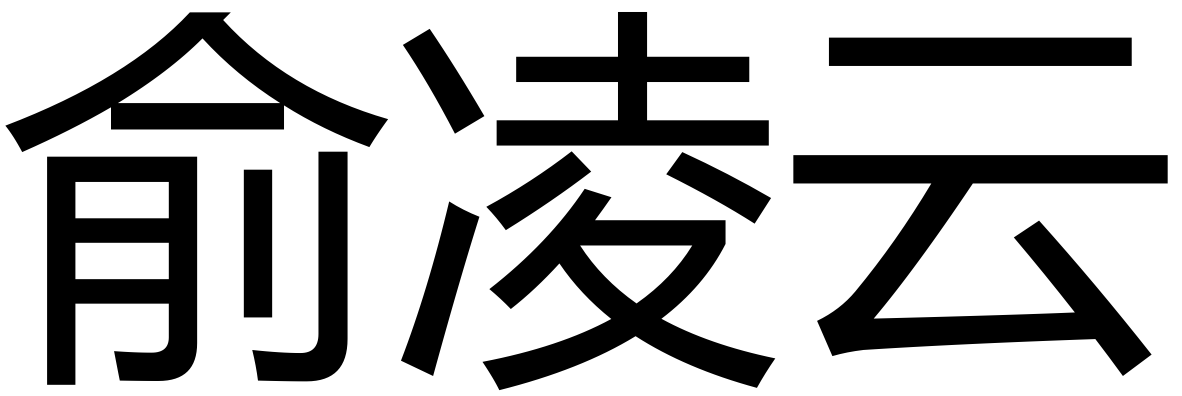}}) are with Xi'an Jiaotong- Liverpool University, China. E-mail:{\{lixiang.zhao17@student $|$ \textls[-10]fuqi.xie20 @student $|$ \textls[-10]lingyun.yu@\}.xjtlu.edu.cn}. L. Yu is the corresponding author.
\item  T. Isenberg is with Universit{\'e} Paris-Saclay, CNRS, Inria, LISN, France. E-mail: given\_name.family\_name@inria.fr.
\item H.-N. Liang (\raisebox{-.5pt}{\includegraphics[height=6pt]{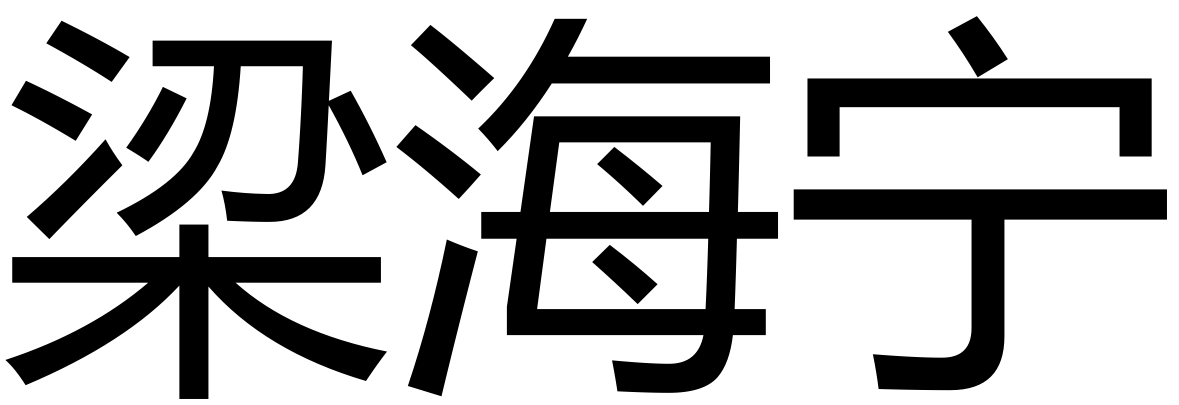}}) is with Hong Kong University of Science and Technology (Guangzhou), China. E-mail: hainingliang@hkust-gz.edu.cn.
}

\shortauthortitle{Biv \MakeLowercase{\textit{et al.}}: Global Illumination for Fun and Profit}

\abstract{%
We propose and study a novel cross-reality environment that seamlessly integrates a monoscopic 2D surface (an interactive screen with touch and pen input) with a stereoscopic 3D space (an augmented reality HMD) to jointly host spatial data visualizations. This innovative approach combines the best of two conventional methods of displaying and manipulating spatial 3D data, enabling users to fluidly explore diverse visual forms using tailored interaction techniques. Providing such effective 3D data exploration techniques is pivotal for conveying its intricate spatial structures---often at multiple spatial or semantic scales---across various application domains and requiring diverse visual representations for effective visualization. To understand user reactions to our new environment, we began with an elicitation user study, in which we captured their responses and interactions. We observed that users adapted their interaction approaches based on perceived visual representations, with natural transitions in spatial awareness and actions while navigating across the physical surface. Our findings then informed the development of a design space for spatial data exploration in cross-reality. We thus developed cross-reality environments tailored to three distinct domains: for 3D molecular structure data, for 3D point cloud data, and for 3D anatomical data. In particular, we designed interaction techniques that account for the inherent features of interactions in both spaces, facilitating various forms of interaction, including mid-air gestures, touch interactions, pen interactions, and combinations thereof, to enhance the users' sense of presence and engagement. We assessed the usability of our environment with biologists, focusing on its use for domain research. In addition, we evaluated our interaction transition designs with virtual and mixed-reality experts to gather further insights. As a result, we provide our design suggestions for the cross-reality environment, emphasizing the interaction with diverse visual representations and seamless interaction transitions between 2D and 3D spaces.

} 

\keywords{Spatial data, immersive visualization, cross reality, interaction techniques.\vspace{-1pt}}

\CCScatlist{ 
 \CCScat{K.6.1}{Management of Computing and Information Systems}%
{Project and People Management}{Life Cycle};
 \CCScat{K.7.m}{The Computing Profession}{Miscellaneous}{Ethics}
}

\teaser{%
    \centering%
	{\footnotesize\begin{overpic}[width=\linewidth]{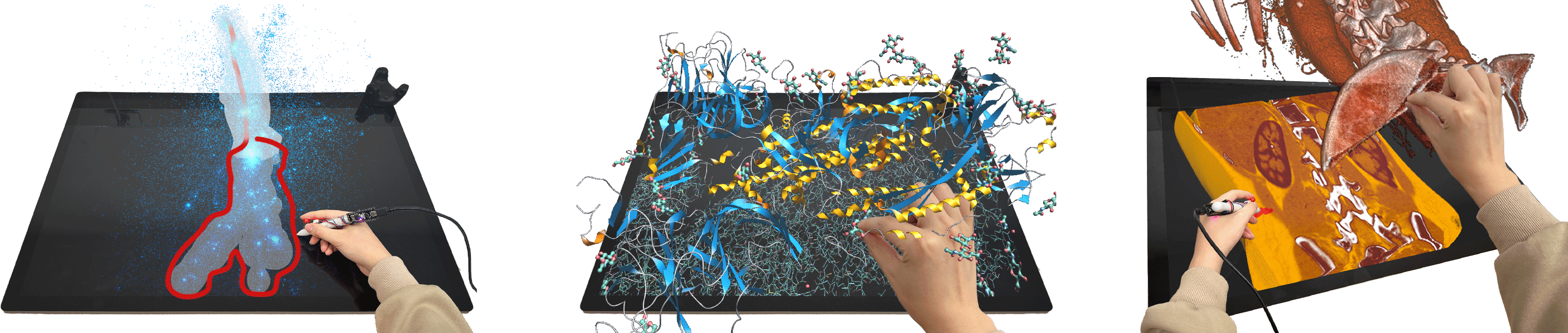}
		\put(1.0,2.0){\textcolor{white}{\textbf{(a)}}}
		\put(38.0,2.0){\textcolor{white}{\textbf{(b)}}}
		\put(73.5,2.0){\textcolor{white}{\textbf{(c)}}}
	\end{overpic}}\vspace{-2pt}
 \caption{Data exploration in \method: (a) select astronomical points, (b) navigate molecular visualization, (c) annotate medical data.}
 \label{fig:teaser}
}

\begin{document}

\firstsection{Introduction}

\maketitle

As computing and simulation technologies have advanced, the complexity of 3D structures within scientific datasets, alongside their contextual information, has grown substantially. Approaches to interactively explore 3D data have emerged to help users understand complex spatial structures across various disciplines. These datasets often encompass elaborate details, comprising millions of intricate components such as unstructured points in astronomical simulations \cite{springel:2008:aquarius,Springel:2005:SimulationsOT}, flood simulation\cite{boorboor:2023:submerse} or biological structures spanning multiple scales \cite{Garrison:2022:TOV}. To gain a deeper understanding of such complex data, researchers often need to selectively focus on different regions. They may adjust visualization parameters to filter the data, emphasize key features, or alter data representations to highlight specific data attributes. This exploration can involve switching between 2D and 3D representations or displaying a combination of both for an in-depth analysis of data structures.  
Abstractocyte \cite{Mohammed:2018:Abstractocyte}, \eg, is an interactive system designed to assist biologists in exploring the morphology of astrocytes using various levels of visual abstraction, while simultaneously analyzing neighboring neurons and their connectivity. This tool allows users to manipulate a visualization widget of 2D abstraction space to smoothly transition between the different 3D and 2D abstraction levels, to identify morphological features for cells of interest. 
Furthermore, various visual representations, such as ball-and-stick models, are employed in molecular visualization to display 3D structures with different data features\cite{O'donoghue:2010:VMS,kutak:2023:state-of-art-Molecular,van:2011:IMV}.

A common approach for supporting researchers to observe diverse combinations of data features, whether in 2D \cite{Becker:1987:BS,Buja:1991:IDV,Wills:2008:LDV} or in 3D \cite{Meuschke:2017:CVV, Schafer:2024:InVADo}, is through multiple linked views, facilitating connections among different representations. An alternative approach is to integrate 2D and 3D visual representations within a unified view \cite{hong:2024:2D3D}. Achieving a seamless interpretation of visual transitions and interaction with the data in such combined representations, however, is challenging---especially when dealing with complex data that contains a lot of detail. These challenges prompted us to explore a different display environment for presenting complex 3D data, one that integrates various visual representations and enables users to work with and seamlessly switch between different representations in a single setting.

To this end, we introduce \method---a cross-reality (CR) environment \cite{riegler:2020:CVVIC} that merges a monoscopic interactive 2D surface (an interactive screen with touch and pen input) with a stereoscopic 3D space (an augmented reality HMD). We can visualize data simultaneously on both the 2D surface and in the 3D space such that the depiction on both displays remains in sync. Rather than imposing restrictions on which visual representations should be displayed in each space, we approach our setup as an integrated system in which the user decides where to display the data and how to interact with it. The integration of both views facilitates smooth transitions for data visualization between the 2D surface and the 3D space, ensuring a continuous situational awareness. 
With our CR \method environment, we establish a platform that is capable of accommodating a wide range of data types that can be displayed and interactively explored as needed. While creating many possibilities for visualization and interaction design, our setup also poses questions on how to best organize the data exploration.

A central question when hosting data visualizations within \method is how interaction techniques should be designed, ensuring that users can seamlessly continue their tasks, without having to consider how interactions should be performed in the respective other display space. Touch-enabled screens facilitate direct input on the 2D screen that typically shows projected views of the spatial 3D data; and many interaction techniques have been developed for such settings (\eg, for navigation \cite{Martinet:2012:ISM, Song:2016:HPA, Wang:2019:ATD, Yu:2010:FI3D}, selection \cite{Olwal:2008:RTP, Yu:2012:ESA, Yu:2016:CEE}, positioning \cite{Butkiewicz:2019:MTP, Strothoff:2011:TriangleCursor}). 
While the exploration of 3D spatial visualizations via a 2D view can pose challenges (\eg, it often requires users to rotate the data to explore its 3D aspects), due to the ubiquity of monoscopic screens, people have already developed a solid mental concept of how interactions are performed on the 2D screen. At the same time, however, they have also established a solid understanding of how gesture interactions are performed in the 3D space. When a visualization spans both of these spaces as with \method, the question arises whether people can maintain their spatial awareness and are able to seamlessly transition between the different interaction techniques in the hybrid environment.
To clarify, ``spatial awareness'' in our context refers to users' understanding of both the spatial position of interactions within the CR environment and the spatial status of the data, including its structure, orientation, and position, which guides them in performing specific actions.

Another crucial question is, when users explore spatial data, how they react as they engage with diverse data representations within a CR environment. We can assume that people decide how interactions should be executed based on their perception and understanding of the specific data structures. When selecting subsets of unstructured point cloud data, \eg, individuals may circle around them to define a range, whereas when selecting string-like structures they may brush along the structure. Any interaction pattern, however, may undergo additional alterations when the data transitions from 3D space to the 2D surface or vice versa. When selecting point cloud data which is rendered partially on the 2D screen and partially in 3D space, for instance, people may realize at some point during the selection action they can no longer circle around the target points in the depth direction. Consequently, the question arises if people maintain or change their interaction strategies when a visual representation is used to depict data across both spaces.

Other important considerations when merging 2D and 3D display spaces and constructing a CR environment include the challenge of effectively presenting 2D/3D visualizations and determining the appropriate level of detail. We can often choose from a diverse set of visual representations that each focus on a particular set of features or contain various scales or visual abstractions, which in our case may be partially rendered stereoscopically in 3D space and partially projected on the 2D screen. We thus ask: how should we represent the data such that users gain a comprehensive understanding of it? 
Moreover, understanding the best specific setup of our CR environment (in particular, the optimal angle or angle range for the 2D surface) is crucial to facilitate the user's interaction experience. In summary, we contribute
\begin{itemize}[nosep]
\item a CR environment that combines a monoscopic 2D surface with a stereoscopic 3D space to facilitate joint spatial data exploration;
\item a study on user actions and interaction strategies while interacting with spatial data;
\item a design space of how CR supports 3D spatial data analysis across various visual representations, scales, and data abstractions; and
\item three uses cases for demonstrating data manipulation, selection, annotation and measurements in three distinct domains.  
\end{itemize}

\section{Related work}
Our work focuses on three key aspects of CR environments: the environment itself, data visualization, and interaction techniques employed within these CR settings. We review related work on these topics next.

\subsection{Cross-reality environments}
The reality-virtuality continuum, as conceptualized by Milgram and Kishino\cite{milgram:1995:reality-virtualitycontinuum}, provides a framework for data analysis and visualization techniques. It spans from visual analytics on 2D displays, to stereoscopic visualizations with immersive technologies, including augmented reality (AR), augmented virtuality (AV), and virtual \mbox{reality} (VR). Cross-reality \cite{riegler:2020:CVVIC} emphasizes seamless integrations and \mbox{transitions} among these visualization environments, offering users effective visual and algorithmic assistance tailored for maximal cognitive and perceptual suitability based on data, tasks, and user requirements.

Recent surveys have extensively explored cross-reality/virtuality environments from various perspectives, such as design and human factors\cite{ccoltekin:2020:ERS}, visualization and interaction techniques \cite{besanccon:2021:state-of-art} and collaborative analysis\cite{sereno:2020:CWAR}. Fr{\"o}hler \etal~\cite{frohler:2022:survey_of_Cross-Virtuality} and Auda \etal~\cite{auda:2023:scoping} conducted comprehensive reviews of existing work, classifying the literature regarding different stages in the reality-virtuality continuum, visualization and view transition techniques, collaboration, visualization and visual analytics techniques, evaluation methods, and application domains. These opportunities and challenges have been recognized by many researchers and extensively discussed in conference workshops\cite{Liang:2023:MRworkshopVR, Liang:2023:CRworkshopISMAR,riegler:2020:CVVIC}.

The main purposes of CR environments for domain research span from observation to collaborative analysis. An initial CR prototype was created by Kijima and Ojika \cite{kijima:1997:transition} based on a projective see-through HMD (PHMD) and 2D monitor, demonstrating how \textbf{object manipulation} can be performed using keyboard and mouse. 
The most obvious benefit of combined 2D and 3D environments is the inherent display environments for \textbf{observing 2D and 3D data representations}. Seraji and Stuerzlinger \cite{seraji:2022:hybridaxes} introduced an immersive visual analytics tool that allows users to conduct analysis at either end of the reality-virtuality continuum. It demonstrates that users can experience a lower cognitive load while viewing both 2D and 3D representations from two environments and task-switching between these virtuality modes.
CR environments are also employed for \textbf{displaying and connecting different views}. 
Reipschl{\"a}ger \etal~\cite{reipschlager:2020:PAR} proposed the combination of large interactive displays with personal head-mounted AR for displaying information in various views to facilitate data analysis. Their approach demonstrates how CR can be designed to address challenges encountered in solely large displays, such as perception, multi-user support, and managing data density and complexity.
Such environments also show great benefits in making the best use of 2D and 3D interfaces and providing users with familiar ways \textbf{perceiving, creating and manipulating 3D contents}. 
Reipschl{\"a}ger and Dachselt presented an augmented design workstation \cite{Reipschlager:2019:DesignAR}, which seamlessly integrated an interactive surface displaying 2D views with a stereoscopic AR HMD, demonstrating how this combined space facilitates 3D model creation.  
For facilitating complex data exploration, decision making, and \textbf{collaborative analysis}, 
Butscher \etal \cite{butscher:2018:CTO} investigated the combination of AR and touch-sensitive tabletops for collaborative multi-dimensional data analysis through 3D parallel coordinates, using established touch interactions with visualizations anchored to the tabletop.

All these purposes for using CR heavily inspired the design of our own CR environment, including the selection of devices, layout, and setup. Typical such devices encompass 2D desktops, tablet/mobile devices, tabletop, wall displays, CAVEs, as well as AR and VR HMDs. We discuss these devices and the implication of their use in detail related to our intended visualization tasks and domains in \autoref{sec:discussion}.

\subsection{Visualization and its tasks in immersive environments}

Immersive environments provide a stereoscopic view to convey complex 3D structural arrangements, with great potential in exploring many types of scientific data. Molecular data, \eg, often consists of densely packed 3D structures with intricate internal detail that spans multiple scales---ma\-king it challenging to comprehend the contextual information of the data. Alharbi \etal \cite{alharbi:2021:nanotilus}, \eg, introduced a gui\-ded-tour generator for immersive environments to navigate and communicate mul\-ti-scale, crowded, scientifically accurate 3D models. Similarly, point clouds such as astronomical simulations typically exhibit complexities: 3D occlusions, non-uniform feature density, or intricate data shapes. Zhao \etal \cite{zhao:2023:metacast} proposed target- and con\-text-aware selection techniques for users to select sub-regions based on their understanding of the 3D structures yet without the need for high input precision. 

A key aspect of these visualizations lies in their emphasis on spatial attributes and their positions within the data context. When exploring or creating data of this nature, researchers often consider these spatial features displayed in different views or using different representations. That is where CR comes into play. In medical imaging, \eg, detailed views that focus on key regions are crucial while maintaining the overall data context. Coffey \etal~\cite{coffey:2011:sliceWIM} addressed this issue by presenting 3D medical data in the air alongside a detailed stereoscopic 3D view. This 3D data was complemented, moreover, by a 2D overview presented on a table, enabling users to interact with the data using familiar 2D touch gestures. 
Furthermore, Sereno \etal \cite{Sereno:2022:HTT} proposed a spatial selection technique for 3D point data, wherein selections are performed on a tablet, while a stereoscopic view is provided by an AR HMD. With AR showing an overview of the point distribution, users can perform selections on the tablet and precisely control the selection in depth.

When CR is used to expand the limited display space, various visual representations can overlay the presented information to indicate connections or provide additional context. Langner \etal \cite{langner:2021:marvis}, \eg, introduced a conceptual framework that extends a 2D scatterplot displayed on a mobile device with superimposed 3D trajectories shown in AR. Similarly, Reipschl{\"a}ger \etal \cite{reipschlager:2020:PAR} proposed a system that augments data on large displays via AR. Satriadi \etal \cite{Satriadi:2022:ASM, Satriadi:2022:TGD} envisioned how to present multivariate data around physical scale models such as tangible globes, with relevant data attributes being displayed on and around the display or tangible interfaces. There are usually no strict constraints on how data should be displayed within and across multiple views, as different views may show different aspects. The crucial issue is that the provided information should be spatially linked to facilitate easy interpretation. 
This point is particularly important in our context. When using multiple visual representations or abstractions to illustrate complex spatial data, it is crucial to ensure that users consistently perceive and understand the spatial relationships and features of the data.

When transitioning a data visualization across various spaces, a central design principle is to ensure that users understand this transformation and can seamlessly continue their tasks, without losing focus. For this purpose, Schwajda \etal \cite{schwajda:2023:TGD} developed and evaluated different variants of transformation to seamlessly transition graph visualizations from 2D to 3D representations and from a 2D surface to 3D AR space, facilitating the development of a mental model in both environments.
Fr{\"o}hler \etal's survey \cite{frohler:2022:survey_of_Cross-Virtuality} introduced various vi\-su\-al\discretionary{/}{}{/}view transition techniques, including portal, fade, and off-screen transition. These techniques guide users when a visualization shifts along the reality-virtuality continuum, such as moving from reality to AR. Lee \etal \cite{Lee:2022:ADS} presented a design space for data visualization transformations between a 2D screen and 3D AR, along with the interactions that facilitate this transformation. Their focus primarily lies on abstract data, such as transforming between 2D and 3D scatterplots, histograms, and parallel coordinate plot extrusions. To the best of our knowledge, there has been limited research addressing visual transitions for spatial data visualization. 
In our work we thus target smooth visual transitions of spatial data visualizations between 2D and 3D spaces, with an emphasis on preserving the intrinsic 3D structures and spatial distribution to maintain continuous situational awareness in CR environments.

\subsection{Spatial/touch interaction in cross-reality}
Drawing from the CR environments we discussed, typical input techniques include touch input, mid-air gestures, pen input, tangible and haptic interactions, as well as input through mobile devices or HMD controllers. This input predominantly occurs ``inside'' the stereoscopic view of 3D data presented in the air, on the 2D visualization displayed on a 2D interface, or directly with the mobile devices. Jackson \etal \cite{jackson:2013:lightweight}, \eg, proposed a prop-based \textbf{tangible interface} to control visualizations of thin 3D fiber structures. Fr{\"o}hlich \etal \cite{frohlich:2000:TCM} created a cubic tangible input device to precisely manipulate the slice with data visualization on a stereoscopic display. In the realm of model design or 3D content creation, multiple interactions facilitate the checking a 3D design, while creating and revising on 2D surfaces. DesignAR\cite{Reipschlager:2019:DesignAR}, \eg, allowed users to create and refine 3D models on a 2D surface with \textbf{touch and pen input}, while manipulating the virtual model with \textbf{mid-air gestures} in 3D. Similarly, Mockup \cite{de:2013:mockup} used sketching tools to construct models on a tabletop, extruding the sketches from the 2D surface to 3D space using mid-air gestures. Hybridaxes\cite{seraji:2022:hybridaxes} demonstrated how to transition 2D or 3D data from a display to AR, allowing users to switch interfaces using \textbf{free-hand interaction or controllers}. Wu \etal \cite{wu:2020:megereality} proposed interactions that harnesses physical affordances to assist digital interaction in AR with hand gestures.

Interestingly, regardless of where these interactions occur, people typically have a solid mental concept of how interactions should be performed on a selected interface. On mobile/touch interfaces, \eg, users usually use two touches for scaling, and one touch for either $x$- and $y$-translation or trackball rotation \cite{Yu:2010:FI3D}. Inspired by touch-based interaction, users tend to use two pinch gestures in the air to zoom content in or out \cite{Benko:2010:MPI}. As users perform a pinch gesture, their fingers naturally come into physical contact, creating explicit haptic feedback to reinforce the virtual action \cite{wu:2020:megereality}. Thus, Lubos \etal \cite{Lubos:2014:TCB} allow users to touch a 3D point cloud in mid-air and transform it with pinch gestures. 

To maximize the benefits of CR environments, interactions can be performed on familiar interfaces or interfaces best suited for the task. The Interactive slice WIM \cite{coffey:2011:sliceWIM}, for instance, projected a data overview on the table and allows users to interact with the 3D data through familiar touch interaction. Similarly, L{\'o}pez \etal \cite{lopez:2015:TAU} allowed viewers to use touch-based interactions to navigate and control the visualization on a monoscopic tablet, while observing data on a large stereoscopic display. In these cases, however, it is crucial to define how 2D interactions can be effectively mapped to the 3D visualization tasks. 

In conclusion, we saw that most studies typically regard the CR environment as comprising distinct environments, each with its inherent and tailored interaction paradigm. It is thus interesting to investigate seamless interactions across different levels of virtuality---which is what we do. If users perceive the entire CR as an integrated environment, there is likely a substantial potential for interactions across the diverse interfaces to facilitate continuous actions, in a way that eases also the mental connection between the different output spaces.

\section{\method: A cross-reality environment}
\label{sec:CRenvironment}
\begin{figure}[t]
    \centering
    \includegraphics[width=1\linewidth]{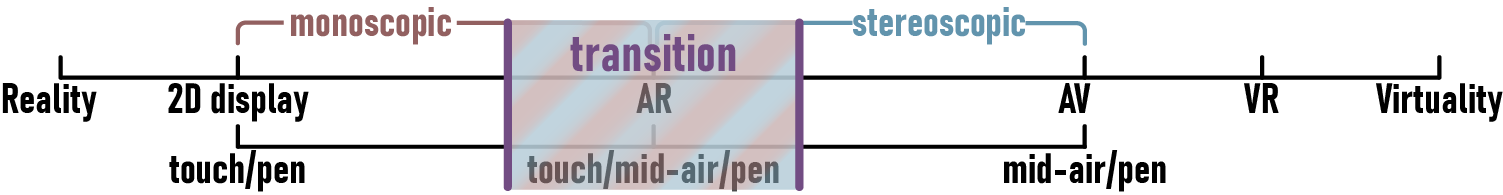}
    \caption{\method's placement within the reality-virtuality continuum.}
    \label{fig:eli:RVC}
\end{figure}

Before we detail our experiments, we first introduce our CR environment \method (for its placement in the reality-virtuality continuum see \autoref{fig:eli:RVC}) with its camera settings, input techniques, interaction devices, and general implementation. As we do so, we also discuss the key considerations drove the design. Our focus at this stage lies solely on creating an integrated visualization environment for displaying spatial data, without yet considering specific interaction designs.

\method comprises two dedicated display areas: a monoscopic 2D interactive surface (Microsoft Surface Studio) and a stereoscopic 3D space (Microsoft HoloLens). We designed it to accommodate 3D spatial data and its associated information. As an integrated (hybrid) visualization space, the data can be positioned and manipulated anywhere within the CR environment---above, on, or below the surface, or spanning across it. To emphasize 3D structures with different data features, we want the visual representations that we use to show it to be able to take various forms, including 2D or 3D representations, various scales and forms of visual abstraction \cite{Viola:2018:PCA,Viola:2020:VA}, or combined forms such as the ball-and-stick representation in molecular visualization\cite{kutak:2023:state-of-art-Molecular}. These requirements of having multiple forms of representation---later to be adjusted to the specific depiction location---set our approach apart from others
as most existing CR applications fully transition data from one visual format to its 3D counterpart when moving it across different spaces (\eg, converting a 2D node-link diagram into a 3D visualization\cite{schwajda:2023:TGD}).
In addition, we not only focus on how data or views should be presented, arranged, and transformed in CR but we also provide a perspective view of the spatial data within the whole environment to ensure that viewers correctly perceive the 3D structures and their spatial distribution, to be able to maintain a continuous situational awareness while transitioning the data between both display spaces.

\begin{figure}
    \centering
    \includegraphics[width=1\linewidth]{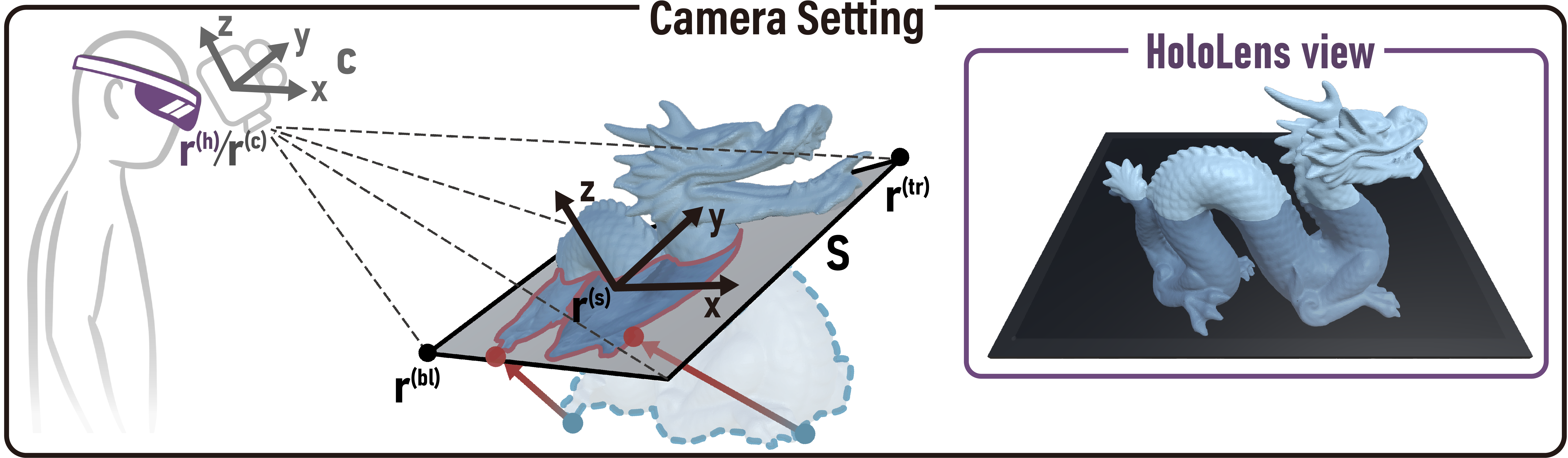}
    \caption{The configurations of the AR camera and the Surface camera $\mathbf{c}$ are depicted, illustrating how the virtual content below the Surface (marked by a blue dotted line) is projected onto the Surface (a red solid line). The HoloLens view shows what users perceive in \method.}
    \label{fig:camSetting}
    \vspace{-1ex}
\end{figure}

\newlength{\lineskiplimitbackup}%
\setlength{\lineskiplimitbackup}{\lineskiplimit}%
\setlength{\lineskiplimit}{-\maxdimen}%
\textbf{Camera settings.}
In our design, regardless of the viewing angle, the data rendered on the surface and through the Hololens seamlessly combines into a cohesive 3D representation.
To achieve this effect, we developed a rendering algorithm that aligns the visualization content displayed on the 2D surface with the AR visualization. We set two virtual cameras in our system, (1) an AR HMD camera rendering the 3D view and (2) a Surface camera $\mathbf{c}$ for rendering 2D view (\autoref{fig:camSetting}). 
For the AR HMD camera, we employ a perspective projection, a standard method used in AR/VR applications.
For the Surface camera $\mathbf{c}$, inspired by the Fishtank VR concept\cite{Ware:1993:FishTankVR}, we first align $\mathbf{c}$ to the HMD's position and then adopt an oblique perspective projection. 
This way we dynamically calculate $\mathbf{c}$'s parameters---position $\mathbf{r}^{(c)}$, orientation, and projection matrix $\mathbf{m}^{(c)}$---in each frame based on the AR HMD camera's position $\mathbf{r}^{(h)}$ and surface $\mathbf{s}$'s center position $\mathbf{r}^{(s)}$ (we describe more detail on the camera configuration in \autoref{sec:app:camera} and also share an open-source simulator of the camera setup in our supplemental material).

\textbf{Touch/Pen/Mid-air Interactions.}
Microsoft's API Surface Studio captures multi-touch input, while we can obtain mid-air input from the AR HMD. For precise interaction needs such as selection and an\-no\-ta\-tion, however, we want to augment touch and mid-air input from the Surface Pen---both on the surface \emph{and} in mid-air. While the Surface Pen input can easily be captured on the Surface, it is normally not detected when used in the air. To overcome this limitation, we attached an Arduino board to the pen, allowing us to detect events when users press the physical button on the pen without it resting on the surface.

\textbf{Devices and Implementations.}
A $28$-inch Microsoft Surface Studio (637\,mm\,\texttimes\,438\,mm; 4,500\,\texttimes\,3,000\,px) serves as the 2D surface of our \method CR setup, which captures both pen and multi-touch input. It can be adjusted smoothly from a vertical position to a nearly horizontal orientation, the latter resulting in a slight inclination of $\approx$\,20\textdegree\ from the horizontal.
As AR HMD we used Microsoft's HoloLens~2 (2,048\,\texttimes\,1,080\,px per eye, 52\textdegree\ FoV), equipped with spatial tracking and gesture recognition.  
We connected both to a PC (Intel Core\texttrademark{} i9, 3.5\,GHz, 64\,GB RAM, GeForce RTX3090, 24\,GB video memory), the AR HMD via holographic remoting. 
We attached a physical button to the Surface pen, along with the Arduino nano board on the back so that users could initiate actions both in 3D space and on the 2D surface. When the Arduino detects a button press, we compute the pen position based on the user's index finger position---tracked by the HoloLens using computer vision---, to which we add a constant offset to arrive at the pen tip. The resulting pen precision was sufficient for our prototype implementation. For more precise input, however, we could explore advanced technologies such as \href{https://www.logitech.com/en-us/promo/vr-ink.html}{Logitech's VR Ink Pilot Edition}. 

We developed our prototype implementations with C\# in the Unity 3D engine and ran on both the Surface and the HoloLens. We realized the communication between both display spaces through Universal Windows Platform sockets. 
After starting the program on the Hololens, we calibrate the setup by aligning the two coordinate systems via QR code tracking and manually manipulating the anchor point, as in past work \cite{Reipschlager:2019:DesignAR}.
After calibration, we employ HTC Vive trackers to follow the position and orientation of the Surface Studio.  
\section{Elicitation study}
\label{sec:elicitation}
To explore potential interaction designs for \method, we conducted an elicitation study to gauge users' reactions to this new environment. Similar to Wang \etal \cite{Wang:2024:UPI}, our study encouraged participants to propose any interactions they could imagine. We captured their responses and interactions, subsequently incorporating this feedback into our final interaction techniques design.

\subsection{Study setup}
\label{subsec:study}
Based on \method that we just introduced in \autoref{sec:CRenvironment}, we had the following goals: (G1) to determine whether an optimal position (or range) exists for the 2D surface to facilitate the observation of 3D data and to enhance user interaction; (G2) to understand how users perceive information within a CR environment; and (G3) to observe the participants' interactions and strategies while they complete exploration tasks with 3D spatial data, as well as to identify the physical locations of where the interactions occur. We pre-registered our study (\href{https://osf.io/avxr9}{\texttt{osf\discretionary{}{.}{.}io\discretionary{/}{}{/}avxr9}}) and received IRB approval for the protocol (XJTLU University Research Ethics Review Panel, \textnumero~20240201174957).

\textbf{\textit{Participants.}} We recruited eight voluntary participants from the local university, with ages ranging from $24$ to $41$ years old (mean 26.5, SD 5.5). The participants' past VR/AR experience included weekly use (3\texttimes), annual use (3\texttimes), and no past experience (2\texttimes). On average, they had 5.3 years of experience in the visualization or hu\-man-com\-pu\-ter interaction fields. One participant was left-handed. All participants had normal or cor\-rec\-ted-to-nor\-mal vision with no color deficiency, ensuring a clear ability to perceive the data visualization and colors in our study.

\textbf{\textit{Datasets.}} To represent a variety of application domains we used five datasets with a diverse set of features (more detail in \autoref{sec:app:ElicitationDataset}):
\begin{description}[nosep,leftmargin=1.5em,labelindent=0em,leftmargin=!,labelindent=!,itemindent=!,font=\normalfont\itshape]
\item[Structured molecular data:] a spike protein from the SARS-CoV-2 virus\cite{Nguyen:2021:MIT}, reconstructed from the electron microscopy images. 
\item[Volume visualization data:] MRI volume data with multiple 2D slices.
\item[Unstructured point cloud data:] three point datasets, each with different features: (1) a synthetic semi-spherical shell of particles; (2) an N-body simulation \cite{springel:2008:aquarius}; and (3) a Millennium-II data subset \cite{Springel:2005:SimulationsOT}.
\end{description}

\textbf{\textit{Task and Procedure.}}
After providing their written informed consent, we asked participants to comfortably sit on a chair $\approx$\,0.5\,m above the floor. We placed the surface screen on the table, $\approx$\,0.75\,m from the floor, with an initial angle of $\approx$\,20\textdegree\ to the horizontal. Participants could adjust the chair height if desired. We then asked them to complete the three tasks described below. Importantly, for both manipulation and selection tasks, we did not show any feedback to the participants on their proposed interactions. We only provided them with a data context with highlighted (exploration or selection) targets and asked them to imagine how actions should be performed according to the given tasks. 

\textbf{Task 1.} 
We instructed the participants to adjust the screen angle to achieve a suitable position for an optimal observation of the 3D structures. In addition, we reminded them of the importance of considering interaction on both the 2D Surface and in 3D space for a comprehensive understanding. We also informed the participants that they could re-adjust the screen angle whenever needed, throughout the study.

\textbf{Task 2.}
We asked the participants to manipulate the five spatial datasets. In all manipulation tasks, we provided them with a target object and a target position. Specifically, we asked them to:
(1) translate data displayed below and above the surface within each display space;
(2) translate data displayed below and above the surface to the other display space;
(3) translate data displayed partially in both spaces to a target position above or below the surface; and
(4) rotate and scale data located in 3D space, within a 2D surface, or partially in both spaces.

\textbf{Task 3.}
We asked the participants to select a given highlighted target object or highlighted regions of interest.

We recorded the screen angle chosen by participants, their actions (incl.\ head/hand position, orientation), and device input events for our analysis (which we make available at \href{https://osf.io/avxr9}{\texttt{osf.io/avxr9}}). After completing the three tasks, we conducted semi-structured interviews with them to discuss their thoughts and suggestions on the interaction.\vspace{-1pt}

\subsection{Findings}
\vspace{-1pt}Our participants demonstrated that the integrated 3D data visualization in our environment allowed them to gain a comprehensive understanding of the 3D content. They mainly concentrated on the 3D visualization in the AR space and attempted to view it from various angles. Their decision on where to interact with the data was influenced by the location of key information such as target data or target location. Their preference for direct manipulation methods remained consistent, however, such as touching the data on the surface or brushing it in the air. Below we present our findings on surface angle, visualization techniques, and interaction techniques tailored to our CR environment.\vspace{-1pt}

\subsubsection{Surface angle}
\vspace{-1pt}From the initial $\approx$\,20\textdegree\ incline w.r.t. the horizontal table, some participants made slight adjustments at the beginning of the study, with angles ranging from 18.2\textdegree\ to 22.6\textdegree\ (mean 21.0\textdegree). After the initial setup, however, they did not make any further adjustments, despite being informed that they could do so. During the interviews, they said that the initial setting provided adequate space for observing stereoscopic 3D data through the AR HMD, while also offering a convenient position and orientation for interacting with data presented on the surface.

We also found that, when comparing our setup with a past stereoscopic multi-touch system\cite{Butkiewicz:2011:MTE, Butkiewicz:2019:MTP}, a notable difference in display orientation: Butkiewicz et al.'s screen was nearly perpendicular to the table. This difference initially seems strange because Butkiewicz et al.\cite{Butkiewicz:2011:MTE, Butkiewicz:2019:MTP} had also relied on innovative interaction techniques using two-finger pinch gestures on the touch screen for tasks such as positioning the cursor under the display for 3D exploration and selection. The only difference appears to be that they had employed an auto-stereo\-sco\-pic touch display, while we used an AR HMD.
A main reason for the difference then appears to be the additional 3D space visible through the AR HMD, \emph{decoupled from the touchscreen}, which substantially reduced people's reliance on the 2D surface. In the recorded data, in tasks where the visualization spanned both two spaces we saw that $\geq$67.5\% of the time was spent on observing 3D data in AR space. Users apparently preferred the stereoscopic AR space and only referred to the 2D surface when needed.
Our environment also encouraged users to actively move around to find a good view, rather than adjusting the surface angle. All these factors diminished the importance and constraints of the surface angle. Moreover, the $\approx$\,20\textdegree\ angle is also appropriate for design work\cite{Reipschlager:2019:DesignAR} as it provides stability and physical support from the base, enabling users to interact confidently and securely. 

\begin{figure}[t]
    \centering
    \includegraphics[width=1\linewidth]{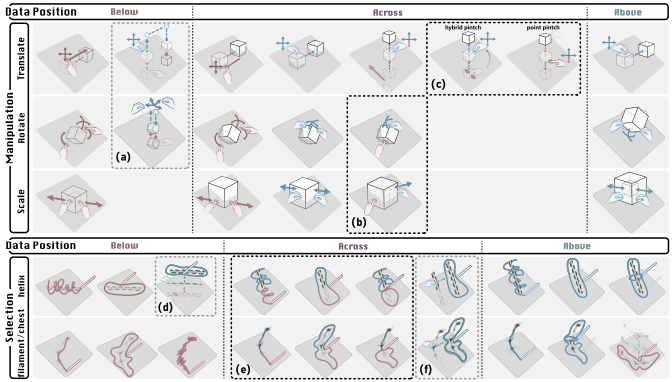}
    \caption{Design space for interaction techniques for two visualization tasks: data manipulation and selection. \textcolor{colorbelow}{Red}: interactions on 2D surface; \textcolor{colorabove}{Blue}: interactions in 3D space. \belowS, \acrossS, and \aboveS: positions of the target data/location. (a), (d), (f): move data above the surface for interaction. (b) (c) and (e): interaction transitions across both spaces.}
    \label{fig:eli:DS}
    \vspace{-1.5ex}
\end{figure}

\subsubsection{Visualization in CR}
In addition to the Fish Tank VR view \cite{Ware:1993:FishTankVR} on the 2D surface, we also asked participants to compare it with a static orthogonal view and a 2D slice view as alternatives.
All participants expressed a strong preference for the perspective Fish Tank VR view, highlighting its effectiveness in creating a seamless 3D experience. Several participants recognized the significance and necessity of 2D slice view on the surface, particularly for tasks such as adjusting 2D image slices or marking MRI volume data.
Furthermore, they also noted that the see-through AR HMD allowed them to look through the stereoscopic rendering directly to the 2D surface visualization. The resulting experience thus combined a comprehensive 3D data understanding with the ability to observe projected information, and the environment is flexible in how and where the information should be presented.

\subsubsection{User interaction in CR}
\textbf{\textit{Interactions in 2D or 3D display spaces.}}
When observing our participants engaging in either of the two individual methods of displaying and manipulating spatial 3D data---projected on the 2D surface or stereoscopically in 3D space---we noticed that they have a well-established mental concept of how an interaction should be performed. As we show in \autoref{fig:eli:DS} (Manipulation, \belowS), when participants manipulated data presented below the surface, they used one finger to pan the visualization and two fingers to rotate and scale it on the surface. Conversely, when they saw data rendered in the AR HMD above the surface (\autoref{fig:eli:DS}, Manipulation, \aboveS), they used one or two hands to grab and manipulate the data directly. Similarly, for the selection tasks, when the data was presented below the surface (\autoref{fig:eli:DS}, Selection, \belowS), participants preferred to brush or circle data on the surface. When the data was entirely above the screen (\autoref{fig:eli:DS}, Selection, \aboveS), however, brushing and circling interactions were performed in 3D space.

Interestingly, some participants also used the benefits of the environment to aid their understanding of the data structure or to make more precise inputs. When both the target data and target location were below the surface, for instance, some participants would grab it out and put it back underneath to position the target. They explained in the interview that this extra action would provide them with more precise control over the depth in the positioning. Another example is that, to make precise selections, some participants would pull the data entirely out of the surface so that they can see the whole dataset stereoscopically or, alternatively, push it entirely below the surface for precise selection input. We highlighted these interactions in \autoref{fig:eli:DS}(a) and \autoref{fig:eli:DS}(d,f).

\textbf{\textit{Interactions across two display spaces.}}
Two types of tasks also required interaction across both spaces: if the data was visualized across both spaces or if it needed to be translated from one to the other. We show these actions in \acrossS in \autoref{fig:eli:DS} (Manipulation, Selection).

In the rotation and scaling tasks with a visualization spanning both spaces, participants performed actions on either the surface, in 3D space, or using both spaces for different tasks. They mainly made this decision based on where they perceived the key information. In addition, 4 participants favored using both spaces, most of them with a notable strategy: they (3\texttimes) used one finger to designate a ro\-ta\-tion\discretionary{/}{}{/}sca\-ling center on the surface, while using their other hand to execute the rotation or scaling actions around the selected center (\autoref{fig:eli:DS}(b)). In the interviews, these participants mentioned for them that this approach was a more precise and secure method of manipulation.

In the selection tasks in which we placed the data across both spaces, we also observed intriguing patterns. Most participants (6\texttimes) approached our environment as an integrated system and used consistent selection strategies across both spaces. For example, participants followed the 3D structure and brushed along the data in 3D space, then continued brushing the rest along the structure projected on the 2D surface (\autoref{fig:eli:DS}(e, left)). Or they drew a large lasso across both spaces to enclose the data (\autoref{fig:eli:DS}(e, middle)). These selection strategies remained consistent even when they physically reached the surface and noticed that they could no longer follow the spatial structure in the depth direction. 

The most interesting finding is that participants maintained their mental model when interacting with data across both display spaces. Some employed a mixed approach, brushing data in 3D space and circling data on the 2D surface, or vice versa (\autoref{fig:eli:DS}(e, right)), which indicates that they maintained spatial awareness and seamlessly transitioned between different interaction techniques in the hybrid environment.

In tasks requiring transition across both spaces, we observed two distinct transition methods (\autoref{fig:eli:DS}(c)). First, when the data was situated below the surface and needed to be moved out, some participants initially employed 2D pinch gestures to pull it out. Once their two fingers naturally came into physical contact, they seamlessly transitioned to 3D pinch gestures to continue the action. They used the same method to put the data back in its original position. Second, some participants pointed at the data on the 2D surface, assuming that it would automatically approach their finger position. They then continued the manipulation using 3D pinch gestures to pull it out. In the AR space, all participants uniformly used 3D pinch gestures, which facilitated positioning of the data---whether in 2D or 3D space---in a rapid manner.

\textbf{\textit{Interactions on visual representations.}}
Participants employed a consistent approach in manipulating all three datasets. They exhibited, however, varied approaches for selecting data with different representations in our environment.
For the structured surface data, participants selected it by adhering to its 3D structure, such as wrapping a protein helix with a helical stroke. Conversely, for the unstructured point cloud data, they attempted to enclose it with a freeform lasso, such as drawing a lasso around the astronomical point cloud data. We observed that participants relied on their established mental models to determine how the data should be selected, based on their perception and understanding of the specific data structures. Therefore, most participants did not alter their selection strategies when interacting across both spaces---even though their interactions were physically limited to reach the visualization below the surface: they believed that through continuous input they could successfully select data within the CR environment.

\section{\method for domain-specific use}
\label{sec:Applications}
Based on these findings, we then focused on developing \method sample applications tailored to three distinct domains: astronomical point cloud analysis, molecular visualization, and medical anatomy imaging---each with its unique challenges for interactive visualization. Based on the optimal angle for the 2D surface identified in the elicitation study (G1), we set the angle of the surface at a 21\textdegree\ incline w.r.t. the horizontal table. We use the Fishtank VR view on the Surface camera and a perspective projection in AR HMD camera to ensure a seamless visualization (G2). In the following description we highlight, in particular, specific realizations within \method for the given type of data or domain. We encourage readers to watch more detailed figures in \autoref{fig:app:cases} and the supplemental video for a demonstration.

\subsection{Astronomical point cloud visualization}
\label{sec:app:pointcloud}
Astronomical simulation datasets typically comprise billions of spatial points\cite{springel:2008:aquarius,Springel:2005:SimulationsOT}. 
Researchers often need to navigate in 3D to obtain a clear view of the structures to select and explore the regions of interest.
The task of data selection becomes paramount in this context, serving as a critical step in data visualization and exploration \cite{Wills:1996:S5W,besanccon:2021:state-of-art}.
While much research has been dedicated to developing selection techniques for 2D surfaces \cite{Yu:2012:ESA,Yu:2016:CEE,perelman:2022:visualtransitions,Chen:2020:LassoNet,Lucas:2005:DE3} and in 3D space \cite{zhao:2023:metacast,Roberto:2020:HMS,Franzluebbers:2022:VRP,Krüger:IntenSelect+:2024,Rongkai:2023:EGA}, these methods are predominantly designed for a single display. 
In 3D space, although users can clearly see the spatial data distribution, limitations persist in their ability to precisely delineate selection regions. Conversely, the 2D surface facilitates accurate input for target inclusion but struggles with depth prediction, posing a challenge for selecting specific data regions.
\method bridges this gap by supporting data observation in AR, affording users a comprehensive understanding of data density distribution and context, while also facilitating precise data selection on a 2D surface. As an integrated environment---users are empowered with the flexibility to decide both \emph{where} the selection occurs (whether on the 2D surface, within 3D space, or spanning both) and \emph{how} these selections are made (via a freeform lasso or a brush). 

Based on the interaction techniques detected from the elicitation study (G3, \autoref{fig:eli:DS}), we developed two new spatial selection techniques to facilitate a seamless transition from 3D selection to 2D selection, and vice versa. These techniques are based on the context- and target-aware selection metaphors CloudLasso\cite{Yu:2012:ESA}, WYSIWYP\cite{Wiebel:2012:WYSIWYP}, and MeTABrush\cite{zhao:2023:metacast}. These methods analyze the density distribution and data features within the local area of user interaction so that users can identify and select critical features of interest, without the need for precise input. 
Our first new technique, \textit{BrushWYP}, draws inspiration from the user interactions illustrated in \autoref{fig:eli:DS}(e, left), MetaBrush\cite{zhao:2023:metacast}, and WYSIWYP\cite{Wiebel:2012:WYSIWYP}. It enables users to trace the string-like shape of 3D point cloud data by brushing over these structures in the 3D space and continuing to brush the rest along the structure on the 2D surface. 
We designed our second new technique, \textit{BrushLasso}, based on \autoref{fig:eli:DS}(e, right), MetaBrush\cite{zhao:2023:metacast}, and CloudLasso\cite{Yu:2012:ESA}. This method allows users to begin their selection by brushing over the data in the 3D space and draw a lasso on the 2D surface to enclose the targeted points.
Beyond the ability to transition between two spaces, both techniques are flexible in that they also allow users to brush the target points on the 2D surface and in 3D space, or encircle the target points on the 2D surface. 
Moreover, similar to our previous works \cite{Yu:2012:ESA, Yu:2016:CEE, zhao:2023:metacast}, subtraction can be achieved using region-based techniques. A potential action to activate subtraction could be, for instance, to turn the Surface pen over and to rub the end of the pen over the selected data.

\setlength{\lineskiplimit}{-\maxdimen}%
Technically similar to MeTACAST\cite{zhao:2023:metacast}, we leveraged a continuous density field $\rho(\mathbf{r})$ to represent the particle density at location $\mathbf{r}$. We pre-compute the density of the field at all nodes $i$ of the regular $128$$\times$$128$$\times$$128$ 3D grid box $B$ that covers the dataset, denoted as $\rho(\mathbf{r}^{(i)})$, offline on GPU. We use the modified Breiman kernel density estimation with a fi\-nite-sup\-port adaptive Epanechnikov kernel \cite{Ferdosi:2011:DEM}. This approach allows us to apply our selection not only to point clouds but also to volumetric data, which samples a scalar field that represents important data aspects in a visually salient way (not limited to density).

\setlength{\lineskiplimit}{\lineskiplimitbackup}%
\textbf{BrushWYP.}
We employ MeTABrush to allow users to select structures in 3D space by direct tracing along 3D structures. When the input transitions from the 3D space to 2D surface, however, the direct brushing on the target location is constrained. We address this issue with a modified WYSIWYP method, enabling users to select points below the surface by identifying the correct depth of the point of interest (POI). 

\begin{figure}
\centering
            {\begin{overpic}[width=\columnwidth]{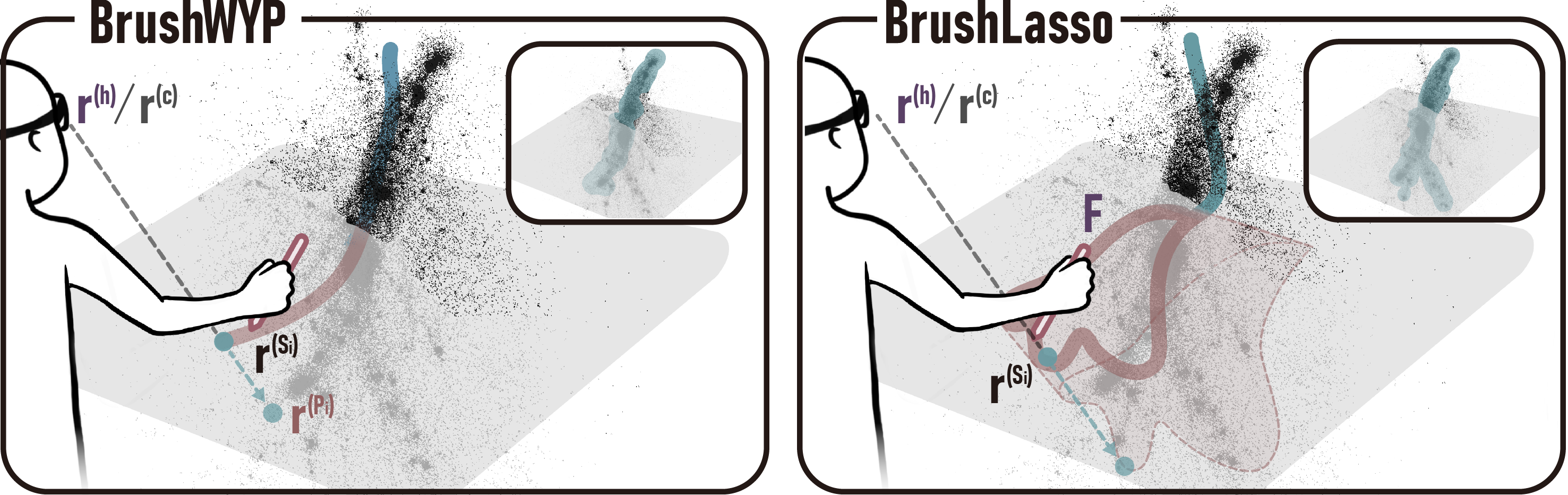}
            \put(1.5,2){\textcolor{black}{(a)}}
                  \put(52,2){\textcolor{black}{(b)}}
        \end{overpic}}
    \caption{Point cloud data selections: (a) BrushWYP, (b) BrushLasso.}
    \label{fig:selectiontech}
    \vspace{-1ex}
\end{figure}

\setlength{\lineskiplimit}{-\maxdimen}%
Our algorithm initiates by sampling points along the input stroke on the surface, denoted as $\mathbf{r}^{(s)}=\{\mathbf{r}^{(s_0)}, \mathbf{r}^{(s_1)}, ... \mathbf{r}^{(s_n)}\}$, as well as above the surface, represented by $\mathbf{r}^{(a)}=\{\mathbf{r}^{(a_0)}, \mathbf{r}^{(a_1)}, ... \mathbf{r}^{(a_m)}\}$. For each sampled point $\mathbf{r}^{(s_i)}$ in $\mathbf{r}^{(s)}$, we project a ray from $\mathbf{r}^{(s_i)}$ toward the direction of  $\mathbf{r}^{(s_i)}-\mathbf{r}^{(h)}$, where $\mathbf{r}^{(h)}$ denotes the position of the AR HMD (\autoref{fig:selectiontech}(a)). Then we traverse along each emitted ray in fixed steps and search for the POI $\mathbf{r}^{(p_i)}$ that exhibits the maximum density along the ray. This approach is different from the original \mbox{WYSIWYP} \cite{Wiebel:2012:WYSIWYP}, which identifies the highest jump of accumulated scalar value along the ray. After traversing $\mathbf{r}^{(s)}$, we get a list of POIs, $\mathbf{r}^{(p)}=\{\mathbf{r}^{(p_0)}, \mathbf{r}^{(p_1)}, ... \mathbf{r}^{(p_n)}\}$. Subsequently, we obtain the input array of MeTABrush as $\mathbf{r}^{(i)}=\{\mathbf{r}^{(p)}, \mathbf{r}^{(a)}\}$ and select the target points with the MeTABrush method. 

\setlength{\lineskiplimit}{\lineskiplimitbackup}%

\textbf{BrushLasso.}
We integrate MeTABrush and CloudLasso to provide users with the ability to brush target points in mid-air and encircle points on the surface through a single, seamless input. The original CloudLasso selects point cloud clusters of high density that fall within the input lasso. Directly merging both methods, however, could lead to disconnected selection volumes---one above the surface and several isolated ones below the surface---which might not align with the user's intention of making a continuous selection from the 3D space to the 2D surface. To solve this issue, we implemented a modified CloudLasso method that ensures that the selection volume is smooth and continuous.

\setlength{\lineskiplimit}{-\maxdimen}%
This algorithm initiates by sampling points along the input stroke on the surface. For each sampled point $\mathbf{r}^{(a)}$ above the surface, we calculate an initial volume of interest (VOI) $V_{\mathrm{init}}$ with MeTABrush algorithm \cite{zhao:2023:metacast}. We then remove the parts below the screen and keep only the VOI above the surface $V_{\mathrm{a}}={V_{\mathrm{init}}}\cap{V_{\mathrm{3D}}}$, where ${V_{\mathrm{3D}}}$ is the space above the surface.
We then connect the sampled points on the surface, $\mathbf{r}^{(s)}$, to form a lasso, $\textbf{L}$ (\autoref{fig:selectiontech}(b)).
We then map the particle coordinates to the view coordinates of the surface camera, $\textbf{c}$, using the model-view transformation. Similar to CloudLasso \cite{Yu:2012:ESA}, we then compute the lasso frustum $\textbf{F}$ based on the first-level binding stage. We can then obtain the VOI below the surface $V_{\mathrm{b}}={\mathrm{F}}\cap{V_{\mathrm{2D}}}$, where ${V_{\mathrm{2D}}}$ is the space below the surface. Finally, we combine both VOIs to get the interconnected VOI, $V_{\mathrm{CR}}={V_{\mathrm{b}}}\cup {V_{\mathrm{a}}}$. We compute the initial density threshold  $\rho_{0}$ as:\vspace{-2ex}

\setlength{\lineskiplimit}{\lineskiplimitbackup}%
\begin{equation}
    \rho_{0}=\frac{1}{N_{V_{CR}}} \sum_{n=1}^{N_{V_{CR}}} \rho(\mathbf{r}^{(n)}),
\end{equation}

\setlength{\lineskiplimit}{-\maxdimen}%
\vspace{1ex}\noindent where $N_{V_{\mathrm{CR}}}$ is the number of grid-nodes, $\mathbf{r}^{(n)}$, inside of the combined VOI, $V_{\mathrm{CR}}$. 
We select the volume $V$ with density $\rho$ above $\rho_{0}$ within $V_{\mathrm{CR}}$:\vspace{-2ex}

\setlength{\lineskiplimit}{\lineskiplimitbackup}%
\begin{equation}
    V=\{\mathbf{r} \,\vert\, \mathbf{r} \in B,\,  \mathbf{r} \in V_{CR},\,  \rho(\mathbf{r})>\rho_{0}\},
\end{equation}
We generate the iso-surface with Marching Cubes based on the density threshold $\rho_{0}$. Both resulting selection techniques facilitate a seamless and natural transition between 2D and 3D selections.

\subsection{Molecular visualization}
\label{subsec:Molecular}
Molecular visualization is a field rich with diverse data representations, each designed to highlight various aspects of molecular structures and interactions. These include space-filling diagrams, ball-and-stick models, ribbon models, licorice visualization, backbone, and surface visualizations, etc. Given the complexity of molecular interactions, it is common to use these representations as a combination to provide a comprehensive view of the molecules.
For biologists and chemists, gaining a deep understanding of how drug molecules interact within larger molecular structures is essential. They require insights into the precise location and distribution of these molecules to assess their interaction and affinity. Traditionally, they need to select specific features or regions of interest from biological sequences and link these on a 3D visualization rendered on a 2D display. The task of selecting spatial features and observing the dynamic behavior of molecules through 2D display, however, presents significant challenges. Even within a pure 3D environment, maintaining spatial awareness can be a struggle for users, given the inherent complexity of the involved datasets.

\begin{figure}
    \centering
            {\begin{overpic}[width=\columnwidth]{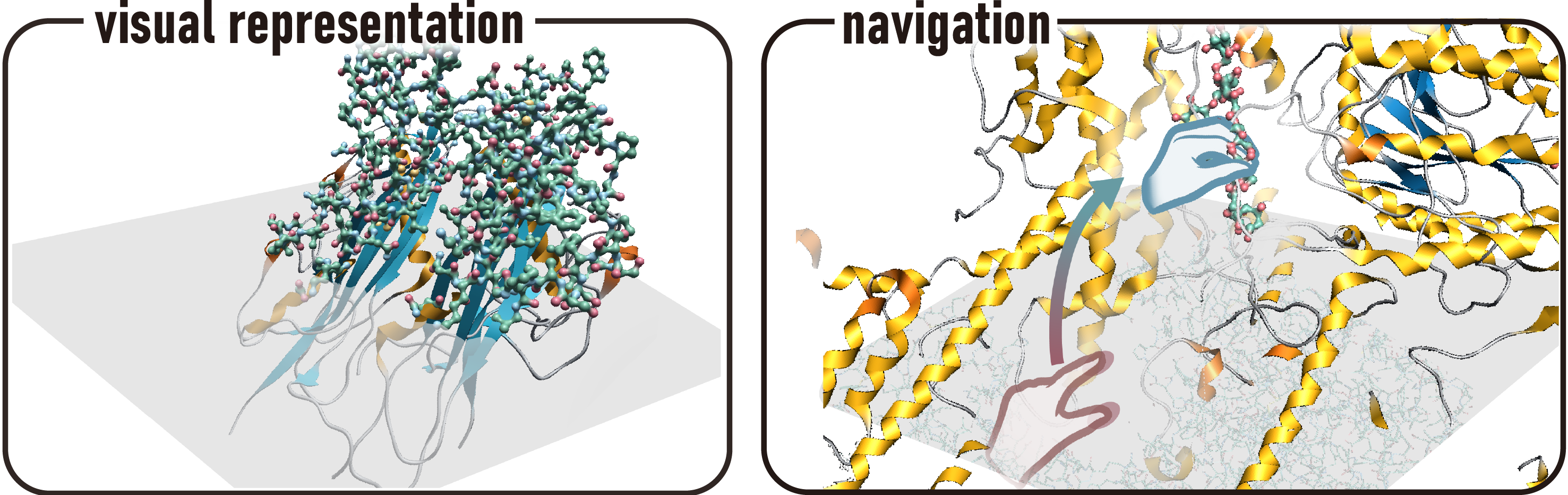}
            \put(1.2,2){\textcolor{black}{(a)}}
                  \put(50,2){\textcolor{black}{(b)}}
        \end{overpic}}\vspace{-2pt}
    \caption{Proteins visualized with UnityMol\cite{doutreligne:2014:unitymol}. PDB ID: (a) \href{https://doi.org/10.2210/pdb4FPQ/pdb}{4fpq}, (b) \href{https://doi.org/10.2210/pdb8RFE/pdb}{8rfe}. Users can ``grab'' the 3D visualization directly to check the local regions.}\vspace{-2pt}
    \label{fig:ProteinAbstraction}
    \vspace{-1ex}
\end{figure}

With \method we address this need by merging the distinct display spaces to facilitate a concurrent visualization of different representations and abstractions. As we illustrate in \autoref{fig:ProteinAbstraction}, \eg, we can show the hyperball representation\cite{chavent:2011:GAA} stereoscopically in AR space for a detailed view of the molecule, while a ribbon model is rendered on the surface for context. Alternatively, we can put a ribbon diagram in AR space, complemented by a licorice diagram on the surface. \method allows users to combine these representations based on their preferences and exploration needs. 
In addition, we enable them to adjust visual representations in the specific regions as needed, both in AR space and on the surface (\autoref{fig:ProteinAbstraction}(a)). Users can also directly grab the visualization to check details of local regions, as we show in \autoref{fig:ProteinAbstraction}(b).

We enable users to interact with data rendered below the surface using familiar touch interactions or above the surface through gesture interactions, as illustrated in \autoref{fig:eli:DS} (manipulation, \belowS and \aboveS). 
In addition, inspired by the findings in the elicitation study as illustrated in \autoref{fig:eli:DS}(c), we developed two seamless interaction transition techniques to enable users to move data across two spaces. The first method supports users to employ a familiar 2D pinch gesture on the touch surface to ``pull'' the data to the surface level (akin to Hancock et al. \cite{Hancock:2010:SST}), then continue with a 3D pinch gesture to ``grab'' it further into the air above the surface, as illustrated in \autoref{fig:eli:DS}(c, left) and \autoref{fig:ProteinAbstraction}(b). Conversely, performing this gesture in reverse enables users to seamlessly return the data to below the surface. This approach supports users to directly interact with and manipulate regions of interest, whether transferring them from the 3D space directly onto the 2D surface for extended analysis or relocating them to any other area within the CR environment for closer inspection. It is worth noting, however, that 2D pinch gestures are typically associated with zooming interactions. The second method thus allows users to press the screen to ``push'' data further into the depth of the surface or to ``pull'' it to the outside. The 3D pinch interaction in AR HMD remains, as we illustrate in \autoref{fig:eli:DS}(c, right). While both methods provide seamless translations across two spaces, we also implemented scaling and rotation: users can point at a particular feature on the 2D surface and use 3D pinch gestures to rotate or scale the data around that chosen point, as illustrated in \autoref{fig:eli:DS}(b).

\subsection{Medical anatomical visualization}
Medical imaging plays a central role in many healthcare practices for diagnosis, treatment planning, and patient care. 
In particular, 3D anatomical visualization has been used for treatment planning for various medical procedures in radiology\cite{jadhav:2022:Md-cave} as well as for teaching\cite{Yu:2022:VeLight}. It often relies on volume rendering to show internal structures in detail. While 2D slices are commonly used and are particularly effective for many diagnostic purposes, 3D visualizations are often necessary for addressing complex cases such as those involving intricate fractures. Physicians also frequently interact with both 2D slices and 3D visualizations for a range of tasks, including assessing injuries and devising treatment plans\cite{pfeiffer:2018:imhotep}. When planning surgical interventions for complicated orthopaedic injuries, for example, the integration of 2D and 3D visualizations allows surgeons to accurately visualize and navigate the affected areas \cite{Benmahdjoub:2023:EAV}. This combined approach significantly enhances the precision of both planning and executing surgical procedures.

\begin{figure}
    \centering
                {\begin{overpic}[width=\columnwidth]{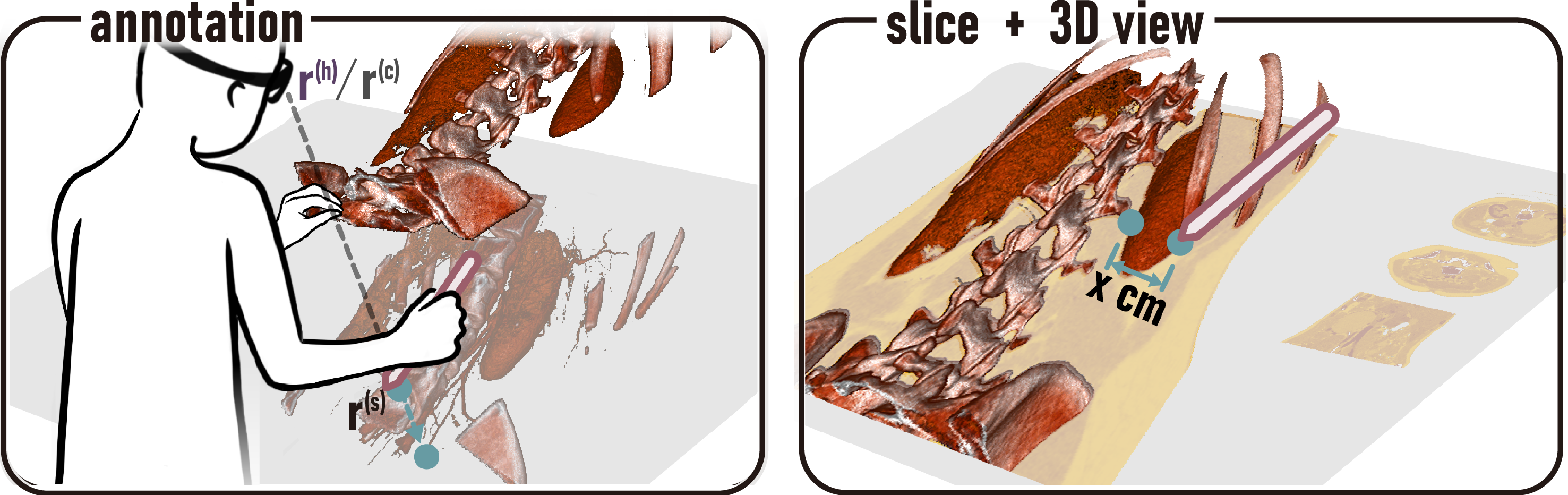}
            \put(1.2,2){\textcolor{black}{(a)}}
                  \put(52,2){\textcolor{black}{(b)}}
        \end{overpic}}\vspace{-2pt}
    \caption{(a) Annotation on 2D surface while grabbing 3D data; (b) 3D visualization superimposed on 2D slices to facilitate distance measurement.}\vspace{-2pt}
    \label{fig:MedicalCuttingplane}
\end{figure}

Prior work has explored the use of 2D displays\cite{Lundstrom:2011:MTT} and 3D environments \cite{Sousa:2017:VRRRRoom} for exploring 3D medical visualizations within clinical research tasks. Particularly relevant is Slice WIM \cite{coffey:2011:sliceWIM}, which uses a stereoscopic display and a multi-touch table to present both an overview and detailed views of 3D medical data, and projects 2D slices onto a wall or table display. For \method, in contrast, we use an integrated environment that allows users to view 2D slices directly on the surface, while also observing stereoscopic renderings superimposed on these slices. Furthermore, users can use an interactive Surface pen for precise annotation and distance measurements. Moreover, with our method selected 2D slices can be saved and set aside on the surface, facilitating quick observation and navigating to specific local structures (\autoref{fig:MedicalCuttingplane}(b)).

With this example we want to highlight that domain experts often need to engage in precise interactions, based on their observations of 3D structures---our environment being able to fulfill these requirements. Physicians can make annotations and measure distances directly on the 2D touch surface, while simultaneously viewing the 3D representation in AR space. To provide a clear view of the 2D slices and facilitate accurate interactions on the surface, we implement a ``lifting'' feature. It is activated with a 3D pinch gesture, which temporarily elevates the 3D volume visualization away from the 2D slice on the touch surface. This separation facilitates unobstructed annotation or marking. Once these tasks are complete, releasing the pinch gesture returns the 3D volume visualization to its original position on the screen, seamlessly integrating the new annotations with the remaining data. 

\setlength{\lineskiplimit}{-\maxdimen}%
In addition, with \method interaction can happen anywhere within CR, including making annotations directly on the 2D slice (on the surface), marking features on the 3D visualization (above the surface), and selecting features that appear in depth but are rendered below the surface. We realized the latter interaction based on the WYSIWYP principle\cite{Wiebel:2012:WYSIWYP}. To identify the depth of a ROI, our method detects a significant change in the accumulated scalar value along the ray. This ray originates from the contact point on the surface, denoted as $r^{(s)}$, and projects it in the direction of $r^{(s)}-r^{(h)}$, as shown in \autoref{fig:MedicalCuttingplane}(a).
Our integrated environment thus offers users a seamless and precise interaction with medical 3D data for a detailed immersive examination.

\setlength{\lineskiplimit}{\lineskiplimitbackup}%

\section{Evaluation}
\label{sec:evaluation}
We ran two evaluations with domain experts, both pre-registered (\href{https://osf.io/avxr9}{\texttt{osf\discretionary{}{.}{.}io\discretionary{/}{}{/}avxr9}}) and IRB-approved (same as before). First, we evaluated \method's usability with domain experts, focusing on its use in the domain. Second, we assessed the interaction with VR\discretionary{/}{}{/}AR\discretionary{/}{}{/}MR experts.

\subsection{Evaluation with domain experts}
\label{subsec:evaluation_bio}
We presented SpatialTouch to four domain experts: two biologists and one astronomer from the local university, and one doctor from a local hospital. All experts had >10 years of professional experience.

With the biologists, our demonstration focused on the SARS-CoV-2 virus spike protein\cite{Nguyen:2021:MIT} (\autoref{fig:teaser}(b)).
The domain experts appreciated that the 2D and 3D perspective views offer a thorough understanding of the molecular data. They were highly interested in transitioning the visualization between 2D and 3D spaces, finding it intuitive to grab the visualization out from the 2D screen to check its 3D structure.
They mentioned several scenarios where \method could significantly benefit their research. First, the stereoscopic rendering in 3D space enhances molecular structure comprehension. It enables them to estimate distances between structures and gain a better understanding of molecular interactions and alignments. 
Second, 3D space provides a flexible interaction platform with more degrees of freedom. One expert noted that ``this environment would be particularly useful for data analysis. I can analyze two molecular datasets on the surface and grab them to the 3D space to test various angles for fitting them together and observing their interaction. Afterward, I can return them back to the surface for further analysis.''
Third, the integration of 2D surface and 3D space combines their unique strengths, enhancing usability. Linking 2D representations on the surface with 3D visualizations in AR space, for instance, facilitates simultaneous editing of legends and observation of molecular interactions. Users can create and compare multiple copies within the environment.
Moreover, the experts discussed how \method could change their approach to data selection. Currently, they select data based on amino acid sequences. This process is cumbersome when the target region is spread over multiple sequence segments. To optimize the ligand affinity, for instance, they need to check the surrounding amino acids and select the pieces one by one from the sequence. Thus, they look forward to intelligent spatial selection techniques that facilitate efficient selections across both 2D and 3D representations to streamline this process.

The astronomer specializes in computational astrophysics, planetary system dynamics, few-body systems, and star cluster dynamics. After seeing the N-body simulation visualizations (\autoref{fig:teaser}(a)) in \method, the expert mentioned that 3D stereoscopic rendering enhances the understanding of data structures, providing an intuitive grasp of spatial relationships and dynamic processes within the data. Its capability is particularly beneficial for exploring simulations of physical phenomena and identifying specific data patterns. Conversely, the 2D surface can be used for coding to perform complex data analysis tasks. The combination of both visual interfaces can allow users to leverage their existing data analysis practices, while motivating them to test their hypotheses and observe 3D simulations more easily. The expert did not show a strong preference for whether the selection interaction should occur on the 2D surface or in 3D space. Instead, they highly valued the flexibility to select based on where the data feature is perceived.

The doctor's research focuses on orthopaedics. After observing and manipulating the anatomical data visualization (\autoref{fig:teaser}(c)), he expressed strong interest in \method and believed that it would play a significant role in surgical planning, where multi-views of 3D structures, 2D measurements, and precise interaction are required. The smooth and flexible transition between 2D and 3D allows doctors to view medical structures and the surgical plan from different perspectives. In addition, he mentioned that it is often difficult to control the depth at which a needle or a screw is inserted, which can harm neighboring tissues or organs. With \method, they can view specific treatment locations on the 2D interface and test, compute, and plan where and how needles or screws should be inserted in 3D space.

\subsection{Evaluation with VR/AR/MR experts}
\label{subsec:evaluation_VRMR}
We then invited three VR/AR/MR experts with each over three years of experience and daily technology use.
After obtaining their informed consent, we introduced them to \method, in two sessions. First, we focused on the manipulation tasks, including translation, rotation, and scaling. After a brief tutorial, we asked them to use our techniques to manipulate the molecular visualization from \autoref{subsec:Molecular}, comprising a combination of ribbon and licorice diagrams (\autoref{fig:teaser}(b)). We particularly evaluated four interaction transitions---two for translation (\autoref{fig:eli:DS}(c)) and rotation and scaling (\autoref{fig:eli:DS}(b))---and gathered their feedback in interviews. Our second session focused on our two selection techniques, BrushWYP and BrushLasso (\autoref{sec:app:pointcloud}). Following these two sessions, we asked them to fill in a questionnaire to assess and compare accuracy, efficacy, and user experience of both selection and manipulation techniques using 7-point Likert scales. In addition, throughout the study we encouraged them to share any feedback they may have.

The experts showed high engagement with our environment and its interactivity. They actively compared the visual effects and differences between the two visual representations across both spaces. All three experts expressed that they enjoyed the feeling of interacting with data seamlessly throughout the environment, appreciating the intuitive nature of the interactions without experiencing any confusion about how interactions should be performed.
The experts particularly focused on the approaches for transferring data between the two displays. In their comparison of the two translation techniques, they mentioned that both methods provided a natural and fluid movement of data across spaces. They felt, however, the 2D pinch + 3D pinch gesture to be more precise than the 2D pointing + 3D pinch one. One reason was that the 2D pinch gave them precise control of data depth---allowing them to directly correlate the distance between their fingers with depth adjustment of the data below the surface. Depth control with a single finger-pointing gesture, in contrast, was challenging as it depended on visual feedback from below the 2D surface. 
An interesting observation was that all three experts did not realize that the familiar 2D pinch gesture had been repurposed from scaling to facilitate translation from \belowS to \aboveS the surface. They seamlessly adapted to using an integrated 2D pinch + 3D pinch gesture to ``grab'' data across spaces. Moreover, while finding all interaction designs intuitive and easy to remember, the experts suggested that additional icons would be helpful visual cues for interaction, especially for transitions between the two distinct spaces.
All experts also appreciated the design of both selection methods. Two of them favored BrushWYP for its uniform brushing approach from start to finish, aligning with their expectation of how data selection should work---even when transitioning across two distinct spaces. The other expert felt that either method could be effective. In comparing the accuracy of both, all experts felt both methods supported them in selecting the intended target data above and below the surface. 
Interestingly, we did not observe that using a lasso on the 2D screen would enhance their perception of accuracy. This may be because their main challenge was not to achieve precise 2D input but rather the inability to track the data below the surface, unlike in AR space. We solved this issue, however, with our target-aware selection technique.

\section{\www{\method and CR visualization and interaction}}
Based on the design considerations of \method and other CR solutions (\autoref{sec:discussion}), and their potential use in domain research, we can now revisit the initial questions we posed: (1) how to represent data to ensure users understand what they see and (2) how to design interaction techniques that allow users to seamlessly continue their tasks without the need to deliberate on how interactions translate between different display spaces. We experimented with a variety of data representation strategies to elucidate data attributes (\eg, ball-stick, surface, stripes). We also investigated different visualization placements, \aboveS or \belowS the surface, employing various levels of detail and abstraction. We also explored the use of perspective views, orthographic views, or exclusively 2D cutting planes for presenting data content on 2D surfaces. One crucial lesson emerged from this work: it is not viable to impose strict limits on which visual representations should be displayed in each space. Instead, we need to treat CR environments as integrated systems, in which users can decide where data is displayed and how to interact with it.
The focus of our design considerations thus shifted toward managing \textbf{transitions} effectively, breaking down into two essential facets: visualization transformation and interaction transition. 

\textbf{Visualization transformation} involves the seamless migration of data representations between two distinct spaces, such as from a 2D display to 3D space. This process demands thoughtful design to ensure that the essence and clarity of the data are preserved during the transition to not break a user's mental model.
Lee \etal\cite{Lee:2022:ADS} have thoroughly explored visualization transformation approaches that can be optimally designed to support visualization tasks. The transition of 3D spatial data from 2D to 3D is inherently facilitated by the data's existing 3D structure if perspective views are employed. The merging of 2D visualizations on flat surfaces with 3D visualizations in spatial environments to represent data features coherently, however, presents a considerable challenge: visual complexity. This complexity arises as users are presented with a blend of visual representations---2D or 3D, with varied visual elements---tailored to highlight distinct data features. For instance, when data moves from a 2D display to 3D space and the visual representations shift from surfaces and lines to stripes and curves, there is a concern about whether users can maintain their situational awareness. 
This challenge of visual complexity is not unique but is inherent across many environment designs within CR settings, especially when incorporating both situated and embedded visualizations to enhance user understanding of data features. After discussions with domain experts, we identified that a promising approach is to enable users to adjust visual representations locally---irrespective of being in 2D, 3D, or intermediate space---based on their specific needs. 
The shallow-depth area across two spaces, in particular, may be an appropriate place for this kind of adjustment. With \method, users can effortlessly switch between different representations to dynamically examine data features.
This flexibility in adapting visual representations aligns with Schwajada \etal's \cite{schwajda:2023:TGD} findings, who highlight the benefits of user-controlled transformation in enhancing task performance efficiency by catering to the users' requirements.

\textbf{Interaction transition.}
To the best of our knowledge, there has not been a thorough exploration of interaction transition within CR environments. We thus introduce the notion of interaction transition as \emph{the process in which interactions extend across various levels of virtuality within a CR environment, ensuring all actions maintain coherence, thus creating a continuous and seamless interactive experience.}
Several factors influence this process, including the tactile feedback upon touching a physical surface, constraints in interacting with both depth and 2D surfaces, established mental models guiding 2D and 3D interactions, and how visual perception of data features impacts interaction.

Central to our understanding is the goal to preserve a user's \textbf{spatial awareness} to facilitate natural interaction within CR. This goal is twofold: first, we need to ensure that users maintain spatial awareness of their surrounding space (whether 2D or 3D), allowing them to seamlessly continue tasks without needing to reconsider interaction modalities for different display spaces. When using \textit{BrushLasso} (\autoref{fig:selectiontech}(b)) in CR, \eg, users intuitively brush data in 3D and then draw a circle to enclose target points on the 2D surface. Second, it is crucial that users maintain spatial awareness of their data (structure, orientation, position). This awareness guides them in understanding where and how to perform specific actions. Users can trace, for instance, the 3D structure of data with \textit{BrushWYP} (\autoref{fig:selectiontech}(a)) and seamlessly continue their interaction by following the structure onto the 2D surface.

To support such seamless transitions, our techniques need to be context- and target-aware. While a user's intention may be clear in one space, it may become ambiguous when transitioning to another. Considering the previous example, if a user is unable to trace a 3D point cloud on the 2D surface any technique needs to predict their selection intentions. This leads to another critical awareness we need to maintain: \textbf{situational awareness}---users should be able to predict the results of their actions in CR. \autoref{fig:ProteinAbstraction}(b) demonstrates this case---users use two pinch gestures to ``grab'' data from and to the surface. Their fingers naturally come into physical contact, promoting a continuous 3D pinch gesture to ``grab'' data into the 3D space. During this process, users have expected that the subsequent action would extract the data further. Another interaction technique, ``lifting'' (\autoref{fig:MedicalCuttingplane}(a)), allows users to temporarily elevate the 3D volume for annotating a 2D slice, expecting that releasing the pinch gesture returns the data to its original position on the surface once the annotation is complete. 
These interaction designs preserve spatial and situational awareness, thereby ensuring a deep engagement in task completion in the CR environment. Thus, they were highly favored by AR/VR experts in our evaluation.

\textbf{Limitations.} Similar to all visualization environments, SpatialTouch has its intrinsic limitations. Here, we focus on its limitations in hosting spatial visualizations. We discuss additional limitations caused by specific hardware settings in \autoref{sec:app:limitations}.
First, the stereoscopic rendering and the 2D monoscopic rendering may interfere with each other, potentially affecting the users' understanding of the spatial data. In our case, we use the HoloLens optical see-through HMD (OST)---when the 2D screen is too bright, it can interfere with the comprehension of the 3D rendering. Similarly, if a video see-through HMD (VST) is used, the stereoscopic rendering may occlude the 2D monoscopic rendering. In addition, the different visualization luminance between 2D displays and AR HMDs may also lead to a sense of disconnection. Therefore, blending the two renderings to obtain a coherent visualization is a meaningful research question for future development. Various factors should be considered to ensure an optimal color blending method, such as the specific AR device property (OST or VST), brightness, and other rendering-related settings.
Second, 2D rendering is limited by the size of the screen, whereas AR theoretically provides unlimited space for 3D rendering. As a result, parts of the data visualization in the environment may not appear complete. While domain experts tend to focus on the local region of the data during exploration, we did not encounter any issues in the interviews. However, future researchers may need to take this into consideration when large-size visualizations are used in cross-reality interfaces that include fixed-size screens.
Finally, collaborative data exploration is currently limited due to the design of FishTank VR monoscopic rendering on the 2D surface. Although users from different viewing angles can have their own view of the data in AR space, they share the same view rendered on the 2D surface. So only one user is able to see a cohesive 3D representation. A potential solution for the future would be to render several views on the screen at high frequency and use a technique similar to ``shutter glasses,'' allowing each user to see their specific 2D rendering on the shared 2D surface.%

\section{Conclusion}
\label{sec:conclusion}
Cross-reality is thus more than simply merging various levels of virtuality, where data is either positioned at a single space or moved to another place. Instead, it stands as an innovative and integrated environment for data presentation and exploration. Holding this vision, data can appear in any form and at any corner of the environment,  tailored to the specific needs of each domain. Through the interviews with domain experts, we learned that transitioning their data between 2D and 3D spaces significantly motivates them to view and analyze their data from previously unexplored angles. In this light, we shared the design insights we gained through experimenting with various configurations for the CR environment, resulting in our recommendations to respect the users' mental models of the spaces in which they interact as well as of their data.
Moreover, whether for understanding data or completing tasks, it is crucial to allow users to control data transformations, such as transforming data across spaces or change visual representations.  Yet, the most critical aspect to consider is why CR is the appropriate choice for the task and data. This rationale shapes all other design aspects: the choices of environment, visualization, and interaction designs.


\acknowledgments{
This work was supported by the National Natural Science Foundation of China (Grant No. 62272396). The authors would like to thank Faez Khan, Lanlan Han, Thijs Kouwenhoven, and Liang Wang for their valuable comments during the domain expert interviews.
}

\section*{Supplemental material pointers}

We share our additional material at \href{https://osf.io/avxr9}{\texttt{osf\discretionary{}{.}{.}io\discretionary{/}{}{/}avxr9}}. We also make our \method simulator available at \href{https://github.com/LixiangZhao98/Cross-Reality-Environment-SpatialTouch}{\texttt{github.com\discretionary{/}{}{/}LixiangZhao98\discretionary{/}{}{/}Cross-Reality-Environment-SpatialTouch}}
 and the point cloud visualization and density estimation available at \href{https://github.com/LixiangZhao98/PointCloud-Visualization-Tool}{\texttt{github.com\discretionary{/}{}{/}LixiangZhao98\discretionary{/}{}{/}PointCloud-Visualization-Tool}}.

\section*{Figures license/copyright}
We as authors state that all of our figures are and remain under our own personal copyright, with the permission to be used here. We make them available under the \href{https://creativecommons.org/licenses/by/4.0/}{Creative Commons At\-tri\-bu\-tion 4.0 International (\ccLogo\,\ccAttribution\ \mbox{CC BY 4.0})} license and share them at \href{https://osf.io/avxr9}{\texttt{osf.io/avxr9}}.

\bibliographystyle{abbrv-doi-hyperref-narrow}

\bibliography{abbreviations,reference}

\begin{thebibliography}{10}
\renewcommand*{\sfdefault}{PTSansNarrow-TLF}

\bibitem{alharbi:2021:nanotilus}
R.~Alharbi, O.~Strnad, L.~R. Luidolt, M.~Waldner, D.~Kou{\v{r}}il, C.~Bohak, T.~Klein, E.~Gr{\"o}ller, and I.~Viola.
\newblock Nanotilus: Generator of immersive guided-tours in crowded {3D} environments.
\newblock {\em IEEE Trans Vis Comput Graph}, 29(3):1860--1875, 2023. \href{https://doi.org/10/gtnn26}
{doi: \textsf{%
10\discretionary{/}{%
}{/}gtnn26}}


\bibitem{auda:2023:scoping}
J.~Auda, U.~Gruenefeld, S.~Faltaous, S.~Mayer, and S.~Schneegass.
\newblock A scoping survey on cross-reality systems.
\newblock {\em ACM Comput Surv}, 56(4),  article no. 83,  38 pages, 2023. \href{https://doi.org/10/gtnn22}
{doi: \textsf{%
10\discretionary{/}{%
}{/}gtnn22}}


\bibitem{Bach:2018:THM}
B.~Bach, R.~Sicat, J.~Beyer, M.~Cordeil, and H.~Pfister.
\newblock The hologram in my hand: How effective is interactive exploration of {3D} visualizations in immersive tangible augmented reality?
\newblock {\em IEEE Trans Vis Comput Graph}, 24(1):457--467, 2018. \href{https://doi.org/10/gcp924}
{doi: \textsf{%
10\discretionary{/}{%
}{/}gcp924}}


\bibitem{Becker:1987:BS}
R.~A. Becker and W.~S. Cleveland.
\newblock Brushing scatterplots.
\newblock {\em Technometrics}, 29(2):127--142, 1987. \href{https://doi.org/10/gf7bgr}
{doi: \textsf{%
10\discretionary{/}{%
}{/}gf7bgr}}


\bibitem{Benko:2010:MPI}
H.~Benko and A.~D. Wilson.
\newblock Multi-point interactions with immersive omnidirectional visualizations in a dome.
\newblock In {\em Proc.\ ITS}, pp. 19--28. ACM, New York, 2010. \href{https://doi.org/10/cxw789}
{doi: \textsf{%
10\discretionary{/}{%
}{/}cxw789}}


\bibitem{Benmahdjoub:2023:EAV}
M.~Benmahdjoub, A.~Thabit, M.-L.~C. van Veelen, W.~J. Niessen, E.~B. Wolvius, and T.~v. Walsum.
\newblock Evaluation of {AR} visualization approaches for catheter insertion into the ventricle cavity.
\newblock {\em IEEE Trans Vis Comput Graph}, 29(5):2434--2445, 2023. \href{https://doi.org/10/mpq6}
{doi: \textsf{%
10\discretionary{/}{%
}{/}mpq6}}


\bibitem{besanccon:2021:state-of-art}
L.~Besan{\c{c}}on, A.~Ynnerman, D.~F. Keefe, L.~Yu, and T.~Isenberg.
\newblock The state of the art of spatial interfaces for {3D} visualization.
\newblock {\em Comput Graph Forum}, 40(1):293--326, 2021. \href{https://doi.org/10/gjbpxp}
{doi: \textsf{%
10\discretionary{/}{%
}{/}gjbpxp}}


\bibitem{boorboor:2023:submerse}
S.~Boorboor, Y.~Kim, P.~Hu, J.~M. Moses, B.~A. Colle, and A.~E. Kaufman.
\newblock Submerse: Visualizing storm surge flooding simulations in immersive dis\-play ecologies.
\newblock {\em IEEE Trans Vis Comput Graph},  13 pages, 2024.
\newblock To appear. \href{https://doi.org/10/gt2smx}
{doi: \textsf{%
10\discretionary{/}{%
}{/}gt2smx}}


\bibitem{Buja:1991:IDV}
A.~Buja, J.~A. McDonald, J.~Michalak, and W.~Stuetzle.
\newblock Interactive data visualization using focusing and linking.
\newblock In {\em Proc.\ Visualization}, pp. 156--163. IEEE Comp.\ Soc., Los Alamitos, 1991. \href{https://doi.org/10/c7wvqq}
{doi: \textsf{%
10\discretionary{/}{%
}{/}c7wvqq}}


\bibitem{Butkiewicz:2019:MTP}
T.~Butkiewicz, A.~H. Stevens, and C.~Ware.
\newblock Multi-touch {3D} positioning with the {P}antograph technique.
\newblock In {\em Proc.\ I3D},  article no. 13,  9 pages. ACM, New York, 2019. \href{https://doi.org/10/gtnn2m}
{doi: \textsf{%
10\discretionary{/}{%
}{/}gtnn2m}}


\bibitem{Butkiewicz:2011:MTE}
T.~Butkiewicz and C.~Ware.
\newblock Multi-touch {3D} exploratory analysis of ocean flow models.
\newblock In {\em Proc.\ OCEANS}, pp. 746--755--10. IEEE, Piscataway, 2011. \href{https://doi.org/10/gjb2jq}
{doi: \textsf{%
10\discretionary{/}{%
}{/}gjb2jq}}


\bibitem{butscher:2018:CTO}
S.~Butscher, S.~Hubenschmid, J.~M{\"u}ller, J.~Fuchs, and H.~Reiterer.
\newblock Clusters, trends, and outliers: How immersive technologies can facilitate the collaborative analysis of multidimensional data.
\newblock In {\em Proc.\ CHI}, pp. 1--12. ACM, New York, 2018. \href{https://doi.org/10/ggfwqj}
{doi: \textsf{%
10\discretionary{/}{%
}{/}ggfwqj}}


\bibitem{chavent:2011:GAA}
M.~Chavent, A.~Vanel, A.~Tek, B.~Levy, S.~Robert, B.~Raffin, and M.~Baaden.
\newblock Gpu-accelerated atom and dynamic bond visualization using hyperballs: A unified algorithm for balls, sticks, and hyperboloids.
\newblock {\em Journal of computational chemistry}, 32(13):2924--2935, 2011. \href{https://doi.org/10/c3tdp5}
{doi: \textsf{%
10\discretionary{/}{%
}{/}c3tdp5}}


\bibitem{Chen:2020:LassoNet}
Z.~Chen, W.~Zeng, Z.~Yang, L.~Yu, C.-W. Fu, and H.~Qu.
\newblock {LassoNet}: Deep lasso-selection of {3D} point clouds.
\newblock {\em IEEE Trans Vis Comput Graph}, 26(1):195--204, 2020. \href{https://doi.org/10/gmghvt}
{doi: \textsf{%
10\discretionary{/}{%
}{/}gmghvt}}


\bibitem{coffey:2011:sliceWIM}
D.~Coffey, N.~Malbraaten, T.~B. Le, I.~Borazjani, F.~Sotiropoulos, A.~G. Erdman, and D.~F. Keefe.
\newblock Interactive {S}lice {WIM}: Navigating and interrogating volume data sets using a multisurface, multitouch {VR} interface.
\newblock {\em IEEE Trans Vis Comput Graph}, 18(10):1614--1626, 2012. \href{https://doi.org/10/fcds9g}
{doi: \textsf{%
10\discretionary{/}{%
}{/}fcds9g}}


\bibitem{ccoltekin:2020:ERS}
A.~{\c{C}}{\"o}ltekin, I.~Lochhead, M.~Madden, S.~Christophe, A.~Devaux, C.~Pettit, O.~Lock, S.~Shukla, L.~Herman, Z.~Stacho{\v{n}}, P.~Kub{\'{i}}{\v{c}}ek, D.~Snopkov{\'a}, S.~Bernardes, and N.~Hedley.
\newblock Extended reality in spatial sciences: A review of research challenges and future directions.
\newblock {\em ISPRS Int J Geo-Inf}, 9(7),  article no. 439,  29 pages, 2020. \href{https://doi.org/10/ghk9xk}
{doi: \textsf{%
10\discretionary{/}{%
}{/}ghk9xk}}


\bibitem{de:2013:mockup}
B.~R. De~Ara{\'u}jo, G.~Casiez, J.~A. Jorge, and M.~Hachet.
\newblock Mockup {B}uilder: {3D} modeling on and above the surface.
\newblock {\em Comput Graph}, 37(3):165--178, 2013. \href{https://doi.org/10/f4wv2s}
{doi: \textsf{%
10\discretionary{/}{%
}{/}f4wv2s}}


\bibitem{doutreligne:2014:unitymol}
S.~Doutreligne, T.~Cragnolini, S.~Pasquali, P.~Derreumaux, and M.~Baaden.
\newblock Unitymol: interactive scientific visualization for integrative biology.
\newblock In {\em 2014 IEEE 4th Symposium on Large Data Analysis and Visualization (LDAV)}, pp. 109--110. IEEE, 2014. \href{https://doi.org/10/m53v}
{doi: \textsf{%
10\discretionary{/}{%
}{/}m53v}}


\bibitem{Ens:2021:Uplift}
B.~Ens, S.~Goodwin, A.~Prouzeau, F.~Anderson, F.~Y. Wang, S.~Gratzl, Z.~Lucarelli, B.~Moyle, J.~Smiley, and T.~Dwyer.
\newblock Uplift: A tangible and immersive tabletop system for casual collaborative visual analytics.
\newblock {\em IEEE Trans Vis Comput Graph}, 27(2):1193--1203, 2021. \href{https://doi.org/10/ghgt5x}
{doi: \textsf{%
10\discretionary{/}{%
}{/}ghgt5x}}


\bibitem{Ferdosi:2011:DEM}
B.~Ferdosi, H.~Buddelmeijer, S.~Trager, M.~Wilkinson, and J.~Roerdink.
\newblock Comparison of density estimation methods for astronomical.
\newblock {\em Astron Astrophys}, 531,  article no. A114,  16 pages, 2011. \href{https://doi.org/10/cj3s5m}
{doi: \textsf{%
10\discretionary{/}{%
}{/}cj3s5m}}


\bibitem{Franzluebbers:2022:VRP}
A.~Franzluebbers, C.~Li, A.~Paterson, and K.~Johnsen.
\newblock Virtual reality point cloud annotation.
\newblock In {\em Proc.\ SUI},  article no. 14,  11 pages. ACM, New York, 2022. \href{https://doi.org/10/gtn4jk}
{doi: \textsf{%
10\discretionary{/}{%
}{/}gtn4jk}}


\bibitem{frohler:2022:survey_of_Cross-Virtuality}
B.~Fr{\"o}hler, C.~Anthes, F.~Pointecker, J.~Friedl, D.~Schwajda, A.~Riegler, S.~Tripathi, C.~Holzmann, M.~Brunner, H.~Jodlbauer, H.-C. Jetter, and C.~Heinzl.
\newblock A survey on cross-virtuality analytics.
\newblock {\em Comput Graph Forum}, 41(1):465--494, 2022. \href{https://doi.org/10/gtnn2w}
{doi: \textsf{%
10\discretionary{/}{%
}{/}gtnn2w}}


\bibitem{frohlich:2000:TCM}
B.~Fr{\"o}hlich and J.~Plate.
\newblock The {C}ubic {M}ouse: A new device for three-dimensional input.
\newblock In {\em Proc.\ CHI}, pp. 526--531. ACM, New York, 2000. \href{https://doi.org/10/dgg8k8}
{doi: \textsf{%
10\discretionary{/}{%
}{/}dgg8k8}}


\bibitem{Garrison:2022:TOV}
L.~A. Garrison, I.~Kolesar, I.~Viola, H.~Hauser, and S.~Bruckner.
\newblock Trends \& opportunities in visualization for physiology: A multiscale overview.
\newblock {\em Comput Graph Forum}, 41(3):609--643, 2022. \href{https://doi.org/10/gtmk6z}
{doi: \textsf{%
10\discretionary{/}{%
}{/}gtmk6z}}


\bibitem{Hancock:2010:SST}
M.~Hancock, T.~ten Cate, S.~Carpendale, and T.~Isenberg.
\newblock Supporting sandtray therapy on an interactive tabletop.
\newblock In {\em Proc.\ CHI}, pp. 2133--2142. ACM, New York, 2010. \href{https://doi.org/10/dv4k73}
{doi: \textsf{%
10\discretionary{/}{%
}{/}dv4k73}}


\bibitem{hong:2024:2D3D}
J.~Hong, R.~Hnatyshyn, E.~A. Santos, R.~Maciejewski, and T.~Isenberg.
\newblock A survey of designs for combined {2D}+{3D} visual representations.
\newblock {\em IEEE Trans Vis Comput Graph}, 30(6):2888--2902, 2024. \href{https://doi.org/10/gtnn2v}
{doi: \textsf{%
10\discretionary{/}{%
}{/}gtnn2v}}


\bibitem{jackson:2013:lightweight}
B.~Jackson, T.~Y. Lau, D.~Schroeder, K.~C. Toussaint, and D.~F. Keefe.
\newblock A lightweight tangible {3D} interface for interactive visualization of thin fiber structures.
\newblock {\em IEEE Trans Vis Comput Graph}, 19(12):2802--2809, 2013. \href{https://doi.org/10/gh38r2}
{doi: \textsf{%
10\discretionary{/}{%
}{/}gh38r2}}


\bibitem{jadhav:2022:Md-cave}
S.~Jadhav and A.~E. Kaufman.
\newblock {MD-Cave}: An immersive visualization workbench for radiologists.
\newblock {\em IEEE Trans Vis Comput Graph}, 29(12):4832--4844, 2022. \href{https://doi.org/10/gt2smv}
{doi: \textsf{%
10\discretionary{/}{%
}{/}gt2smv}}


\bibitem{kijima:1997:transition}
R.~Kijima and T.~Ojika.
\newblock Transition between virtual environment and workstation environment with projective head mounted display.
\newblock In {\em Proc.\ VR}, pp. 130--137. IEEE Comp.\ Soc., Los Alamitos, 1997. \href{https://doi.org/10/fwtdxw}
{doi: \textsf{%
10\discretionary{/}{%
}{/}fwtdxw}}


\bibitem{Krüger:IntenSelect+:2024}
M.~Krüger, T.~Gerrits, T.~Römer, T.~Kuhlen, and T.~Weissker.
\newblock {IntenSelect+}: Enhancing score-based selection in virtual reality.
\newblock {\em IEEE Trans Vis Comput Graph}, 30(5):2829--2838,  10 pages, 2024. \href{https://doi.org/10/gtn4jm}
{doi: \textsf{%
10\discretionary{/}{%
}{/}gtn4jm}}


\bibitem{kutak:2023:state-of-art-Molecular}
D.~Kut'{\'a}k, P.-p. V{\'a}zquez, T.~Isenberg, M.~Krone, M.~Baaden, J.~By{\v{s}}ka, B.~Kozl{\'\i}kov{\'a}, and H.~Miao.
\newblock State of the art of molecular visualization in immersive virtual environments.
\newblock {\em Comput Graph Forum}, 42(6),  article no. e14738,  29 pages, 2023. \href{https://doi.org/10/kt4n}
{doi: \textsf{%
10\discretionary{/}{%
}{/}kt4n}}


\bibitem{langner:2021:marvis}
R.~Langner, M.~Satkowski, W.~B{\"u}schel, and R.~Dachselt.
\newblock {MARVIS}: Combining mobile devices and augmented reality for visual data analysis.
\newblock In {\em Proc.\ CHI},  article no. 468,  17 pages. ACM, New York, 2021. \href{https://doi.org/10/gksmhm}
{doi: \textsf{%
10\discretionary{/}{%
}{/}gksmhm}}


\bibitem{Lee:2022:ADS}
B.~Lee, M.~Cordeil, A.~Prouzeau, B.~Jenny, and T.~Dwyer.
\newblock A design space for data visualisation transformations between {2D} and {3D} in mixed-reality environments.
\newblock In {\em Proc.\ CHI},  article no. 25,  14 pages. ACM, New York, 2022. \href{https://doi.org/10/kvj4}
{doi: \textsf{%
10\discretionary{/}{%
}{/}kvj4}}


\bibitem{Liang:2023:CRworkshopISMAR}
H.-N. Liang, H.-C. Jetter, F.~Maurer, U.~Gruenefeld, M.~Billinghurst, and C.~Anthes.
\newblock {\em 1\textsuperscript{st} Joint Workshop on Cross Reality}.
\newblock 2023. \href{https://doi.org/10/gtnn2r}
{doi: \textsf{%
10\discretionary{/}{%
}{/}gtnn2r}}


\bibitem{Liang:2023:MRworkshopVR}
H.-N. Liang, L.~Yu, and F.~Liarokapis.
\newblock Workshop: Mixing realities: Cross-reality visualization, interaction, and collaboration.
\newblock In {\em VR and 3DUI Abstracts and Workshops}, pp. 298--300. IEEE Comp.\ Soc., Los Alamitos, 2023. \href{https://doi.org/10/gtnn2q}
{doi: \textsf{%
10\discretionary{/}{%
}{/}gtnn2q}}


\bibitem{lopez:2015:TAU}
D.~L{\'o}pez, L.~Oehlberg, C.~Doger, and T.~Isenberg.
\newblock Towards an understanding of mobile touch navigation in a stereoscopic viewing environment for {3D} data exploration.
\newblock {\em IEEE Trans Vis Comput Graph}, 22(5):1616--1629, 2016. \href{https://doi.org/10/f8ghxc}
{doi: \textsf{%
10\discretionary{/}{%
}{/}f8ghxc}}


\bibitem{Lubos:2014:TCB}
P.~Lubos, R.~Beimler, M.~Lammers, and F.~Steinicke.
\newblock Touching the cloud: Bimanual annotation of immersive point clouds.
\newblock In {\em Proc.\ 3DUI}, pp. 191--192. IEEE Comp.\ Soc., Los Alamitos, 2014. \href{https://doi.org/10/gjb2n8}
{doi: \textsf{%
10\discretionary{/}{%
}{/}gjb2n8}}


\bibitem{Lucas:2005:DE3}
J.~F. Lucas.
\newblock Design and evaluation of {3D} multiple object selection techniques.
\newblock Master's thesis, Virginia Polytechnic Institute and State University, USA, 2005.
\newblock \href{http://hdl.handle.net/10919/31769}{hdl: \textsf{10919/31769}}.

\bibitem{Lundstrom:2011:MTT}
C.~Lundstr{\"o}m, T.~Rydell, C.~Forsell, A.~Persson, and A.~Ynnerman.
\newblock Multi-touch table system for medical visualization: Application to orthopedic surgery planning.
\newblock {\em IEEE Trans Vis Comput Graph}, 17(12):1775--1784, 2011. \href{https://doi.org/10/bzwxj8}
{doi: \textsf{%
10\discretionary{/}{%
}{/}bzwxj8}}


\bibitem{Martinet:2012:ISM}
A.~Martinet, G.~Casiez, and L.~Grisoni.
\newblock Integrality and separability of multitouch interaction techniques in {3D} manipulation tasks.
\newblock {\em IEEE Trans Vis Comput Graph}, 18(3):369--380, 2012. \href{https://doi.org/10/c6g7jt}
{doi: \textsf{%
10\discretionary{/}{%
}{/}c6g7jt}}


\bibitem{Meuschke:2017:CVV}
M.~Meuschke, S.~Voss, O.~Beuing, B.~Preim, and K.~Lawonn.
\newblock Combined visualization of vessel deformation and hemodynamics in cerebral aneurysms.
\newblock {\em IEEE Trans Vis Comput Graph}, 23(1):761--770, 2017. \href{https://doi.org/10/f92d8h}
{doi: \textsf{%
10\discretionary{/}{%
}{/}f92d8h}}


\bibitem{milgram:1995:reality-virtualitycontinuum}
P.~Milgram, H.~Takemura, A.~Utsumi, and F.~Kishino.
\newblock Augmented reality: A class of displays on the reality-virtuality continuum.
\newblock In {\em Proc.\ Telemanipulator and Telepresence Technologies}, pp. 282--292. SPIE, Bellingham, 1995. \href{https://doi.org/10/dh8jnv}
{doi: \textsf{%
10\discretionary{/}{%
}{/}dh8jnv}}


\bibitem{Mohammed:2018:Abstractocyte}
H.~Mohammed, A.~K. Al-Awami, J.~Beyer, C.~Cali, P.~Magistretti, H.~Pfister, and M.~Hadwiger.
\newblock Abstractocyte: A visual tool for exploring nanoscale astroglial cells.
\newblock {\em IEEE Trans Vis Comput Graph}, 24(1):853--861, 2018. \href{https://doi.org/10/gcqk6d}
{doi: \textsf{%
10\discretionary{/}{%
}{/}gcqk6d}}


\bibitem{Roberto:2020:HMS}
R.~A. Montano-Murillo, C.~Nguyen, R.~H. Kazi, S.~Subramanian, S.~DiVerdi, and D.~Martinez-Plasencia.
\newblock Slicing-{V}olume: Hybrid {3D}\discretionary{/}{}{/}{2D} multi-target selection technique for dense virtual environments.
\newblock In {\em Proc.\ VR}, pp. 53--62. IEEE Comp.\ Soc., Los Alamitos, 2020. \href{https://doi.org/10/gjb2j2}
{doi: \textsf{%
10\discretionary{/}{%
}{/}gjb2j2}}


\bibitem{Nguyen:2021:MIT}
N.~Nguyen, O.~Strnad, T.~Klein, D.~Luo, R.~Alharbi, P.~Wonka, M.~Maritan, P.~Mindek, L.~Autin, D.~S. Goodsell, and I.~Viola.
\newblock Modeling in the time of {COVID-19}: Statistical and rule-based mesoscale models.
\newblock {\em IEEE Trans Vis Comput Graph}, 27(2):722--732, 2021. \href{https://doi.org/10/k8sh}
{doi: \textsf{%
10\discretionary{/}{%
}{/}k8sh}}


\bibitem{O'donoghue:2010:VMS}
S.~I. O'Donoghue, D.~S. Goodsell, A.~S. Frangakis, F.~Jossinet, R.~A. Laskowski, M.~Nilges, H.~R. Saibil, A.~Schafferhans, R.~C. Wade, E.~Westhof, and A.~J. Olson.
\newblock Visualization of macromolecular structures.
\newblock {\em Nat Methods}, 7(3):S42--S55, 2010. \href{https://doi.org/10/df36n3}
{doi: \textsf{%
10\discretionary{/}{%
}{/}df36n3}}


\bibitem{Olwal:2008:RTP}
A.~Olwal, S.~Feiner, and S.~Heyman.
\newblock Rubbing and tapping for precise and rapid selection on touch-screen displays.
\newblock In {\em Proc.\ CHI}, pp. 295--304. ACM, New York, 2008. \href{https://doi.org/10/dgz29t}
{doi: \textsf{%
10\discretionary{/}{%
}{/}dgz29t}}


\bibitem{perelman:2022:visualtransitions}
G.~Perelman, E.~Dubois, A.~Probst, and M.~Serrano.
\newblock Visual transitions around tabletops in mixed reality: Study on a visual acquisition task between vertical virtual displays and horizontal tabletops.
\newblock {\em Proc ACM Hum-Comput Interact}, 6(ISS):660--679, 2022. \href{https://doi.org/10/grd27g}
{doi: \textsf{%
10\discretionary{/}{%
}{/}grd27g}}


\bibitem{pfeiffer:2018:imhotep}
M.~Pfeiffer, H.~Kenngott, A.~Preukschas, M.~Huber, L.~Bettscheider, B.~M{\"u}ller-Stich, and S.~Speidel.
\newblock Imhotep: virtual reality framework for surgical applications.
\newblock {\em International journal of computer assisted radiology and surgery}, 13:741--748, 2018. \href{https://doi.org/10/gdj9p9}
{doi: \textsf{%
10\discretionary{/}{%
}{/}gdj9p9}}


\bibitem{Reipschlager:2019:DesignAR}
P.~Reipschl{\"a}ger and R.~Dachselt.
\newblock {DesignAR}: Immersive {3D}-modeling combining augmented reality with interactive displays.
\newblock In {\em Proc.\ ISS}, pp. 29--41. ACM, New York, 2019. \href{https://doi.org/10/gtdktf}
{doi: \textsf{%
10\discretionary{/}{%
}{/}gtdktf}}


\bibitem{reipschlager:2020:PAR}
P.~Reipschl{\"a}ger, T.~Flemisch, and R.~Dachselt.
\newblock Personal augmented reality for information visualization on large interactive displays.
\newblock {\em IEEE Trans Vis Comput Graph}, 27(2):1182--1192, 2021. \href{https://doi.org/10/ghgt5w}
{doi: \textsf{%
10\discretionary{/}{%
}{/}ghgt5w}}


\bibitem{riegler:2020:CVVIC}
A.~Riegler, C.~Anthes, H.-C. Jetter, C.~Heinzl, C.~Holzmann, H.~Jodlbauer, M.~Brunner, S.~Auer, J.~Friedl, B.~Fr{\"o}hler, C.~Leitner, F.~Pointecker, D.~Schwajda, and S.~Tripathi.
\newblock Cross-virtuality visualization, interaction and collaboration.
\newblock In {\em Proc.\ XR{@}ISS},  article no. 1,  4 pages, 2020.
\newblock Online: \href{https://ceur-ws.org/Vol-2779/paper1.pdf}{\texttt{ceur\discretionary{}{-}{-}ws\discretionary{}{.}{.}org\discretionary{/}{}{/}Vol\discretionary{}{-}{-}2779\discretionary{/}{}{/}paper1\discretionary{}{.}{.}pdf}}.

\bibitem{Satriadi:2022:ASM}
K.~Satriadi, A.~Cunningham, B.~Thomas, A.~Drogemuller, A.~Odi, N.~Patel, C.~Aston, and R.~Smith.
\newblock Augmented scale models: Presenting multivariate data around physical scale models in augmented reality.
\newblock In {\em Proc.\ ISMAR}, pp. 54--63. IEEE Comp.\ Soc., Los Alamitos, 2022. \href{https://doi.org/10/gtnn2p}
{doi: \textsf{%
10\discretionary{/}{%
}{/}gtnn2p}}


\bibitem{Satriadi:2022:TGD}
K.~A. Satriadi, J.~Smiley, B.~Ens, M.~Cordeil, T.~Czauderna, B.~Lee, Y.~Yang, T.~Dwyer, and B.~Jenny.
\newblock Tangible globes for data visualisation in augmented reality.
\newblock In {\em Proc.\ CHI},  article no. 505,  16 pages. ACM, New York, 2022. \href{https://doi.org/10/grv5tm}
{doi: \textsf{%
10\discretionary{/}{%
}{/}grv5tm}}


\bibitem{Schafer:2024:InVADo}
M.~Sch{\"a}fer, N.~Brich, J.~By\v{s}ka, S.~M. Marques, D.~Bedn\'{a}\v{r}, P.~Thiel, B.~Kozlíková, and M.~Krone.
\newblock Invado: Interactive visual analysis of molecular docking data.
\newblock {\em IEEE Trans Vis Comput Graph}, 30(4):1984--1997, 2024. \href{https://doi.org/10/gtnn2n}
{doi: \textsf{%
10\discretionary{/}{%
}{/}gtnn2n}}


\bibitem{schwajda:2023:TGD}
D.~Schwajda, J.~Friedl, F.~Pointecker, H.-C. Jetter, and C.~Anthes.
\newblock Transforming graph data visualisations from {2D} displays into augmented reality {3D} space: A quantitative study.
\newblock {\em Front Virtual Reality}, 4,  article no. 1155628,  19 pages, 2023. \href{https://doi.org/10/kvj9}
{doi: \textsf{%
10\discretionary{/}{%
}{/}kvj9}}


\bibitem{seraji:2022:hybridaxes}
M.~R. Seraji and W.~Stuerzlinger.
\newblock Hybridaxes: An immersive analytics tool with interoperability between {2D} and immersive reality modes.
\newblock In {\em Proc.\ ISMAR-Adjunct}, pp. 155--160. IEEE Comp.\ Soc., Los Alamitos, 2022. \href{https://doi.org/10/gtnn2x}
{doi: \textsf{%
10\discretionary{/}{%
}{/}gtnn2x}}


\bibitem{sereno:2019:SVD}
M.~Sereno, L.~Besan{\c{c}}on, and T.~Isenberg.
\newblock Supporting volumetric data visualization and analysis by combining augmented reality visuals with multi-touch input.
\newblock In {\em EuroVis Posters}, pp. 21--23, 2019. \href{https://doi.org/10/kt5f}
{doi: \textsf{%
10\discretionary{/}{%
}{/}kt5f}}


\bibitem{Sereno:2022:HTT}
M.~Sereno, S.~Gosset, L.~Besançon, and T.~Isenberg.
\newblock Hybrid touch/tangible spatial selection in augmented reality.
\newblock {\em Comput Graph Forum}, 41(3):403--415, 2022. \href{https://doi.org/10/gqq53j}
{doi: \textsf{%
10\discretionary{/}{%
}{/}gqq53j}}


\bibitem{sereno:2020:CWAR}
M.~Sereno, X.~Wang, L.~Besan{\c{c}}on, M.~J. Mcguffin, and T.~Isenberg.
\newblock Collaborative work in augmented reality: A survey.
\newblock {\em IEEE Trans Vis Comput Graph}, 28(6):2530--2549, 2022. \href{https://doi.org/10/gjkq7w}
{doi: \textsf{%
10\discretionary{/}{%
}{/}gjkq7w}}


\bibitem{Rongkai:2023:EGA}
R.~Shi, Y.~Wei, X.~Qin, P.~Hui, and H.-N. Liang.
\newblock Exploring gaze-assisted and hand-based region selection in augmented reality.
\newblock {\em Proc ACM Hum-Comput Interact}, 7(ETRA),  article no. 160,  19 pages, 2023. \href{https://doi.org/10/gt2sm2}
{doi: \textsf{%
10\discretionary{/}{%
}{/}gt2sm2}}


\bibitem{Song:2016:HPA}
P.~Song, X.~Yan, W.~B. Goh, A.~Q. Chen, and C.-W. Fu.
\newblock Hand-posture-augmented multitouch interactions for exploratory visualization.
\newblock In {\em SIGGRAPH ASIA Technical Briefs},  article no. 27,  4 pages. ACM, New York, 2016. \href{https://doi.org/10/gtnn2k}
{doi: \textsf{%
10\discretionary{/}{%
}{/}gtnn2k}}


\bibitem{Sousa:2017:VRRRRoom}
M.~Sousa, D.~Mendes, S.~Paulo, N.~Matela, J.~Jorge, and D.~S.~o. Lopes.
\newblock {VRRRRoom}: Virtual reality for radiologists in the reading room.
\newblock In {\em Proc.\ CHI}, p. 4057–4062. ACM, New York, 2017. \href{https://doi.org/10/mn9z}
{doi: \textsf{%
10\discretionary{/}{%
}{/}mn9z}}


\bibitem{springel:2008:aquarius}
V.~Springel, J.~Wang, M.~Vogelsberger, A.~Ludlow, A.~Jenkins, A.~Helmi, J.~F. Navarro, C.~S. Frenk, and S.~D. White.
\newblock Aquarius project: The subhaloes of galactic haloes.
\newblock {\em Mon Not R Astron Soc}, 391(4):1685--1711, 2008. \href{https://doi.org/10/fsjgzw}
{doi: \textsf{%
10\discretionary{/}{%
}{/}fsjgzw}}


\bibitem{Springel:2005:SimulationsOT}
V.~Springel, S.~D.~M. White, A.~Jenkins, C.~S. Frenk, N.~Yoshida, L.~Gao, J.~F. Navarro, R.~J. Thacker, D.~J. Croton, J.~C. Helly, J.~A. Peacock, S.~Cole, P.~A. Thomas, H.~M.~P. Couchman, A.~E. Evrard, J.~M. Colberg, and F.~R. Pearce.
\newblock Simulations of the formation, evolution and clustering of galaxies and quasars.
\newblock {\em Nature}, 435:629--636, 2005. \href{https://doi.org/10/c3cmxr}
{doi: \textsf{%
10\discretionary{/}{%
}{/}c3cmxr}}


\bibitem{Strothoff:2011:TriangleCursor}
S.~Strothoff, D.~Valkov, and K.~Hinrichs.
\newblock Triangle cursor: Interactions with objects above the tabletop.
\newblock In {\em Proc.\ ITS}, pp. 111--119. ACM, New York, 2011. \href{https://doi.org/10/b6jrqq}
{doi: \textsf{%
10\discretionary{/}{%
}{/}b6jrqq}}


\bibitem{van:2011:IMV}
M.~van Der~Zwan, W.~Lueks, H.~Bekker, and T.~Isenberg.
\newblock Illustrative molecular visualization with continuous abstraction.
\newblock {\em Comput Graph Forum}, 30(3):683--690, 2011. \href{https://doi.org/10/c893rz}
{doi: \textsf{%
10\discretionary{/}{%
}{/}c893rz}}


\bibitem{Viola:2020:VA}
I.~Viola, M.~Chen, and T.~Isenberg.
\newblock Visual abstraction.
\newblock In {\em Foundations of Data Visualization}, chap.~2, pp. 15--37. Springer, Berlin, 2020. \href{https://doi.org/10/gk874c}
{doi: \textsf{%
10\discretionary{/}{%
}{/}gk874c}}


\bibitem{Viola:2018:PCA}
I.~Viola and T.~Isenberg.
\newblock Pondering the concept of abstraction in (illustrative) visualization.
\newblock {\em IEEE Trans Vis Comput Graph}, 24(9):2573--2588, 2018. \href{https://doi.org/10/gd3k7m}
{doi: \textsf{%
10\discretionary{/}{%
}{/}gd3k7m}}


\bibitem{Wang:2024:UPI}
N.~Wang, D.~Zielasko, and F.~Maurer.
\newblock User preferences for interactive {3D} object transitions in cross reality -- {A}n elicitation study.
\newblock In {\em Proc.\ AVI},  article no. 22,  9 pages. ACM, New York, 2024. \href{https://doi.org/10/gt2smz}
{doi: \textsf{%
10\discretionary{/}{%
}{/}gt2smz}}


\bibitem{Wang:2020:TUA}
X.~Wang, L.~Besan{\c{c}}on, D.~Rousseau, M.~Sereno, M.~Ammi, and T.~Isenberg.
\newblock Towards an understanding of augmented reality extensions for existing {3D} data analysis tools.
\newblock In {\em Proc.\ CHI},  article no. 528,  13 pages. ACM, New York, 2020. \href{https://doi.org/10/gm4jj8}
{doi: \textsf{%
10\discretionary{/}{%
}{/}gm4jj8}}


\bibitem{Wang:2019:ATD}
X.~Wang, L.~Besançon, M.~Ammi, and T.~Isenberg.
\newblock Augmenting tactile {3D} data navigation with pressure sensing.
\newblock {\em Comput Graph Forum}, 38(3):635--647, 2019. \href{https://doi.org/10/gjbgt9}
{doi: \textsf{%
10\discretionary{/}{%
}{/}gjbgt9}}


\bibitem{Ware:1993:FishTankVR}
C.~Ware, K.~Arthur, and K.~S. Booth.
\newblock Fish tank virtual reality.
\newblock In {\em Proc.\ CHI}, pp. 37--42. ACM, New York, 1993. \href{https://doi.org/10/cqjwdf}
{doi: \textsf{%
10\discretionary{/}{%
}{/}cqjwdf}}


\bibitem{Wiebel:2012:WYSIWYP}
A.~Wiebel, F.~M. Vos, D.~Foerster, and H.-C. Hege.
\newblock {WYSIWYP}: What you see is what you pick.
\newblock {\em IEEE Trans Vis Comput Graph}, 18(12):2236--2244, 2012. \href{https://doi.org/10/f4ft8h}
{doi: \textsf{%
10\discretionary{/}{%
}{/}f4ft8h}}


\bibitem{Wills:2008:LDV}
G.~Wills.
\newblock Linked data views.
\newblock In C.-h. Chen, W.~H{\"a}rdle, and A.~Unwin, eds., {\em Handbook of Data Visualization}, chap. II.9, pp. 217--241. Springer, Berlin, 2008. \href{https://doi.org/10/dkvn57}
{doi: \textsf{%
10\discretionary{/}{%
}{/}dkvn57}}


\bibitem{Wills:1996:S5W}
G.~J. Wills.
\newblock Selection: 524,288 ways to say ``this is interesting''.
\newblock In {\em Proc.\ InfoVis}, pp. 54--60. IEEE Comp.\ Soc., Los Alamitos, 1996. \href{https://doi.org/10/bjq6dh}
{doi: \textsf{%
10\discretionary{/}{%
}{/}bjq6dh}}


\bibitem{wu:2020:megereality}
S.~Wu, D.~Byrne, and M.~W. Steenson.
\newblock ``megereality'': Leveraging physical affordances for multi-device gestural interaction in augmented reality.
\newblock In {\em CHI Extended Abstracts},  article no. INT008,  4 pages. ACM, New York, 2020. \href{https://doi.org/10/gmkzwj}
{doi: \textsf{%
10\discretionary{/}{%
}{/}gmkzwj}}


\bibitem{Yu:2012:ESA}
L.~Yu, K.~Efstathiou, P.~Isenberg, and T.~Isenberg.
\newblock Efficient structure-aware selection techniques for {3D} point cloud visualizations with {2DOF} input.
\newblock {\em IEEE Trans Vis Comput Graph}, 18(12):2245--2254, 2012. \href{https://doi.org/10/f4fv9z}
{doi: \textsf{%
10\discretionary{/}{%
}{/}f4fv9z}}


\bibitem{Yu:2016:CEE}
L.~Yu, K.~Efstathiou, P.~Isenberg, and T.~Isenberg.
\newblock {CAST}: Effective and efficient user interaction for context-aware selection in {3D} particle clouds.
\newblock {\em IEEE Trans Vis Comput Graph}, 22(1):886--895, 2016. \href{https://doi.org/10/kt5n}
{doi: \textsf{%
10\discretionary{/}{%
}{/}kt5n}}


\bibitem{Yu:2022:VeLight}
L.~Yu, J.~Ouwerling, P.~Svetachov, F.~H.~J. van Hoesel, P.~M.~A. van Ooijen, and J.~Kosinka.
\newblock {VeLight}: A {3D} virtual reality tool for {CT}-based anatomy teaching and training.
\newblock {\em J Vis}, 25(2):293--306, 2022. \href{https://doi.org/10/mpq7}
{doi: \textsf{%
10\discretionary{/}{%
}{/}mpq7}}


\bibitem{Yu:2010:FI3D}
L.~Yu, P.~Svetachov, P.~Isenberg, M.~H. Everts, and T.~Isenberg.
\newblock {FI3D}: Direct-touch interaction for the exploration of {3D} scientific visualization spaces.
\newblock {\em IEEE Trans Vis Comput Graph}, 16(6):1613--1622, 2010. \href{https://doi.org/10/fc2df8}
{doi: \textsf{%
10\discretionary{/}{%
}{/}fc2df8}}


\bibitem{zhao:2023:metacast}
L.~Zhao, T.~Isenberg, F.~Xie, H.-N. Liang, and L.~Yu.
\newblock {MeTACAST}: Target-and context-aware spatial selection in {VR}.
\newblock {\em IEEE Trans Vis Comput Graph}, 30(1):480--494, 2024. \href{https://doi.org/10/gtnn25}
{doi: \textsf{%
10\discretionary{/}{%
}{/}gtnn25}}


\end{thebibliography}
\clearpage

\clearpage

\begin{strip} 
\noindent\begin{minipage}{\textwidth}
\makeatletter
\centering%
\sffamily\bfseries\fontsize{15}{16.5}\selectfont
\vgtc@title\\[.5em]
\large Appendix\\[.75em]
\makeatother
\normalfont\rmfamily\normalsize\noindent\raggedright In this appendix we provide additional images and discussion beyond the material that we could include in the main paper due to space limitations or because it was not essential for explaining our approach.
\end{minipage}
\end{strip}

\appendix

\section{Datasets used in the elicitation study}
\label{sec:app:ElicitationDataset}

\begin{figure}[t]
    \centering
        {\begin{overpic}[width=0.96\columnwidth]{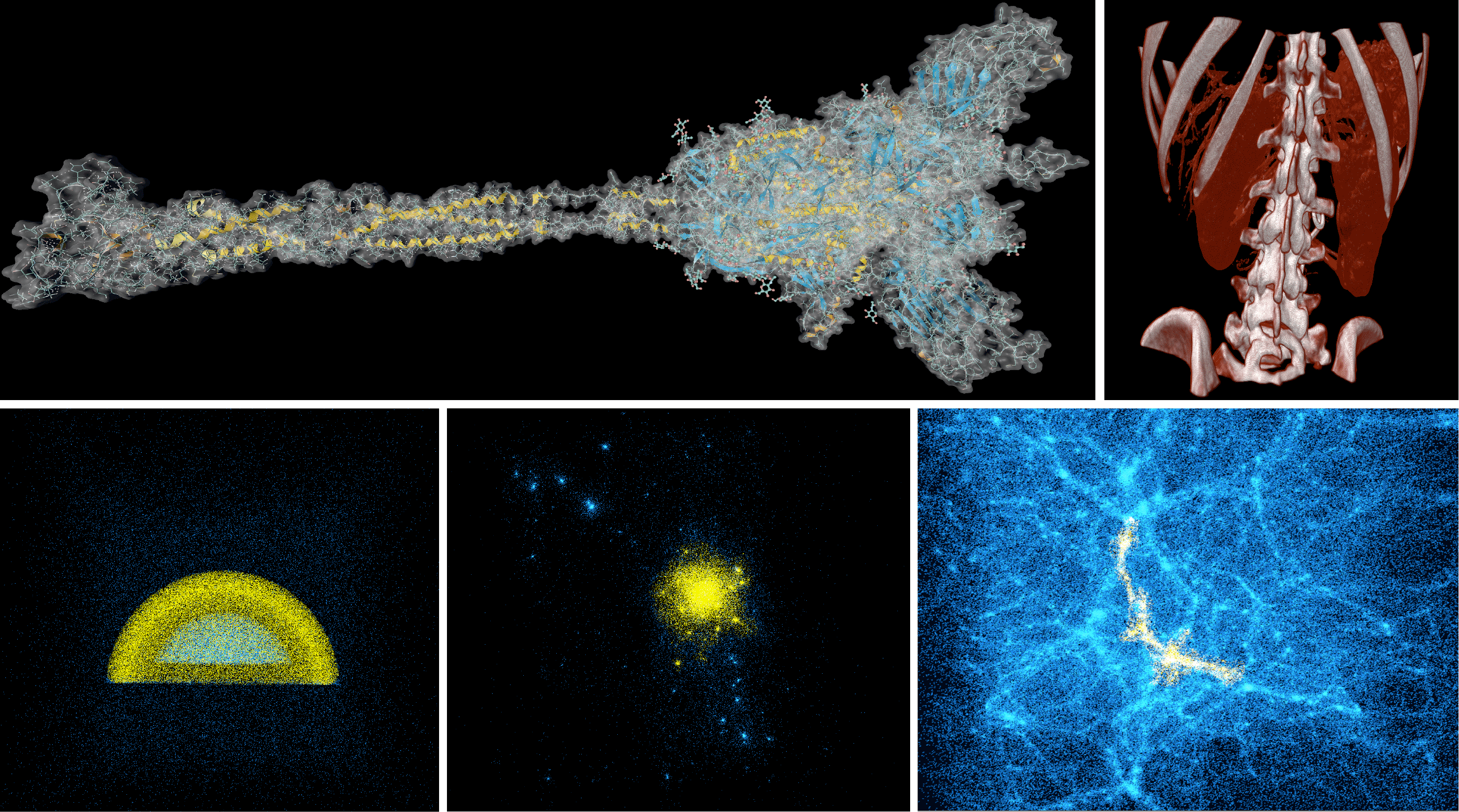}%
            \put(0.6,29){\textcolor{white}{(a)}}
            \put(76,29){\textcolor{white}{(b)}}
            \put(0.6,0.7){\textcolor{white}{(c)}}
             \put(31,0.7){\textcolor{white}{(d)}}
              \put(63,0.7){\textcolor{white}{(e)}}
        \end{overpic}}
        \caption{Elicitation study datasets: (a) Protein data (data from \cite{Nguyen:2021:MIT})
        , (b) MRI multi-slice anatomical data, (c)filament.}
    \label{fig:app:ElicitationDataset}
\end{figure}

\textbf{\textit{Datasets.}} We used five datasets with a diverse set of data features, including:
\begin{description}[nosep,leftmargin=1.5em,labelindent=0em,leftmargin=!,labelindent=!,itemindent=!,font=\normalfont\itshape]
\item[Structured surface data:] The molecular structure data (\autoref{fig:app:ElicitationDataset}(a)) is a spike protein from the SARS-CoV-2 virus\cite{Nguyen:2021:MIT}, reconstructed from the electron microscopy images. It is rendered as a ribbon diagram, which includes lines, sheets, and helix structures.
\item[Volume visualization data:] The MRI volume data (\autoref{fig:app:ElicitationDataset}(b)) comprises multiple slices and facilitates interactive cutting plane exploration to focus on specific regions of the anatomical structures.
\item[Unstructured point cloud data:] We used three point cloud datasets: (1) a synthetic semi-spherical shell of particles  (\autoref{fig:app:ElicitationDataset}(c)) that partially encompasses a half-ball of interfering particles; (2) a cosmological N-body simulation \cite{springel:2008:aquarius} (\autoref{fig:app:ElicitationDataset}(d)) with a vast, densely populated central cluster encircled by numerous smaller clusters; and (3) the Millennium-II data subset \cite{Springel:2005:SimulationsOT} (\autoref{fig:app:ElicitationDataset}(e)), a complex network of filaments connecting high-density clusters.
\end{description}

\section{Design Space for User interaction in CR}
\label{sec:app:DS}
\begin{figure*}[t]
    \centering
    \includegraphics[width=\linewidth]{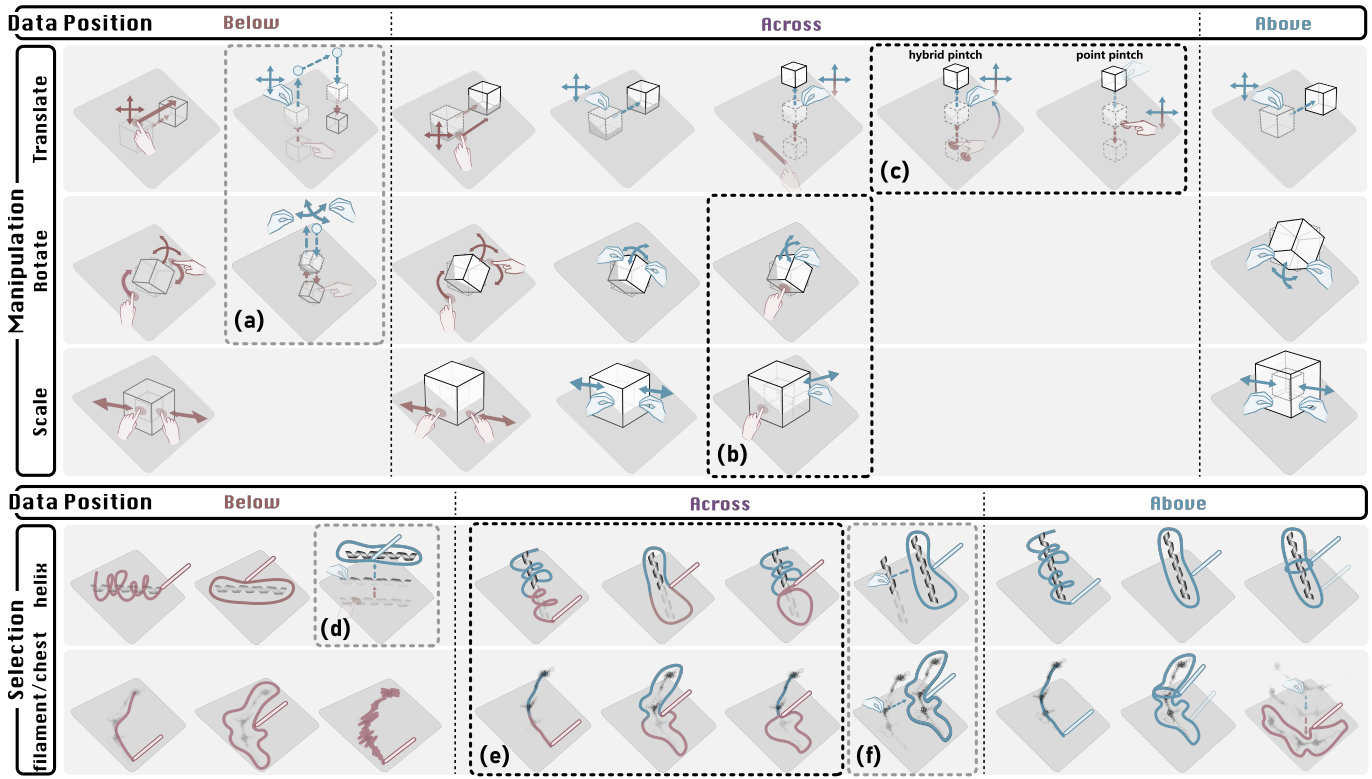}
    \caption{Design space for interaction techniques for two visualization tasks: data manipulation and selection. \textcolor{colorbelow}{Red}: interactions on 2D surface; \textcolor{colorabove}{Blue}: interactions in 3D space. \belowS, \acrossS, and \aboveS: positions of the target data/location. (a), (d), (f): move data above the surface for interaction. (b) (c) and (e): interaction transitions across both spaces.}
    \label{fig:app:DS}
\end{figure*}

To improve the readability of \autoref{fig:eli:DS}, we also include it in this appendix at full page width as \autoref{fig:app:DS}.

\section{Algorithm for Camera Setting}
\label{sec:app:camera}

\setlength{\lineskiplimit}{-\maxdimen}%
To ensure the data visualized on the surface and through the Hololens merge seamlessly into a unified 3D representation, we developed a rendering algorithm that aligns the visual content on the 2D surface with the AR visualization. To achieve this alignment we need to compute the surface position of data located below the surface. The positions of the surface camera and the AR HMD camera are denoted as $\mathbf{r}^{(c)}$ and $\mathbf{r}^{(h)}$, respectively. The configuration process involves the following key steps in each frame:
\begin{enumerate}[nosep,left=0pt .. \parindent]
    \item We translate $\mathbf{c}$ to the HMD's position, in which $\mathbf{r}^{(c)}$=$\mathbf{r}^{(h)}$.
    \item We adjust the forward direction of the $\mathbf{c}$ (negative $z$-axis) to $\mathbf{s}$ perpendicularly, in which the $z$-axis of $\mathbf{c}$ aligns with the $z$-axis of $\mathbf{s}$.
    \item We modify the $x$-axis of $\mathbf{c}$ to align with the $x$-axis of $\mathbf{s}$, maintaining consistency in horizontal orientation between $\mathbf{c}$ and $\mathbf{s}$.    
    \item We compute the surface's center position $\mathbf{r}^{(s)}$, the bottom-left corner $\mathbf{r}^{(bl)}$, and the top-right corner $\mathbf{r}^{(tr)}$ in local coordinates w.r.t.\ $\mathbf{r}^{(c)}$.
    \item The surface $\mathbf{s}$ serves as the near projection plane of $\mathbf{c}$. The distance of $\mathbf{c}$'s far projection plane is denoted by $f$. Then, we subsequently derive the projection matrix $\mathbf{m}^{(c)}$ as:
\end{enumerate}
\setlength{\lineskiplimit}{\lineskiplimitbackup}%
\renewcommand{\arraystretch}{1.7} 
\[\mathbf{m}^{(c)}=
\begin{pmatrix}
\frac{2\mathbf{r}_{z}^{(s)}}{\mathbf{r}_{x}^{(tr)}-\mathbf{r}_{x}^{(bl)}} & 0 &\frac{\mathbf{r}_{x}^{(tr)}+\mathbf{r}_{x}^{(bl)}}{\mathbf{r}_{x}^{(tr)}-\mathbf{r}_{x}^{(bl)}} & 0  \\
0 & \frac{2\mathbf{r}_{z}^{(s)}}{\mathbf{r}_{y}^{(tr)}-\mathbf{r}_{y}^{(bl)}}  &\frac{\mathbf{r}_{y}^{(tr)}+\mathbf{r}_{y}^{(bl)}}{\mathbf{r}_{y}^{(tr)}-\mathbf{r}_{y}^{(bl)}} & 0  \\
0 & 0  &\frac{-(f+\mathbf{r}_{z}^{(s)})}{f-\mathbf{r}_{z}^{(s)}}&\frac{-2f\mathbf{r}_{z}^{(s)}}{f-\mathbf{r}_{z}^{(s)}} \\
0 & 0  &-1 & 0  \\
\end{pmatrix}
\]

\section{Additional Figures for the Application Cases}
\label{fig:app:cases}
In this section, we present figures of three cases discussed in the paper to demonstrate how \method is tailored to three distinct domains.

\subsection{Molecular visualization}
\begin{figure}[t]
    \centering
    \includegraphics[width=\linewidth]{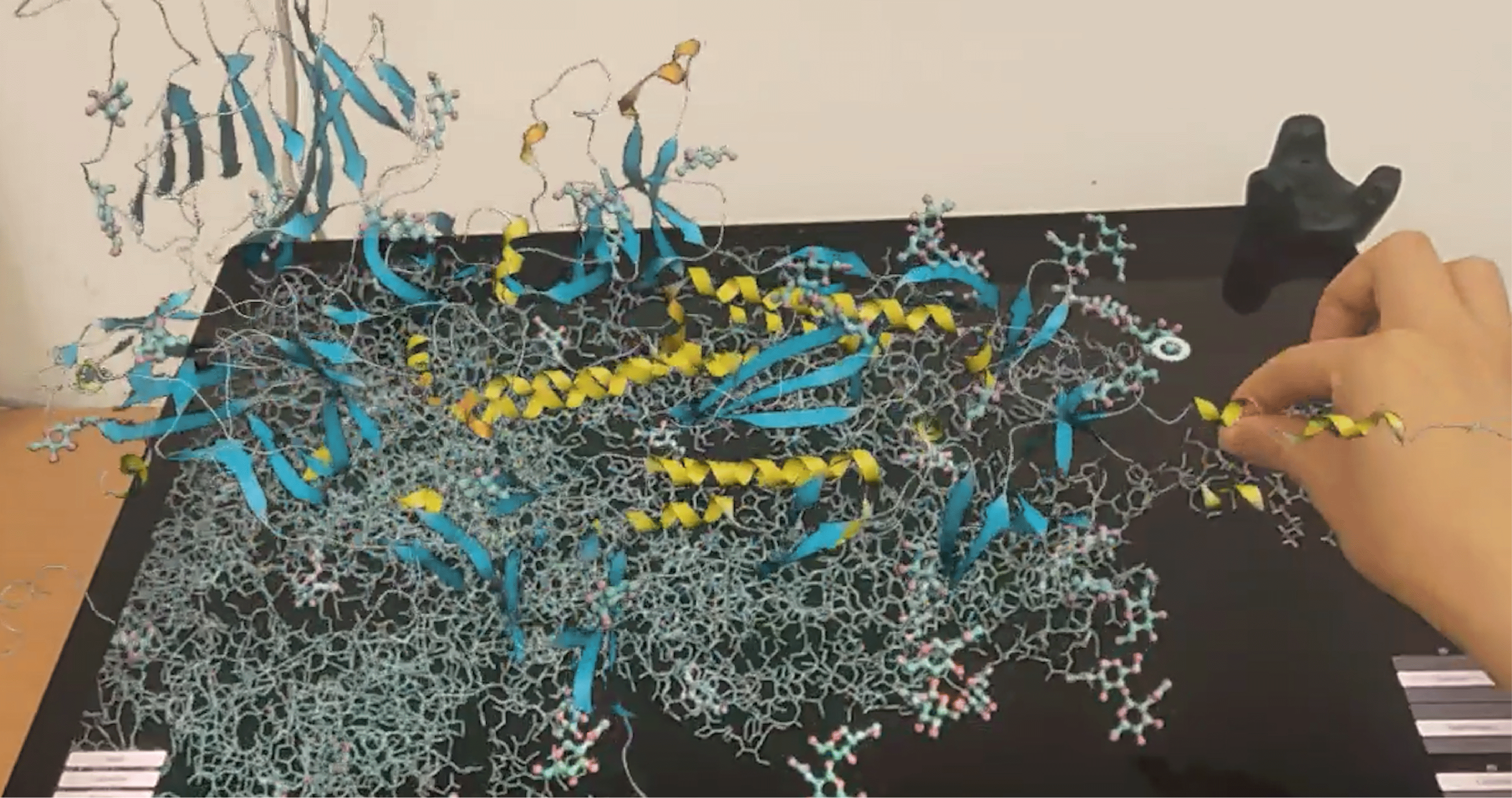}
    \caption{Protein data visualization \cite{Nguyen:2021:MIT} with the ribbon (shown in AR) and the licorice (shown on the surface) visual representations.}
    \label{fig:append:Molecular_VisRep}
     \vspace{-1ex}
\end{figure}

\begin{figure}[t]
\centering  
 \includegraphics[width=0.96\linewidth]{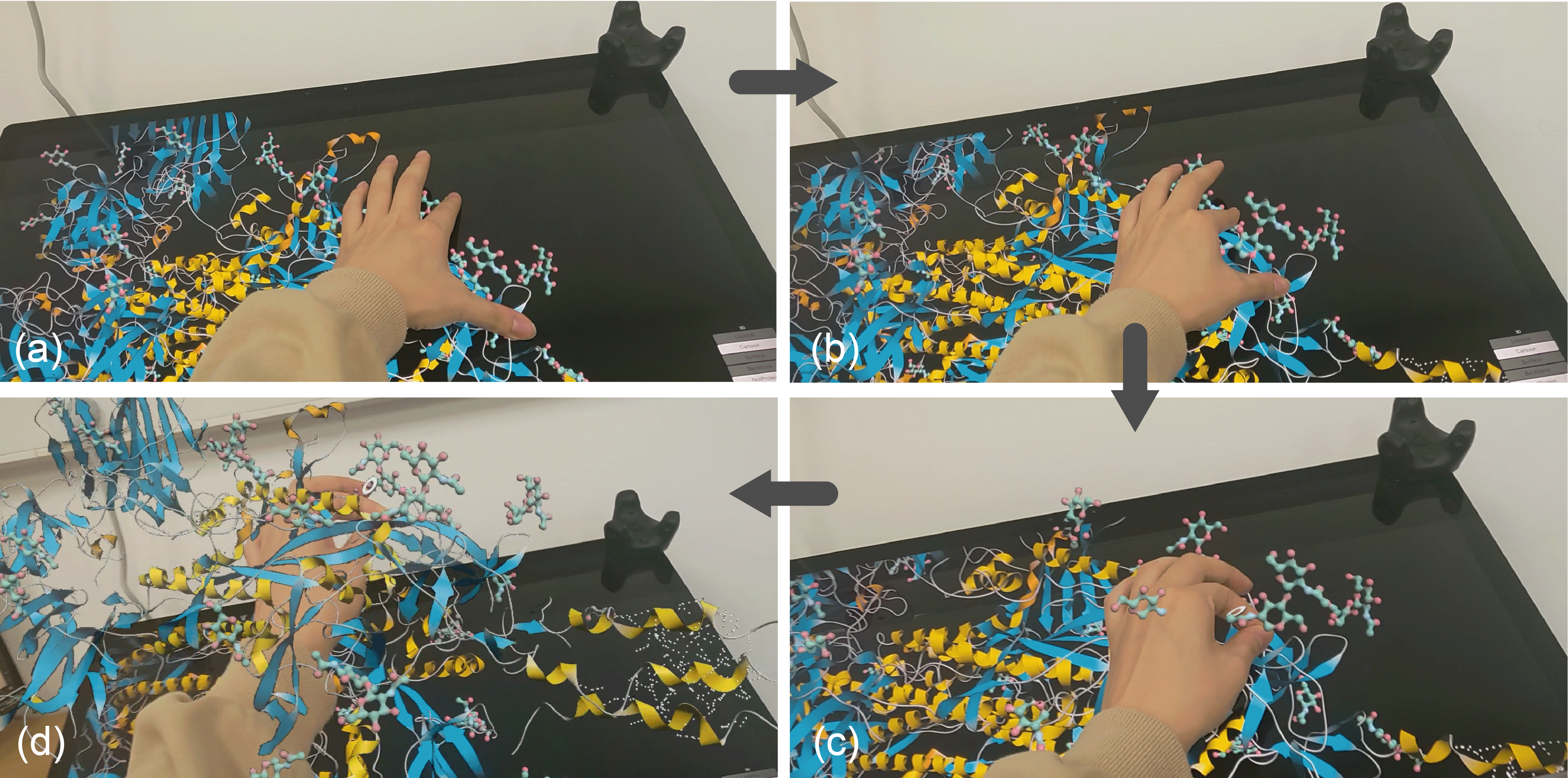}
\caption{The interaction transition technique to move the data visualization from 2D to 3D spaces. (a) start pinching on the surface to ``pull'' the visualization from 2D to 3D. (b) during the pinching, the visualization ``moves towards 3D space'' with the distance between the thumb and index finger decreasing. (c) the two fingers merge and the visualization is pulled up to near the level of the surface. (d) continue to move the visualization by pinching in the air.}
\label{fig:append:InteractionTrans}
\end{figure}

As shown in \autoref{fig:append:Molecular_VisRep}, we merge two distinct display spaces (AR HMD and Surface) to facilitate a concurrent visualization of different representations and abstractions. 
We developed a seamless interaction transition technique to enable users to move data across two spaces. It supports users to employ a familiar 2D pinch gesture on the touch surface to ``pull'' the visualization below the surface to the surface level and continue to use pinch gesture in 3D space to manipulate the visualization (shown in \autoref{fig:append:InteractionTrans}).

\subsection{Medical anatomical visualization}
\begin{figure}[t]
    \centering
    \includegraphics[width=\linewidth]{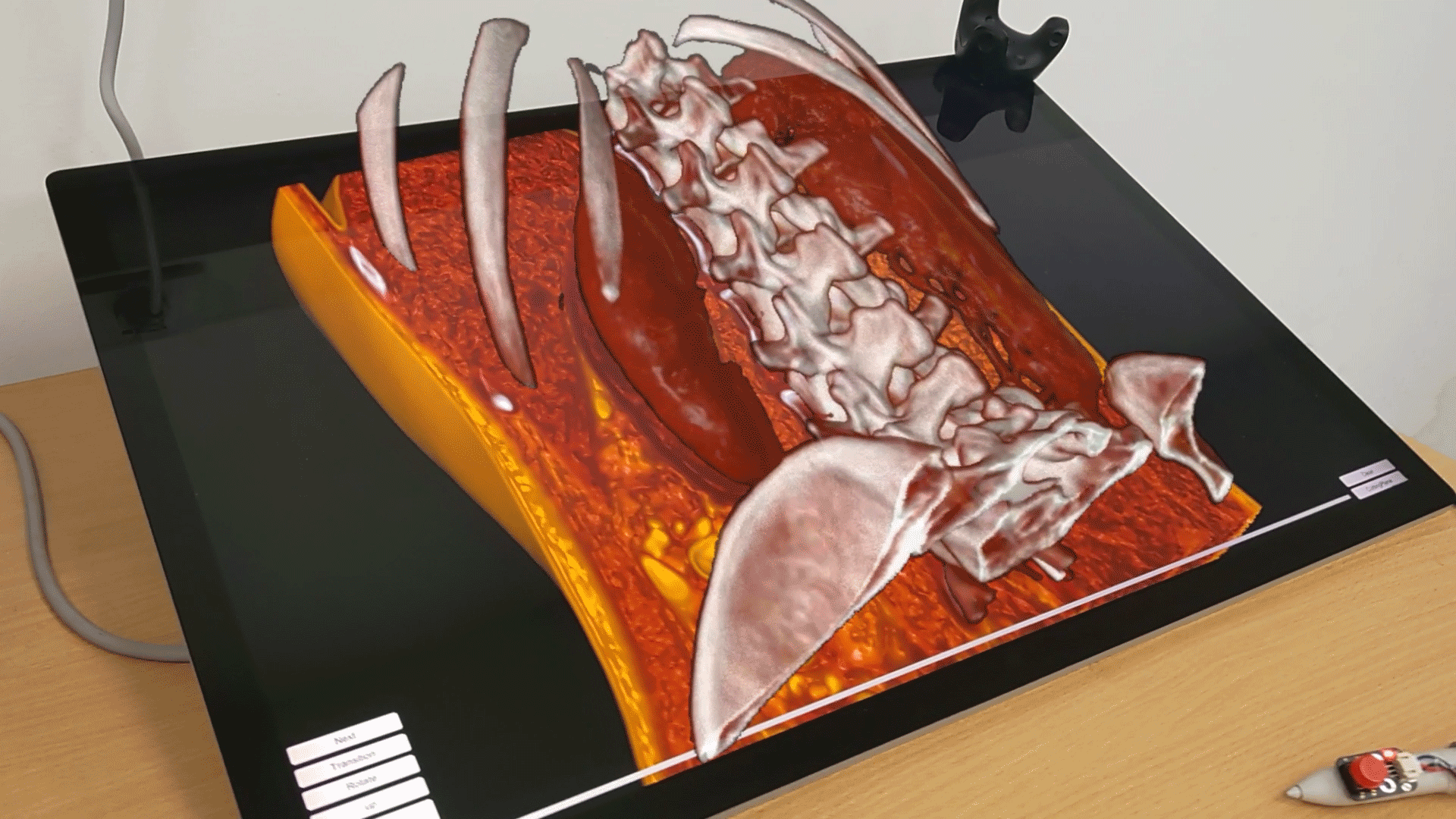}
    \caption{3D anatomic visualization on the surface and AR device with different transfer function settings.}
    \label{fig:append:Medical_3D3D}
     \vspace{-1ex}
\end{figure}

\method allows users to view 2D slices directly on the surface, while also observing stereoscopic renderings superimposed on the slice (\autoref{fig:append:Medical_3D3D}).

\begin{figure}[t]
    \centering
    \includegraphics[width=\linewidth]{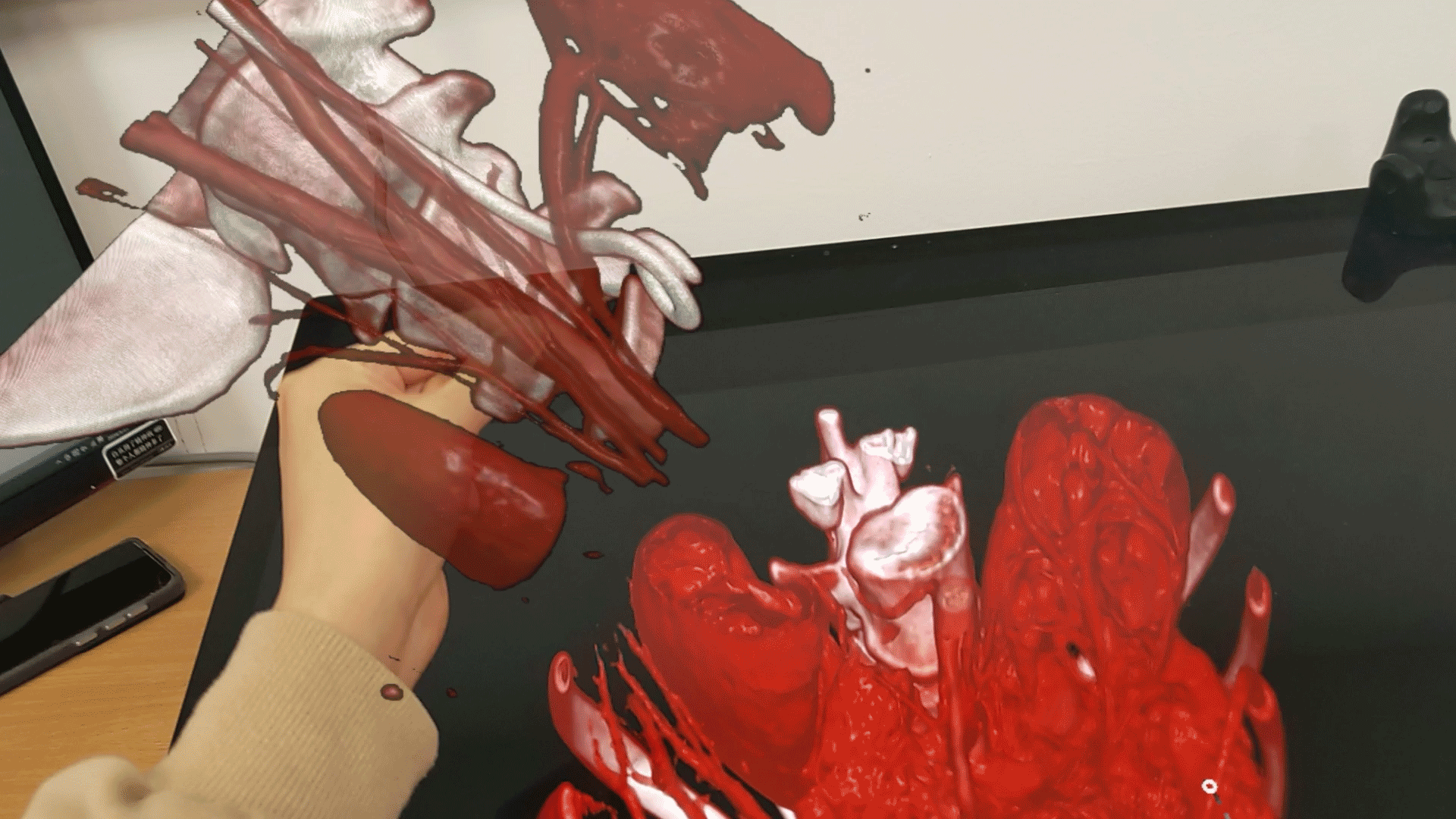}
    \caption{Lifting the stereo rendering with left pinch. }
    \label{fig:append:medical_lifting}
     \vspace{-1ex}
\end{figure}

Users can use the ``lifting'' feature to elevate the 3D volume visualization away from the 2D slice on the touch surface (\autoref{fig:append:medical_lifting}).

\begin{figure}[t]
\centering  

\subfigure{%
        {\begin{overpic}[width=0.48\columnwidth]{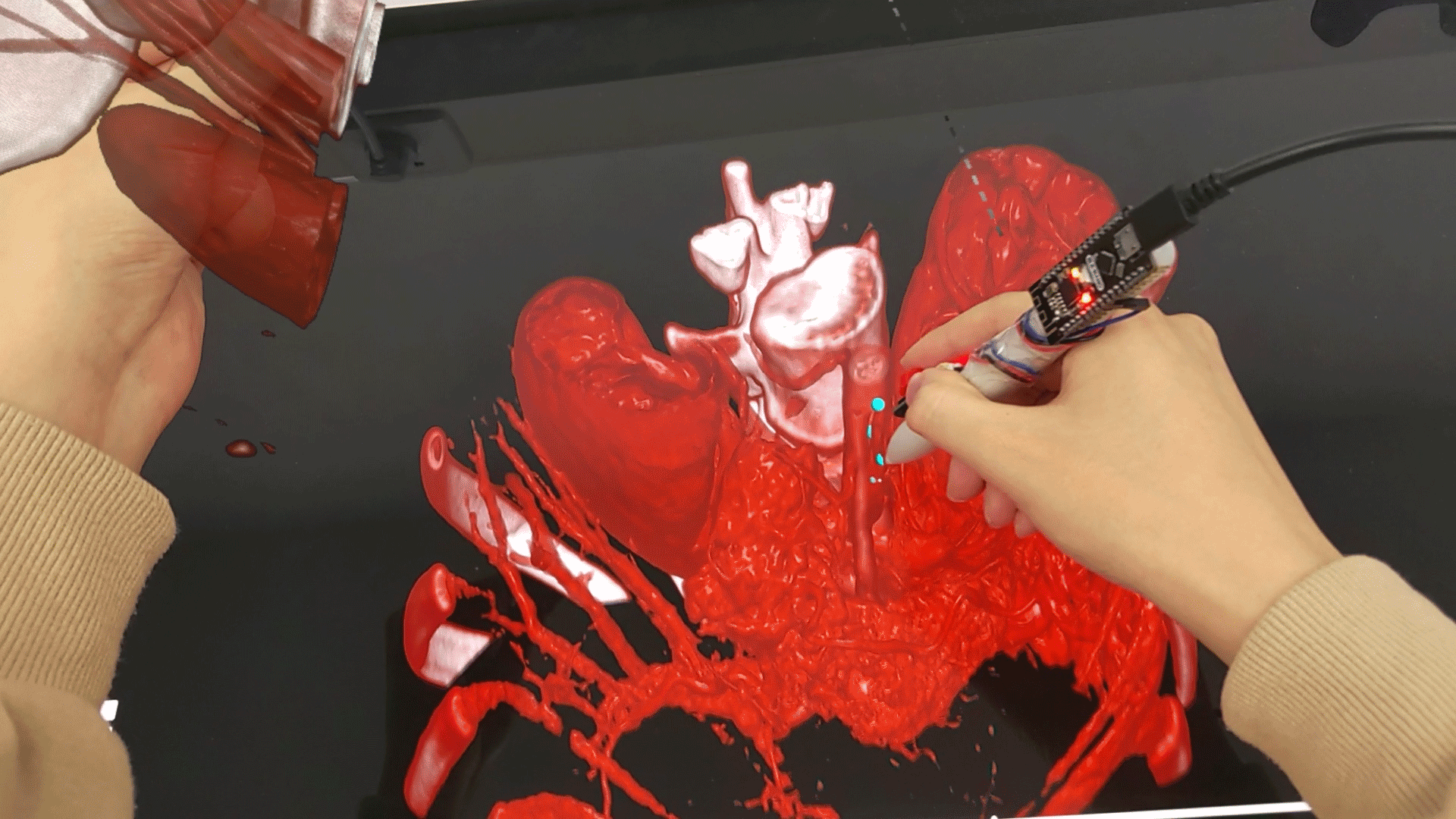}%
            \put(3,3){\textcolor{white}{(a)}}%
        \end{overpic}}
}\hfill%
\subfigure{%
        {\begin{overpic}[width=0.48\columnwidth]{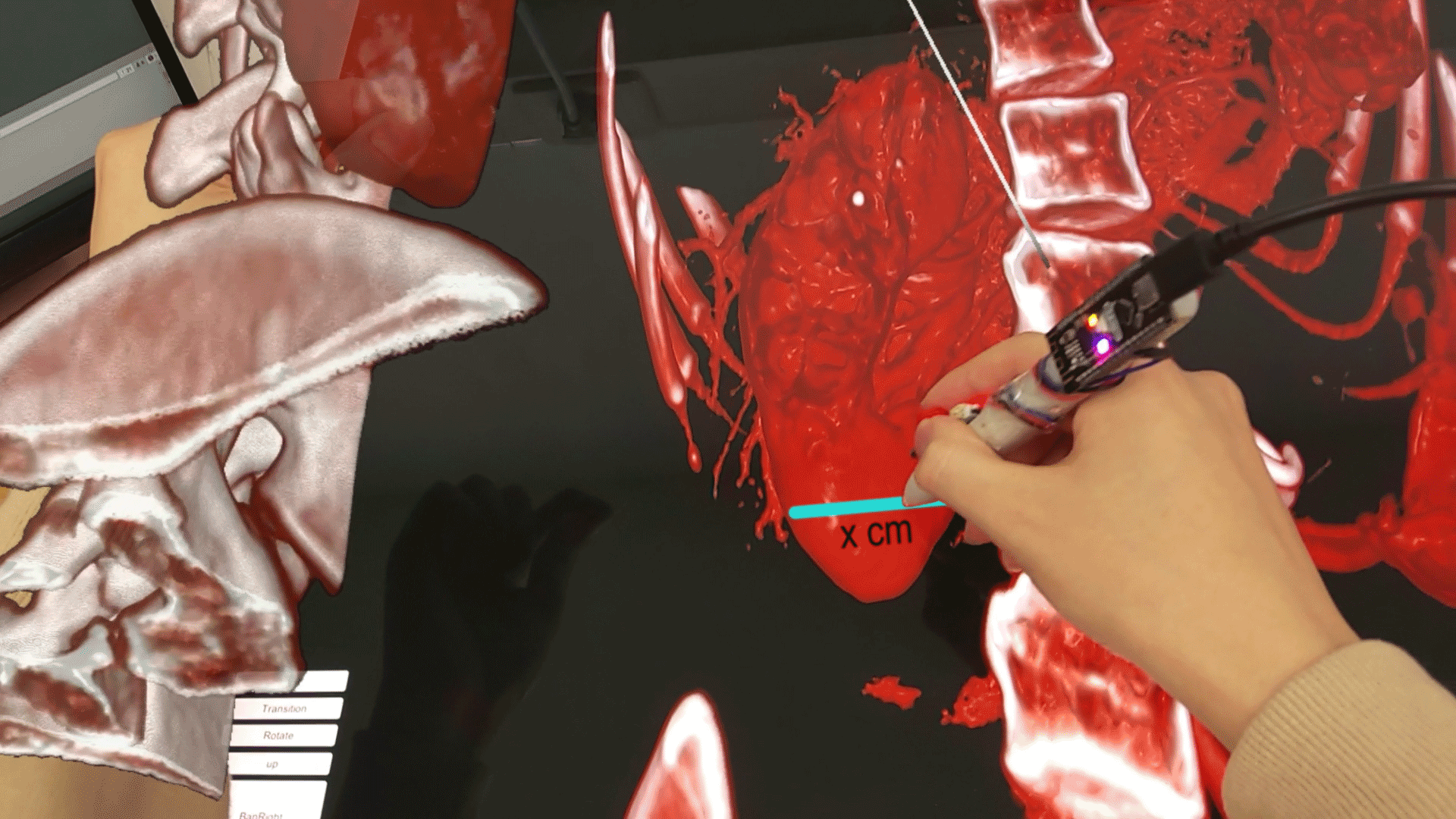}%
            \put(3,3){\textcolor{white}{(b)}}%
        \end{overpic}}
        }
\caption{Lifting the data and (a) annotating the features in depth, (b) measuring the width of the lung with pen on 2D surface.}
\label{fig:append:Medical_markmeasure}
\end{figure}

In this way, users can obtain a clear view of the 2D slices and perform accurate interaction on the surface, such as precise annotation (\autoref{fig:append:Medical_markmeasure} (a)) and distance measurements (\autoref{fig:append:Medical_markmeasure} (b)). \method supports users in marking features on the 3D visualization directly on the 2D slice (on the surface). This method is particularly useful when features appear in depth but are rendered on the surface.

\subsection{Astronomical point cloud visualization}
\begin{figure}[t]
    \centering
    \includegraphics[width=\linewidth]{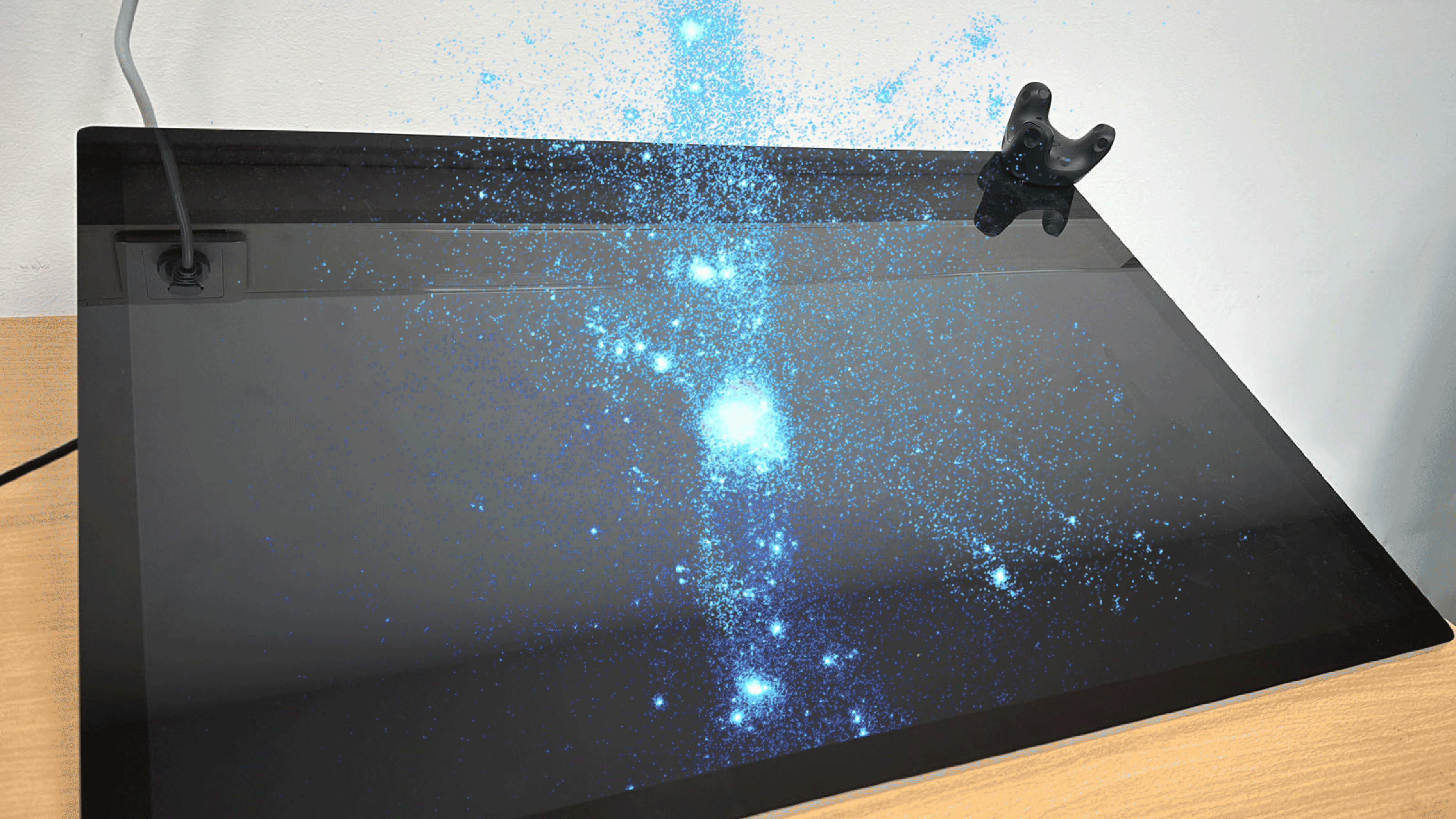}
    \caption{Cosmological N-body simulation visualization \cite{springel:2008:aquarius} across 2D (dark blue) and 3D environments (light blue).}
    \label{fig:append:pointcloud}
     \vspace{-1ex}
\end{figure}

As shown in \autoref{fig:append:pointcloud}, users can view 3D point cloud visualizations in AR space to gain a comprehensive understanding of data density distribution and context, while performing precise data analysis, such as selection or annotation, on a 2D surface.

\begin{figure}[t]
\centering  
\subfigure{%
        {\begin{overpic}[width=0.48\columnwidth]{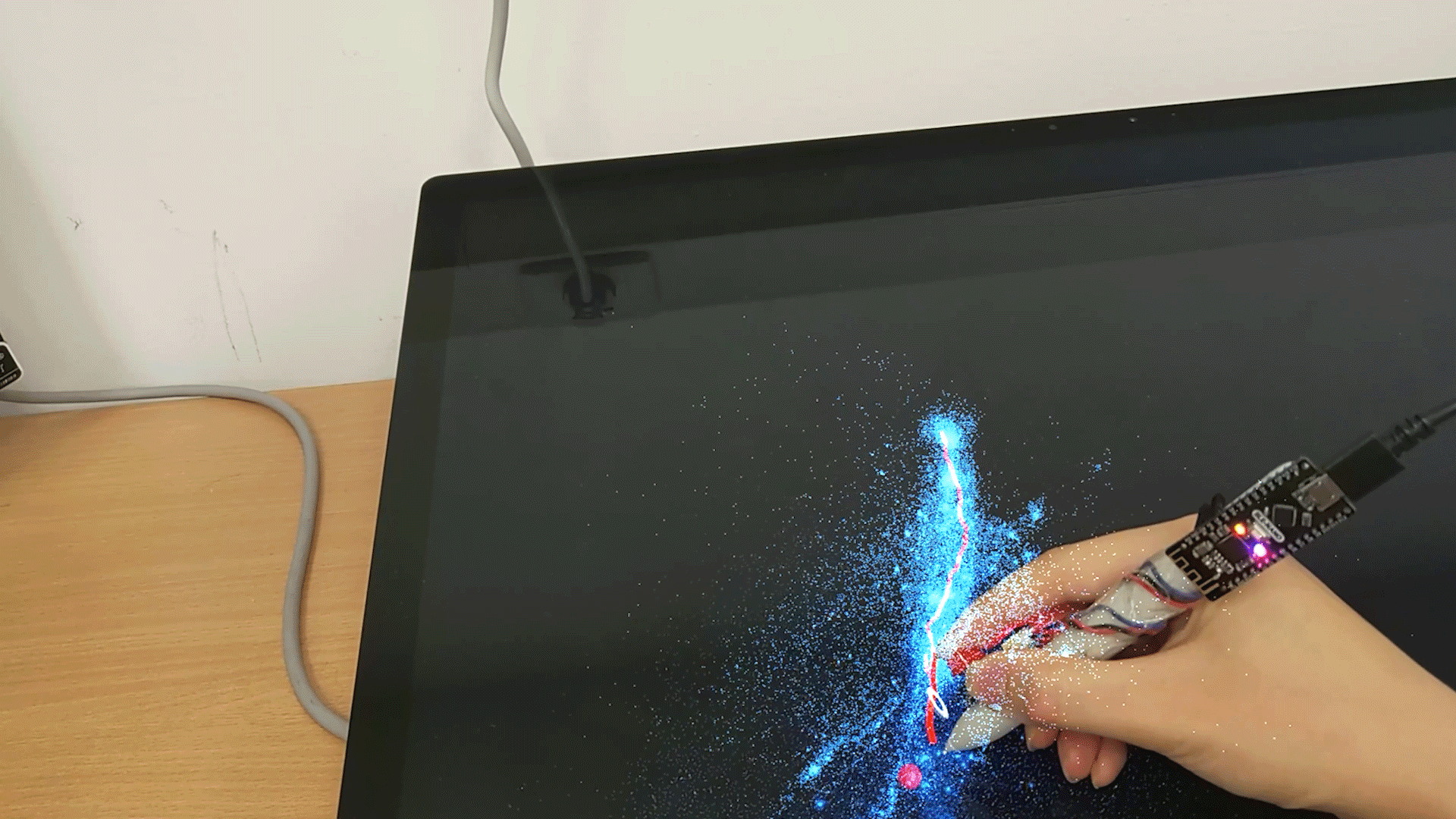}%
            \put(3,3){\textcolor{white}{(a)}}%
        \end{overpic}}
}\hfill%
\subfigure{%
        {\begin{overpic}[width=0.48\columnwidth]{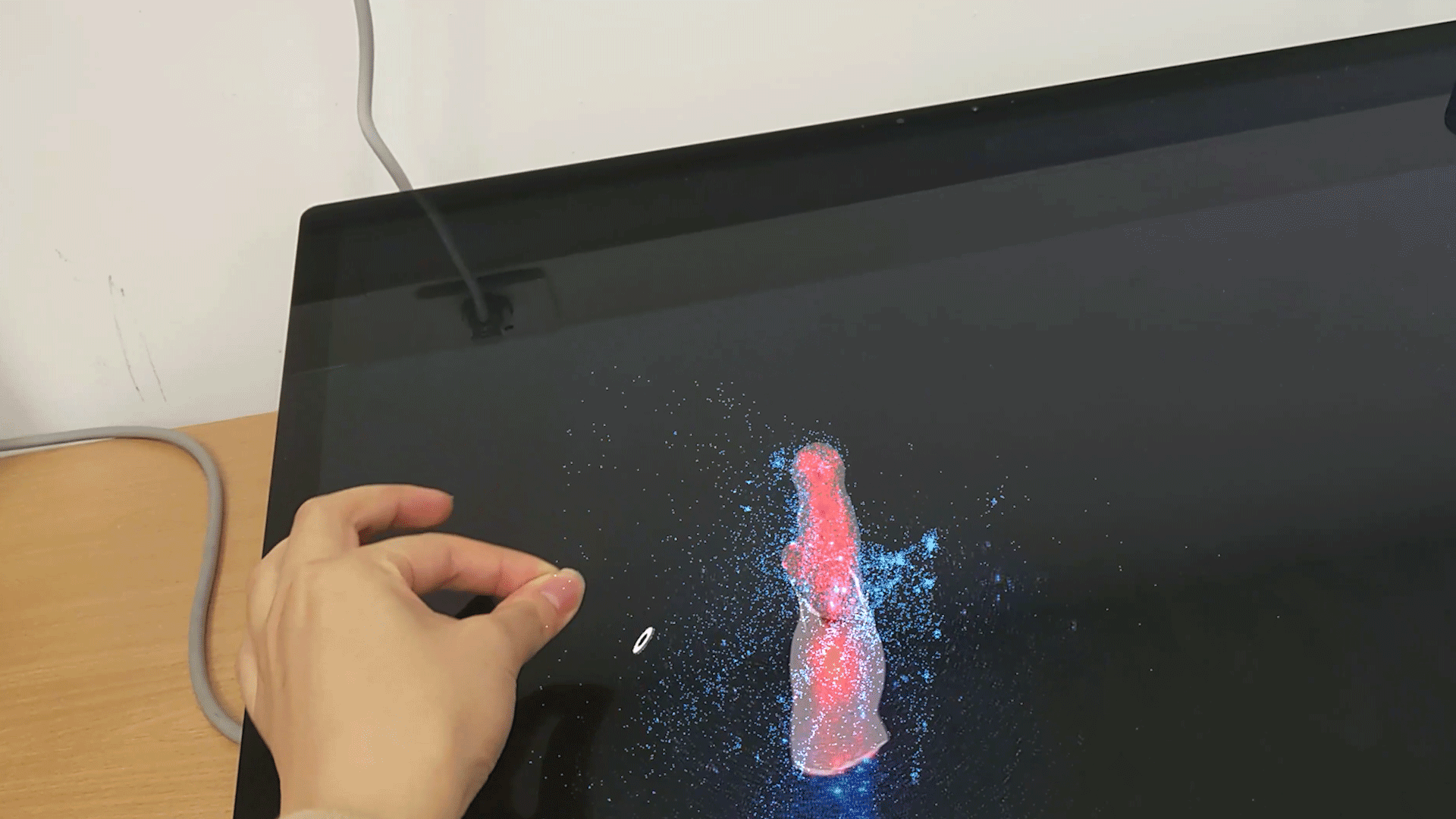}%
            \put(3,3){\textcolor{white}{(b)}}%
        \end{overpic}}
}
\caption{BrushWYP: (1) direct brushing on the target points across two spaces, and (b) the selection result.}
\label{fig:append:BrushWYP}
\end{figure}

We developed two seamless spatial selection techniques for point cloud visualization. With BrushWYP (\autoref{fig:append:BrushWYP}), users can brush over the string-like shape of 3D point cloud data in the 3D space and continue to brush the rest along the structure on the 2D surface.

\begin{figure}[t]
\centering  
\subfigure{%
        {\begin{overpic}[width=0.48\columnwidth]{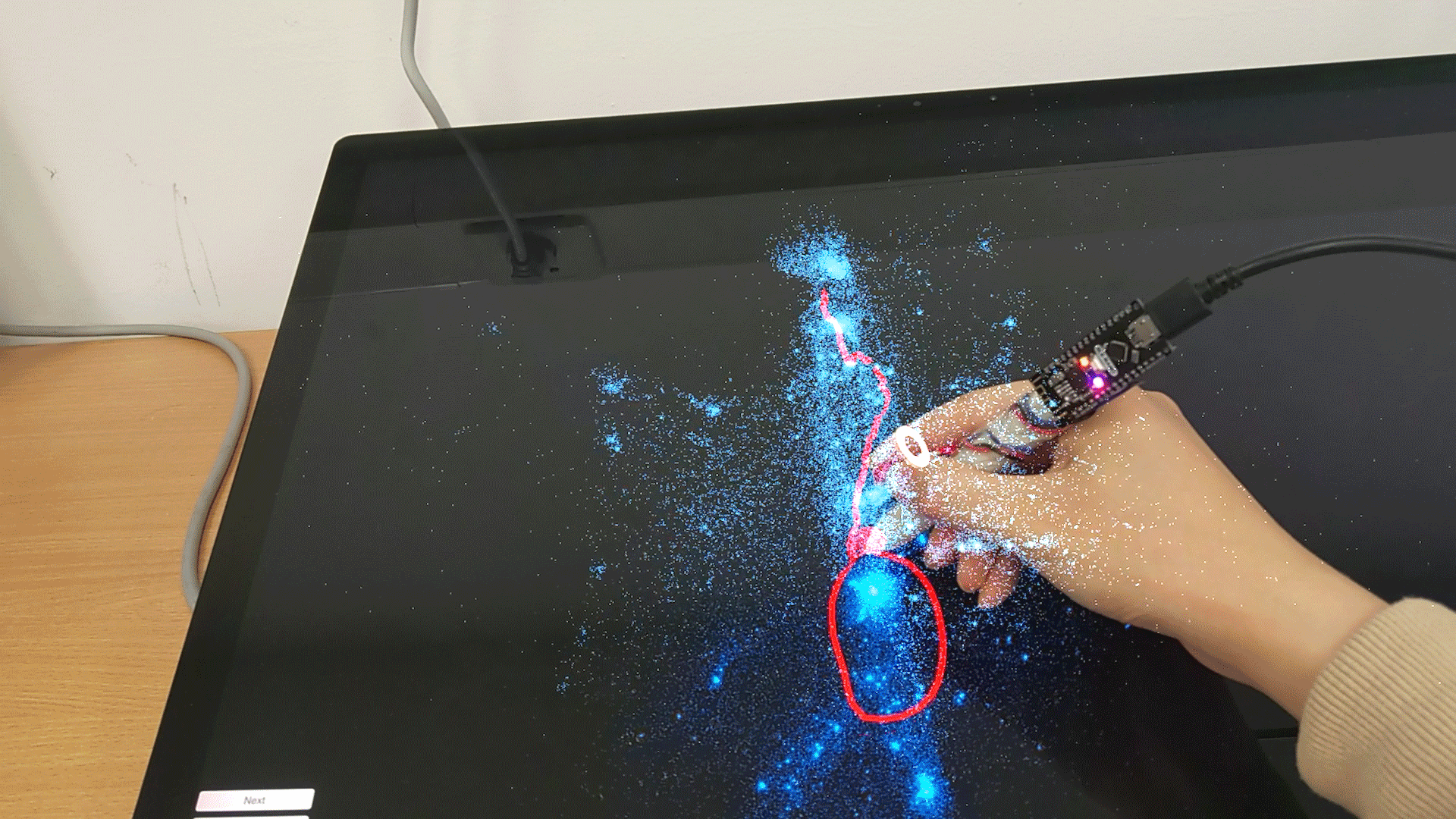}%
            \put(3,3){\textcolor{white}{(a)}}%
        \end{overpic}}
}\hfill%
\subfigure{%
        {\begin{overpic}[width=0.48\columnwidth]{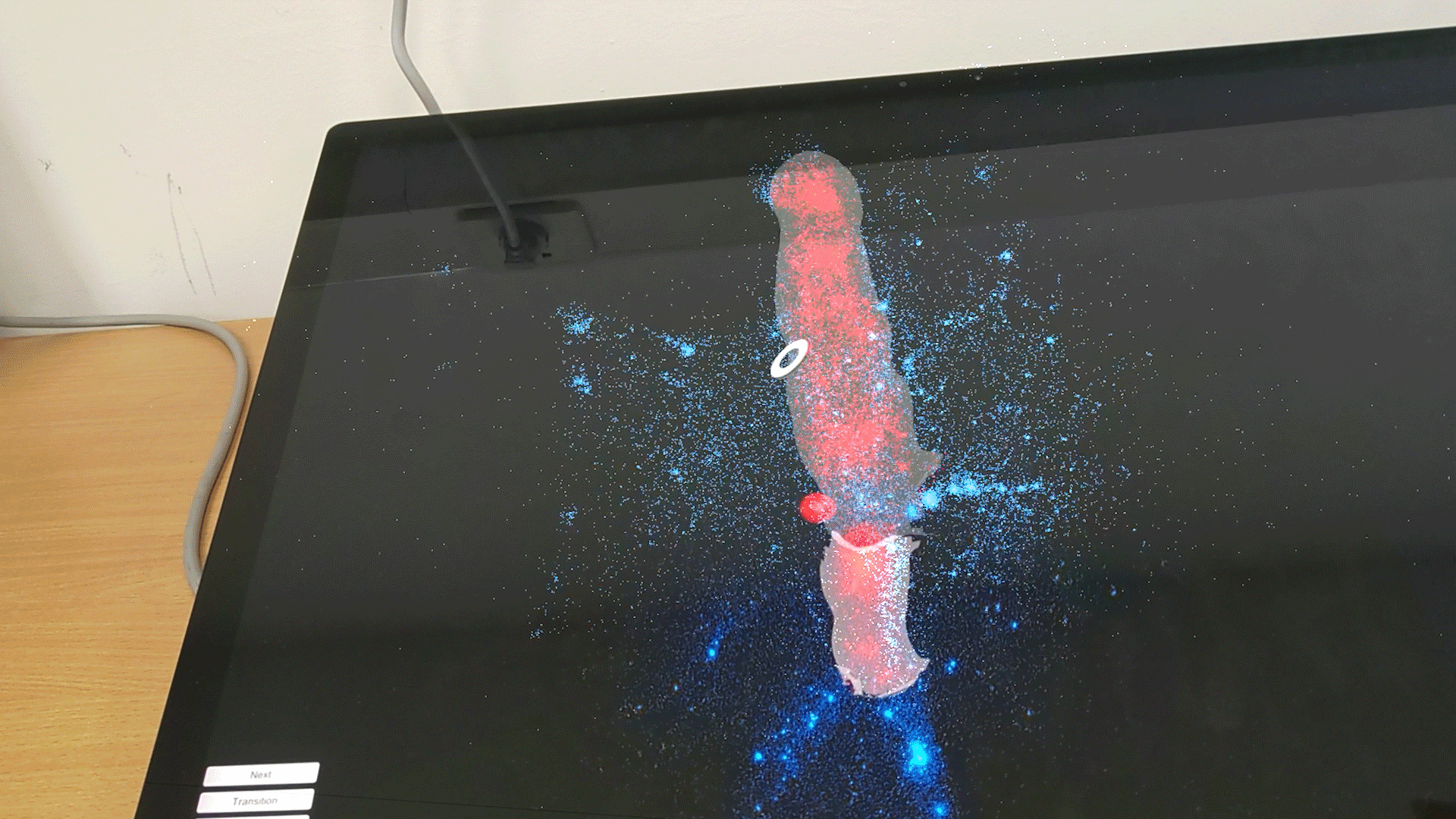}%
            \put(3,3){\textcolor{white}{(b)}}%
        \end{overpic}}

}
\caption{BrushLasso: (1) brushing the target points in mid-air and drawing a lasso on the surface, (b) the selection results.}
\label{fig:append:BrushLasso}
\end{figure}

With BrushLasso (\autoref{fig:append:BrushLasso}), users can brush target points in mid-air and encircle points on the surface through a single, seamless input.

\begin{figure}[t]
\centering  
\subfigure{%
        {\begin{overpic}[width=0.48\columnwidth]{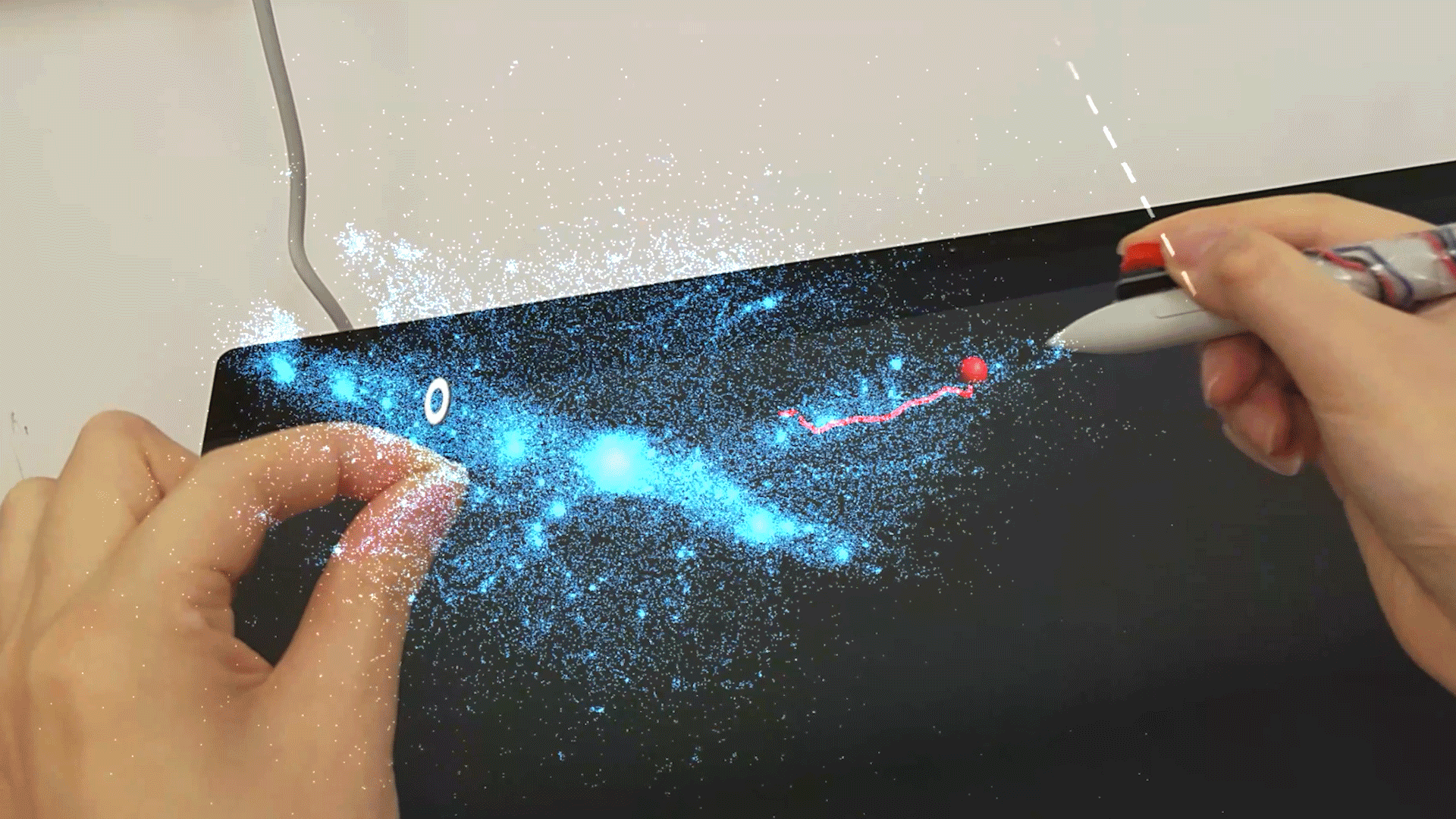}%
            \put(3,3){\textcolor{white}{(a)}}%
        \end{overpic}}
        }\hfill%
\subfigure{%
        {\begin{overpic}[width=0.48\columnwidth]{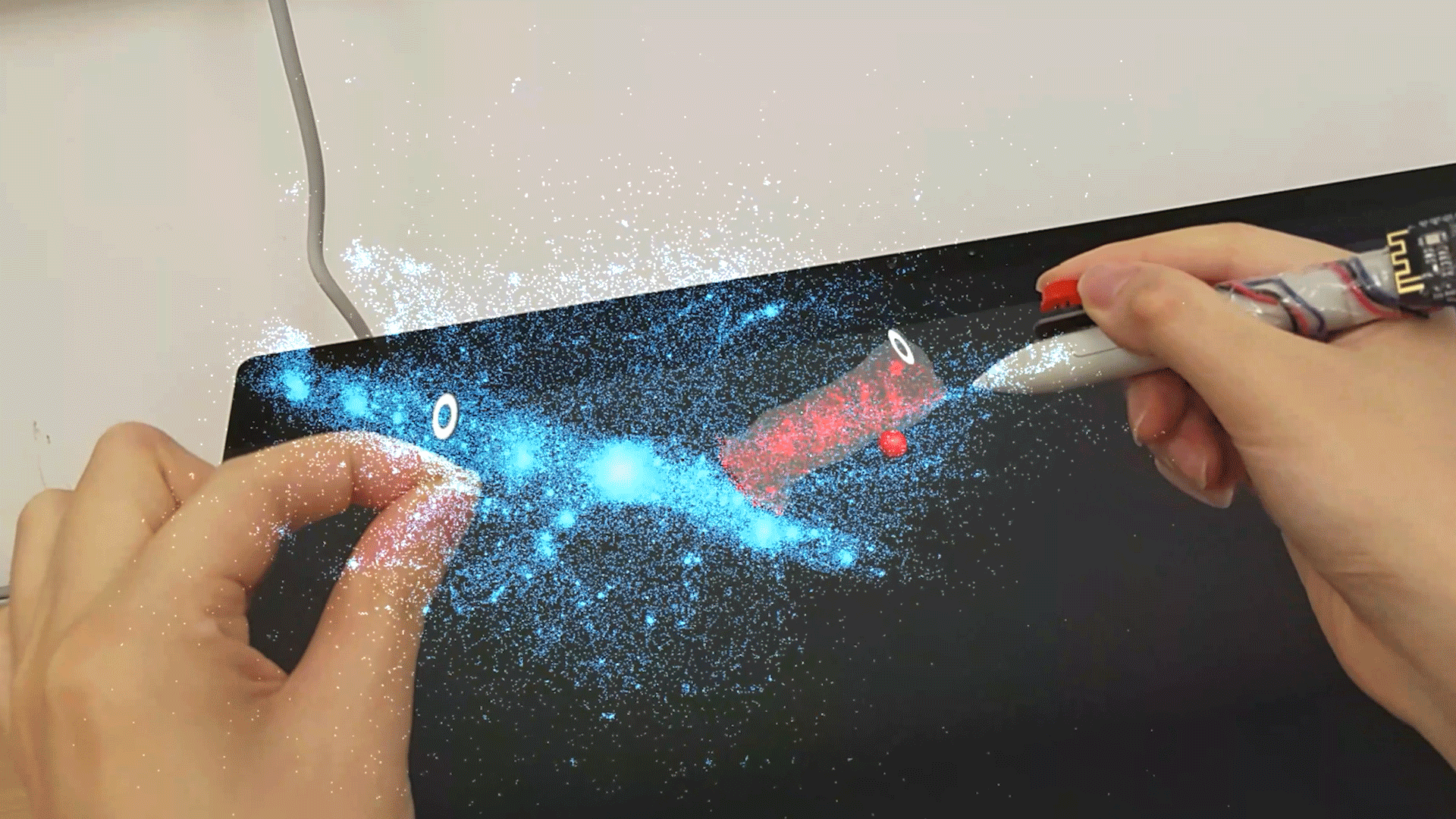}%
            \put(3,3){\textcolor{white}{(b)}}%
        \end{overpic}}
}
\caption{MeTACAST: (1) brushing the target points directly in mid-air, (b) the selection results.}
\label{fig:append:MeTACAST}
\end{figure}

\begin{figure}[t]
\centering  
\subfigure{%
        {\begin{overpic}[width=0.48\columnwidth]{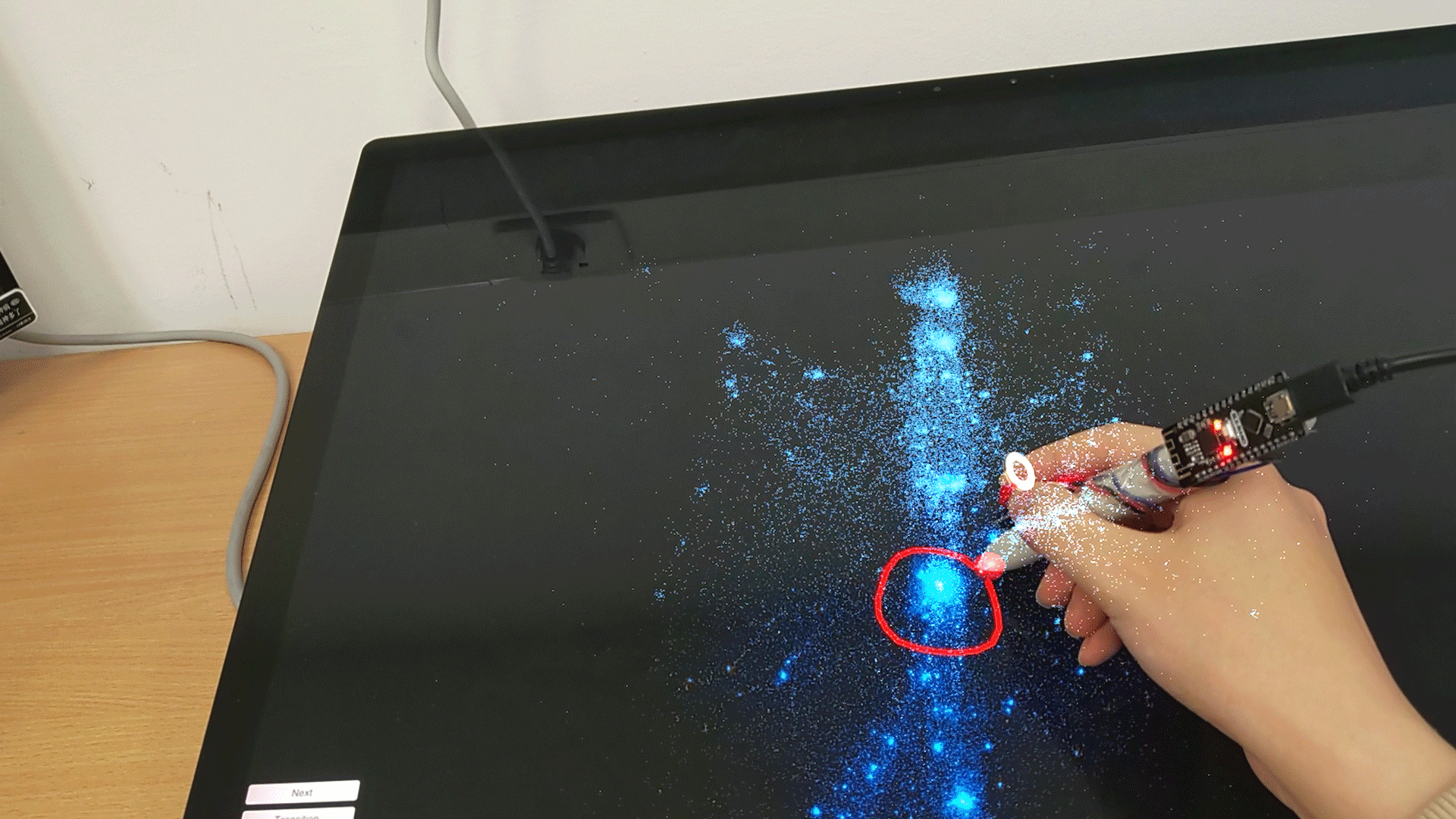}%
            \put(3,3){\textcolor{white}{(a)}}%
        \end{overpic}}
}\hfill%
\subfigure{%
        {\begin{overpic}[width=0.48\columnwidth]{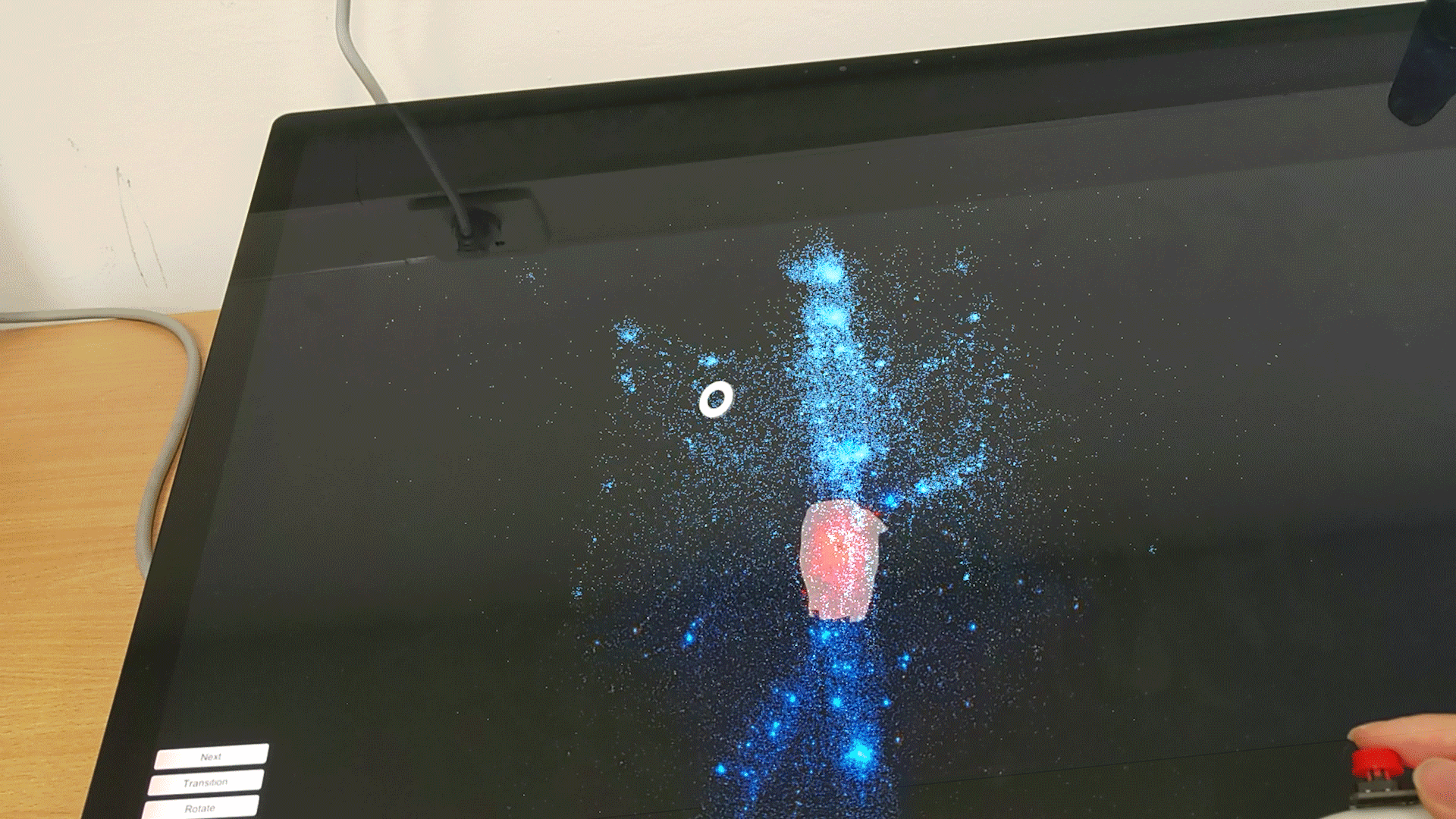}%
            \put(3,3){\textcolor{white}{(b)}}%
        \end{overpic}}
        }
\caption{CloudLasso: drawing a lasso around the target points on the surface, (b) the selection results.}
\label{fig:append:CloudLasso}
\end{figure}

In addition, users are able to brush target points only in mid-air with MeTABrush \cite{zhao:2023:metacast} (\autoref{fig:append:MeTACAST}), or draw a lasso around them on the surface with CloudLasso \cite{Yu:2012:ESA} (\autoref{fig:append:CloudLasso}).

\section{The CR visualization landscape}
\label{sec:discussion}

With \method we developed and studied a new CR environment that merges a monoscopic 2D surface with a stereoscopic 3D space for data visualization and exploration. As we demonstrated, \method facilitates comprehensive 3D spatial data analysis across various visual representations, scales, and data abstractions. This section extends beyond the design of \method to discuss the broader landscape of CR solutions for visualization. We believe this discussion will be useful for future CR designers in creating CR environments for spatial data visualizations and more.


\begin{figure*}[t]
    \centering
    \includegraphics[width=1\linewidth]{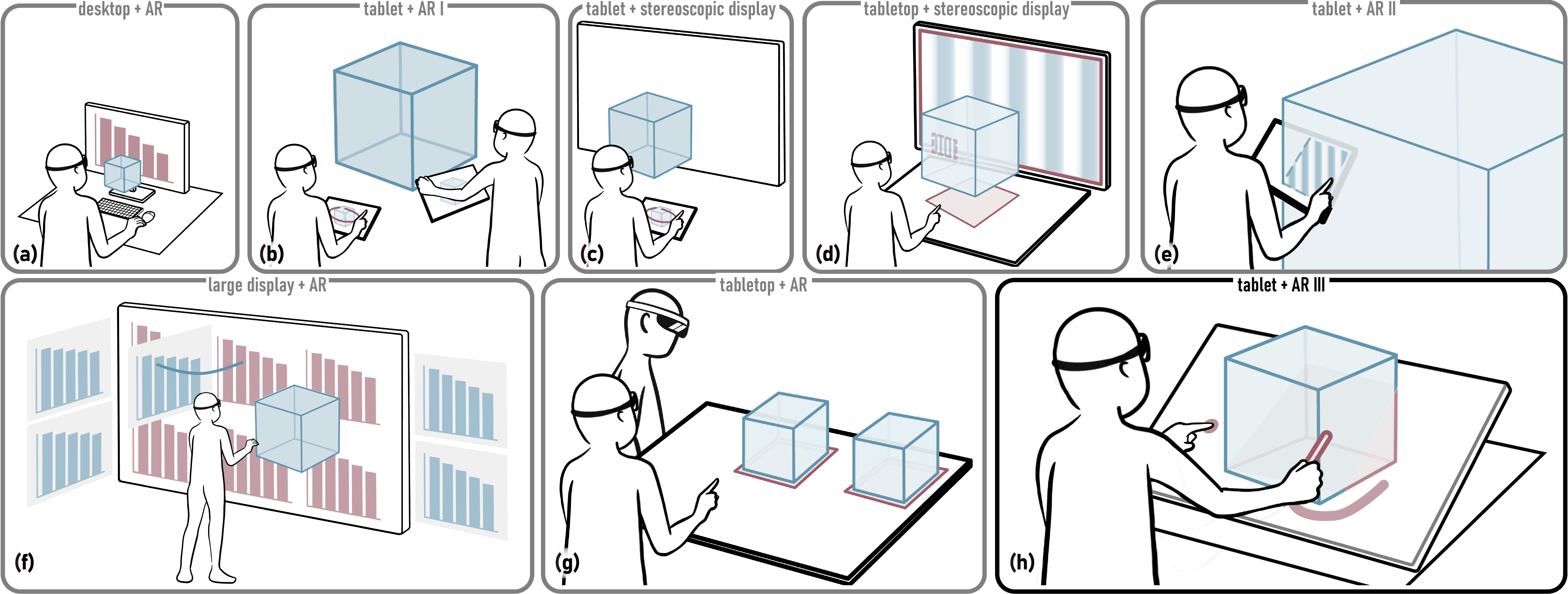}
    \caption{Existing CR environments for visualization: (a) monitor + AR\cite{Wang:2020:TUA, kijima:1997:transition, seraji:2022:hybridaxes}, (b) tablet + AR \cite{sereno:2019:SVD, Sereno:2022:HTT}, (c) tablet + Stereoscopic display\cite{lopez:2015:TAU}, (d) tablet + stereoscopic display\cite{coffey:2011:sliceWIM}, (e) tablet + AR  \cite{Bach:2018:THM, Sereno:2022:HTT, perelman:2022:visualtransitions}
    , (f) large display + AR\cite{reipschlager:2020:PAR},
    (g) tablet + AR\cite{Ens:2021:Uplift}, (h) tablet + AR\cite{Reipschlager:2019:DesignAR}, \method.}
    \label{fig:dis:environment}
    \vspace{-1ex}
\end{figure*}

\subsection{Motivation (why?)}

When considering a CR environment for visualization, the first factor to consider is the underlying motivation as it shapes the entire design.
A key advantage of CR is its ability to \textbf{augment visual representations or show additional information} for enhanced comprehension, data analysis support, or sense-making. The different CR designs we illustrated in \autoref{fig:dis:environment} all exemplify this capability. Additional information may include extra data dimensions, contextual detail, or 3D representations. Examples are an AR plus traditional PC-based data analysis tool for 3D data\cite{Wang:2020:TUA} (\autoref{fig:dis:environment}(a)), techniques for navigating volume data through overview and detail views\cite{coffey:2011:sliceWIM} (\autoref{fig:dis:environment}(d)), embedded AR visualizations for presenting multivariate data\cite{reipschlager:2020:PAR} (\autoref{fig:dis:environment}(f)), and SpatialTouch for presenting 3D visualization and multiple selected 2D slices (\autoref{fig:MedicalCuttingplane}(b)).
Furthermore, understanding the context of data and regions of interest allows users to focus on their interaction and may lead to more precise input, as they gain a clear understanding of the data's relative position and their interaction. In our elicitation study we observed that our participants' actions typically centered on regions of interest and crucial data points. CR environments enhance user engagement by offering \textbf{diverse interaction spaces} with additional information. Users can, \eg, accurately position a viewing window inside a medical structure on a touch table, while they observe both the overview and detail \cite{coffey:2011:sliceWIM} (\autoref{fig:dis:environment}(d)). In \method, users can precisely circle point data on a 2D surface, after assessing its spatial distribution in AR space (\autoref{fig:dis:environment}(h)).
Another key motivation for employing CR is \textbf{collaborative data analysis}. CR empowers multiple collaborators to simultaneously visualize and discuss shared visual content, especially for spatial 3D data. \autoref{fig:dis:environment}(b)  illustrates how users in a shared AR space can interact with a 3D visualization, each manipulating the data independently via a private tablet\cite{Sereno:2022:HTT}. Conversely, \autoref{fig:dis:environment}(g, f) show how display content is shared among collaborators, while AR HMDs provide private views for individual analysis \cite{Ens:2021:Uplift, reipschlager:2020:PAR}. These private views can be merged with the public one as required for sharing insights. A particularly innovative design is the use of a shared AR visualization, while participants have personal tablets directly inside the AR space with exact spatial alignment for individual analysis (\autoref{fig:dis:environment}(e)). 
Furthermore, while not extensively covered in previous work, we demonstrated in \method the use of \textbf{varied visual abstractions} both \belowS and \aboveS the surface to deepen users’ comprehension of data features. 
The surface + AR designs in \autoref{fig:dis:environment}(e, h) are particularly effective for this goal because their unified depiction presents spatial data alongside embedded representations, facilitating an intuitive understanding of complex data.

\subsection{Environment (where?) and dataset (what?)}
As illustrated in \autoref{fig:eli:RVC}, CR environments exploit various levels of virtuality to enhance user comprehension and facilitate task completion. Typically, they rely on monoscopic displays---mobile devices, surface interfaces, or desktop screens---, augmented with one or more stereoscopic AR/VR views (\autoref{fig:dis:environment}). The 2D displays offer multiple perspectives for data visualization, while 3D spaces are often dedicated to spatial data, resulting in a comprehensive data presentation.

For these configurations, it is crucial to create effective connections between contextual information and the visualized data attributes across multiple views as well as within spatial data. Various environments (\autoref{fig:dis:environment}(a--d)) employ approaches such as visual cues (\eg, color, highlighting), linked views, and animated transitions to achieve this coherence. Notably, designs \autoref{fig:dis:environment}(e--h) emphasize a strong connection and precise \textbf{spatial alignment} between AR and display content. Reipschl{\"a}ger and Dachselt\cite{Reipschlager:2019:DesignAR} previously defined three levels of spatial proximity: AR content positioned in front of or behind the display, AR content arranged close to or at the edge of the display, and AR content rendered with no spatial relation to the display. Analyzing these setups, \autoref{fig:dis:environment}(f, g) use the display as a frame of reference for AR content placement. Designs in \autoref{fig:dis:environment}(e, h)---including \method---leverage the spatial nature of 3D data that is rendered in AR space and position the 2D display inside the AR visualization. This approach is particularly effective for demonstrating the relationship between contents in two different spaces, as demonstrated in spatial selection tasks \cite{Sereno:2022:HTT} and the visualization of medical volumes using cutting planes in \method.

\section{Limitations of the specific hardware settings}
\label{sec:app:limitations}
In this section, we discuss the limitations caused by the specific hardware settings (Surface Studio and Hololens HMD) of \method.
First, \method requires high precision in the registration between the 2D surface and the 3D space to ensure accurate alignment between the two views. In our prototype, the HoloLens and Surface Studio run the program separately as two clients, with visualization and interaction states synchronized via the server. Similar to past work\cite{Reipschlager:2019:DesignAR}, after starting the application on Hololens, we need to manually align the two coordinate systems. Thus, although the view synchronization algorithm presented in \autoref{sec:CRenvironment} works accurately, as shown in the supplemental demo, it was time-consuming and difficult to achieve a high level of precision in the coordinate system registration. For future development, we recommend the researchers leverage automatic registration techniques with high precision or develop targeted CR display ecology with a unique coordinate system.
Second, the limited field of View (FOV) and the restricted gesture recognition area of the HoloLens significantly hinder the user's immersive experience. The narrow FOV confines the visualization to a small window in the user's vision, which breaks the sense of immersion as users are constantly reminded of the boundaries of the display. The constrained gesture recognition area means that users need to perform interactions within a small space in front of the body. It increases the cognitive load of users as they need to be conscious of keeping their hands within the detectable zone. We recommend that researchers leverage HMDs or glasses with large FOVs and extensive gesture recognition areas. For the FOV, current mainstream video see-through HMDs, such as the Oculus Quest 3, can sample the real world and map the results onto display devices with a large FOV. However, the video quality of VST HMDs is not yet sufficient for users to clearly observe the data visualizations on 2D surfaces in the real world. We believe that in the future, the image quality of VST HMDs will continue to improve, and issues such as video stream latency and image distortion will be resolved. For gesture recognition, external sensors such as Leap Motion can be introduced to facilitate gesture recognition and tracking.

\setlength{\lineskiplimit}{\lineskiplimitbackup}%

\section*{Figures license/copyright}
We as authors state that all of our figures in this appendix are and remain under our own personal copyright, with the permission to be used here. We make them available under the \href{https://creativecommons.org/licenses/by/4.0/}{Creative Commons At\-tri\-bu\-tion 4.0 International (\ccLogo\,\ccAttribution\ \mbox{CC BY 4.0})} license and share them at \href{https://osf.io/avxr9}{\texttt{osf.io/avxr9}}.

\end{document}